\documentclass[letterpaper]{JHEP3}
\usepackage{array}

\usepackage{nicefrac}
\usepackage{amsmath}

\usepackage{amssymb,amsfonts,mathrsfs}
\usepackage{graphicx}
\usepackage{epsfig}

\usepackage{amsfonts}

\newcommand{\Tr}{\mbox{Tr}}

\newcommand{\BB}{{\cal B}}
\newcommand{\CC}{{\cal C}}
\newcommand{\DD}{{\cal D}}
\newcommand{\EE}{{\cal E}}
\newcommand{\QQ}{{\cal Q}}
\newcommand{\II}{{\cal I}}
\newcommand{\suup}{1}
\newcommand{\sudown}{2}

\newcommand{\rep}[1]{\mathbf{#1}}

\global \long \def \NN{ \mathcal{N}}
\global \long \def \II{ \mathcal{I}}

\def\s{\sigma}
\def\g{\gamma}
\def\t{\tau}
\def\a{\alpha}
\def\b{\beta}
\def\e{\epsilon}

\def\h{\eta}

\def\half{{\frac12}}

\def\IC{\relax\hbox{$\inbar\kern-.3em{\rm C}$}}

\def\IC{{\bf C}}

\def\bea{\begin{eqnarray}}
\def\eea{\end{eqnarray}}
\def\be{\begin{equation}}
\def\ee{\end{equation}}
\def\ba{\begin{align}}
\def\ea{\end{align}}
\def\bse{\begin{subequations}}
\def\ese{\end{subequations}}
\def\1F1{{}_1\!F_1}
\def\2F0{{}_2\!F_0}

\def\a{\alpha}

\def\h3{$\textrm{H}_3^+$}

\def\IC{{\mathbb C}}

\def\Tr{{\rm Tr}}

\def\lbldef#1#2{\expandafter\gdef\csname #1\endcsname {#2}}

\def\href#1#2{#2}

\newcommand{\beq}{\begin{equation}}
\newcommand{\eeq}{\end{equation}}
\newcommand{\ber}{\begin{eqnarray}}
\newcommand{\eer}{\end{eqnarray}}

\def\be{\begin{eqnarray}}
\def\ee{\end{eqnarray}}

\newcommand{\Surface}{ {\cal C}}

\providecommand{\tabularnewline}{\\}

\def\({\left(}
\def\){\right)}
\def\[{\left[}
\def\]{\right]}
\def\<{\langle}
\def\>{\rangle}

\def\p{\Psi}

\def\q{\mathsf q}
\def\p{\mathsf p}
\def\t{\mathsf t}

\def\q{\rho}
\def\p{\sigma}
\def\t{\tau}

\title{Gauge Theories and Macdonald Polynomials}
\preprint{YITP-SB-11-30}
\author{Abhijit Gadde$^{\clubsuit\heartsuit}$, 
Leonardo Rastelli$^\clubsuit$, 
Shlomo S. Razamat$^{\clubsuit\diamondsuit}$, 
and Wenbin Yan$^\clubsuit$
\\
\\
$^\clubsuit$\it C.N. Yang Institute for Theoretical Physics,
\it Stony Brook, NY 11794-3840, USA\\
$^\heartsuit$\it California Institute of Technology,  Pasadena, CA 91125, USA\\
$^\diamondsuit$\it Institute for Advanced Study, Princeton, NJ 08540, USA}

\bigskip

\abstract{We study  the ${\cal N}=2$ four-dimensional superconformal index in various interesting limits, 
such that only states
 annihilated by more than one supercharge contribute.
 Extrapolating from the $SU(2)$ generalized quivers,
  which have a Lagrangian description,
we conjecture explicit  formulae 
for all $A$-type quivers of class ${\cal S}$, which in general do not have one.
We test our proposals against several expected dualities.
The index can always be interpreted as a correlator in a  two-dimensional topological theory, which 
we  identify in each limit as a certain deformation of two-dimensional Yang-Mills theory. 
The structure constants of the topological algebra  are diagonal in the basis of Macdonald polynomials of the holonomies.

 }


\begin{document}
\newpage

\begin{flushleft}
{\it{{  \large 
In  memory of Francis A. Dolan }}}
\end{flushleft}

\

\section{Introduction}

In recent years we have learnt many  surprising facts
about four-dimensional superconformal field theories (SCFTs), a subject
where disparate strands of mathematical physics  come together in  new beautiful ways.
For maximal supersymmetry there is an extraordinarily rich model,
${\cal N}=4$ super Yang-Mills, which is a unique theory given a choice
of gauge group.  Theories with ${\cal N}=2$ superconformal symmetry are even  richer.  
The vast majority of them do not have a weakly-coupled regime nor a conventional Lagrangian description.
This  fact, which may have been suspected since the early days of string dualities,  has taken
center stage after the  more explicit construction of the ${\cal N}=2$ superconformal theories of ``class ${\cal S}$''~\cite{Gaiotto:2009we, Gaiotto:2009hg},
most of which  are not Lagrangian.\footnote{Though very large, class ${\cal S}$  does not cover
the full space of  ${\cal N}=2$ SCFTs.  Counterexamples can be found  {\it e.g.} in~\cite{Gaiotto:2009we, Cecotti:2011rv}.
See \cite{Cecotti:2010fi,Cecotti:2011rv,Alim:2011ae} for the beginning of a classification program for ${\cal N}=2$ $4d$ SCFTs.}

Class ${\cal S}$ theories arise by 
compactification of the six-dimensional $(2, 0)$ theory on a punctured Riemann surface ${\cal C}$.
There is a growing dictionary relating 
four-dimensional quantities with quantities
associated to the surface ${\cal C}$. A basic entry of the dictionary identifies
the exactly marginal couplings
of the $4d$ theory with the complex structure moduli of ${\cal C}$.\footnote{On the other hand, the conformal factor of the metric on ${\cal C}$ is 
irrelevant (in the RG sense) and its memory   lost in the IR SCFT. See \cite{Anderson:2011cz} for a recent holographic check of this fact.}
According to the celebrated AGT conjecture  \cite{Alday:2009aq, Wyllard:2009hg,Mironov:2009by},  
the $4d$ partition functions  on the $\Omega$-background~\cite{Nekrasov:2002qd} and on $S^4$~\cite{Pestun:2007rz}  are computed by Liouville/Toda theory on ${\cal C}$.
An analogous relation exists between the $4d$ superconformal index~\cite{Kinney:2005ej,Romelsberger:2005eg} (which can also be viewed as a supersymmetric partition function on $S^3 \times S^1$) 
and topological quantum field theory (TQFT) on ${\cal C}$ \cite{Gadde:2009kb, Gadde:2010te, Gadde:2011ik}. 
In this paper we continue to explore this last relation.

The superconformal index is a simpler observable than the $S^4$ partition function,
and it should be a good starting point for a microscopic derivation of the $4d/2d$
dictionary from the $6d$ (2,0) theory.  Being coupling-independent, the index
   is computed by a topological  correlator on ${\cal C}$ \cite{Gadde:2009kb}, as opposed to a CFT correlator
as in the AGT correspondence. For the subset of class ${\cal S}$ theories that have a Lagrangian description,
 it can be easily evaluated in the free-field limit, unlike the $S^4$ partition function,
which is sensitive to non-perturbative physics and requires a sophisticated localization calculation \cite{Nekrasov:2002qd, Pestun:2007rz}.

Despite these simplifying features,  the index of  class ${\cal S}$ theories  is still  a very non-trivial observable
with remarkable mathematical structure.
First of all, there is no direct way to compute it  for  the non-Lagrangian  SCFTs, which by definition are  not continuously connected to free-field theories.\footnote{We should mention
that for ${\cal N}=1$ SCFTs obtained as IR points of an RG flow, a prescription to compute the index
 in terms of the UV field content and the charges of the anomaly
free R-symmetry  was put forward by Romelsberger~\cite{Romelsberger:2005eg,Romelsberger:2007ec} 
and recently revisited with more rigor in~\cite{Festuccia:2011ws}. Following the seminal 
work of Dolan and Osborn~\cite{Dolan:2008qi} there have been many checks and implications 
of this conjecture, see {\it e.g.}~\cite{Spiridonov:2008zr,Spiridonov:2009za,Spiridonov:2010hh,Gadde:2010en,Vartanov:2010xj,Spiridonov:2011hf}.}
An indirect route is to use the generalized S-dualities \cite{argyres-2007-0712, Gaiotto:2009we} that relate non-Lagrangian with Lagrangian theories.
This is the strategy  used in \cite{Gadde:2010te} to evaluate the index of the strongly-coupled SCFT with $E_6$
flavor symmetry~\cite{Minahan:1996fg}. In principle this procedure could be carried out recursively to find
the index of all the non-Lagrangian theories, but it suffers from two drawbacks: conceptually, one would
rather use the index to test  dualities, than assume  dualities to compute the index; and
practically, this program gets quickly too complicated to be useful. 

What one should aim for  is  a direct algorithm that applies to all class ${\cal S}$ theories --
one would like to identify and solve the $2d$ TQFT that computes the index.
The first step in this direction has been recently taken in \cite{Gadde:2011ik}: in a limit where
a single superconformal fugacity  is kept (out of the original three)  the $2d$ topological theory is  recognized as the zero-area limit of $q$-deformed Yang-Mills
theory.  In this paper we generalize this result 
 to a two-parameter slice $(q,t)$ of the three-dimensional fugacity space,
 which reduces to the limit considered in  \cite{Gadde:2011ik} for $t=q$.
  We give a fully explicit prescription
 to compute this limit of the index for the most general\footnote{In particular in \cite{Gadde:2011ik} certain overall normalization
 factors were determined only for theories with  special types of punctures. Here we fill this gap and work in complete generality.} $A$-type generalized quiver of class  ${\cal S}$.

 The principle that selects this particular fugacity
 slice  is supersymmetry enhancement, which  leads to simplifications.
 We study systematically the limits where the index receives contributions
only from states annihilated by more than one supercharge.
The $(q,t)$ slice is the most general  limit of this kind sensitive to
the  flavor fugacities associated  to the punctures. We also study another interesting slice $(Q, T)$, where the index  receives contribution
only from ``Coulomb-branch'' operators, which are flavor-neutral, so the flavor dependence is lost.

Let us  briefly outline the strategy of our computation.
It is essential to the 
construction of \cite{Gaiotto:2009we}
that a punctured  surface ${\cal C}$ 
can be obtained, usually in more than one way, by gluing three-punctured spheres (pairs of pants)  with cylinders.
 Different ways to decompose the same surface ${\cal C}$ correspond
to different S-duality frames of the same SCFT. The index is a topological quantity intrinsically
associated to ${\cal C}$ and independent of the  choice of pair-of-pants decomposition.
The three-punctured spheres correspond to isolated  $4d$ SCFTs, which are the elementary building blocks for
all other class ${\cal S}$ theories. To each puncture $I$ is associated a flavor symmetry $G_I \subset SU(k)$ (for  the $A$-type theories that we focus on).
The basic gluing operation joins two maximal punctures  (that is, both with $G=SU(k)$) and corresponds
to gauging the common $SU(k)$ symmetry. It is then sufficient to give an expression for the index associated
to the three-punctured spheres. Taking  for   illustration the SCFT associated to the sphere with three maximal punctures,  its index
is  some function
 ${\mathcal I}_{q,p,t}({\mathbf a_1},{\mathbf a_2},{\mathbf a_3})$
 of the three superconformal fugacities $(q,p,t)$ and of the $SU(k)$ flavor fugacities
$\mathbf{a}_I= (a_I^{1}, \dots a_I^{k})$ at each puncture. 
The topological nature of the index
is very constraining. It guarantees
the existence of a complete basis of functions
$\{ f^\lambda_{\{q,p,t\}}(\mathbf{a} )  \}$, where $\lambda$ labels  irreducible $SU(k)$ representations,
such that ${\mathcal I}_{q,p,t}({\mathbf a_1},{\mathbf a_2},{\mathbf a_3})$
 has the diagonal expansion
\be\label{res}
{\mathcal I}_{q,p,t}({\mathbf a_1},{\mathbf a_2},{\mathbf a_3}) =
\sum_{\lambda} C(q,p,t)_{\lambda\lambda\lambda}\;
f^\lambda_{\{q,p,t\}}({\mathbf a_1})\,f^\lambda_{\{q,p,t\}}({\mathbf a_2})\,
f^\lambda_{\{q,p,t\}}({\mathbf a_3})\,.
\ee 
The left-hand-side is a priori unknown --
except in the  $SU(2)$ case,
where it is the index of the free hypermultiplet theory. The idea is to focus on 
 the explicit  $SU(2)$ expression, write it  in the form (\ref{res}), and try to extrapolate the answer to general $SU(k)$.
This program succeeds  for the two-dimensional slice $(q,0,t)$ in fugacity space,
where  the functions $f_{q,t}^\lambda({\mathbf a})$ 
turn out to be closely related to a well-studied family of symmetric polynomials, the Macdonald polynomials,
which are defined for all root systems.  One is led to a compelling general conjecture that passes many tests. 
The extension 
 to the three-dimensional fugacity space must be possible but is not entirely  straightforward,
as the  basis  that diagonalizes the structure constants  is expected to consist of symmetric functions of an  elliptic
kind, which are less understood. We comment on this generalization in our conclusions.

The  TQFT that computes the index turns out to be a deformation of two-dimensional Yang-Mills theory.
For $t=q$ Macdonald polynomials reduce to Schur polynomials and
the TQFT can be related to the zero-area limit of  $2d$  $q$-deformed Yang-Mills theory~\cite{Aganagic:2004js,
Buffenoir:1994fh,Klimcik:1999kg},  which can also be viewed   as an analytic continuation
of Chern-Simons theory on ${\cal C} \times S^1$ away from integer rank. 
For the more general $(q,0,t)$ slice the TQFT appears to be
closely related to the ``refined'' version of Chern-Simons theory recently discussed in~\cite{Aganagic:2011sg}.

\

The rest of the paper is organized as follows. In section~\ref{indexsec} we review the definition
of  ${\mathcal N}=2$ superconformal index, paying special attention
to the parametrization of the superconformal fugacities. 
 In section~\ref{TQFTsec}  we review the 
 TQFT  structure of the index and describe the strategy of our computation.
 In section~\ref{limitssec} we define 
interesting limits of the index characterized by enhanced supersymmetry.
In section~\ref{betasec} we apply our strategy to the simplest limit, the fugacity slice $(0, 0, t)$. In this case
the functions that diagonalize the structure constants are proportional to   Hall-Littlewood polynomials.
We conjecture an explicit general expression and present a wide range 
of checks of our proposal. In particular we make contact with the results of \cite{Benvenuti:2010pq,Hanany:2010qu}:
we show that for genus-zero quivers the HL index is equivalent to the Hilbert series of the Higgs branch 
and test this equivalence in several examples.
 In section~\ref{qsec} we consider the slice  $(q, p, q)$ (which is in fact independent of $p$).
 This is 
limit of the index previously considered  in~\cite{Gadde:2011ik}.  The relevant symmetric functions
are proportional to Schur polynomials.
We generalize
the results of~\cite{Gadde:2011ik} and give explicit expressions valid for arbitrary punctures.
In section~\ref{higgssec} we combine and generalize the results 
of sections~\ref{betasec} and~\ref{qsec}. We consider the fugacity slice $(q, 0, t)$
and conjecture an expression for the index associated
to the general three-puncture sphere in terms of Macdonal polynomials,
equation (\ref{macgen}). This is our main result.
In section~\ref{coulsec} we consider
an index that counts Coulomb-branch operators.  Amusingly we are able to give a ``physics proof''
of Macdonald's constant term identities.
We conclude in section~\ref{discsec} with a discussion of our results and  speculations
on a few open questions.
Several appendices supplement the text with technical details and reference material.

\

{\it
As we were finalizing our draft  we  learnt about the tragic death of Francis A. Dolan.
Dolan and Osborn's beautiful  results on superconformal representation theory and on the index
were  a  direct influence and inspiration for our work. This paper
 intersects Francis'  interests 
in  so many ways, that we knew 
he would be one of our most demanding readers, and
aspired
to be up to   the standards he set. This paper is dedicated to his memory.
}

\

\section{The  ${\cal N}=2$ superconformal index}\label{indexsec}

The superconformal index~\cite{Kinney:2005ej} encodes the information about the protected spectrum of a SCFT
that can be obtained from representation theory alone. It is evaluated by a trace formula,
of the schematic  form
\be \label{schem}
{\cal I}(\mu_i)  =\Tr(-1)^F\, e^{-\mu_i T_i}\,e^{-\beta\, \delta }\, ,  \qquad \delta =  2\left\{{\mathcal Q},{\mathcal Q}^\dagger\right\}\,,
\ee
where ${\cal Q}$ is the supercharge ``with respect to which'' the index
is calculated and $\{ T_i \}$ a complete set of 
generators that commute with ${\cal Q}$ and with each other.
 The trace is over the states of the theory on $S^{d-1}$  (in the usual radial quantization).
By standard arguments,  states with $\delta \neq 0$ cancel pairwise, so the index counts
  states  with $\delta = 0$   (the  ``harmonic representatives'' of the cohomology classes of  ${\cal Q}$)
and  it is independent of $\beta$. From the index  one can  reconstruct the spectrum of short multiplets,
up to the equivalence relations that set to zero the combinations of short multiplets that may a priori recombine into long ones \cite{Kinney:2005ej}.
\begin{table}
  \begin{centering}
  \begin{tabular}{|c|c|c|c|c|c|c|}
  \hline
${\mathcal Q}$ &$SU(2)_1$&$SU(2)_2$&$SU(2)_R$&$U(1)_r$& $\delta$ & Commuting $\delta$s \tabularnewline
  \hline
  \hline
$  {\mathcal Q}_{{\suup }-}$ &$-\half$& $0$& $\;\;\;\half$&$\half$& $\delta_{{\suup }-}=  E-2j_1-2R-r$
 & $\delta_{{\sudown}{+}}$,\quad $\tilde \delta_{{\suup}\dot{+}}$,\quad $\tilde \delta_{{\suup}\dot{-}}$ \tabularnewline
  %
  \hline
$ {\mathcal Q}_{{\suup}+}$ &$\;\;\half$& $0$& $\;\;\;\half$&$\half$& $\delta_{{ \suup}+}=  E+2j_1-2R-r$
 & $\delta_{{\sudown}{-}}$,\quad $\tilde \delta_{{\suup}\dot{+}}$,\quad $\tilde \delta_{{\suup}\dot{-}}$ \tabularnewline 
 %
   \hline
$ {\mathcal Q}_{{\sudown}-}$ &$-\half$& $0$& $-\half$&$\half$& $ \delta_{{ \sudown}-}= E-2j_1+2R-r$
 & $\delta_{{\suup}{+}}$,\quad $\tilde \delta_{{\sudown}\dot{+}}$,\quad $\tilde \delta_{{\sudown}\dot{-}}$  \tabularnewline
  %
  \hline
  ${\mathcal Q}_{{\sudown}+}$  &$\;\;\half$& $0$& $-\half$&$\half$& $\delta_{{  \sudown}+}=  E+2j_1+2R-r$
 & $\delta_{{\suup}{-}}$,\quad $\tilde \delta_{{\sudown}\dot{+}}$,\quad $\tilde \delta_{{\sudown}\dot{-}}$ \tabularnewline 
 \hline
 %
 %
  \hline
  $\widetilde {\mathcal Q}_{{\suup}\dot{-}}$ &$0$&$-\half$&  $\;\;\;\half$&$-\half$& $\tilde \delta_{{\suup}\dot{-}} = E-2j_2-2R+r$
 & $\tilde \delta_{{\sudown}\dot{+}}$,\quad $\delta_{{\suup}{+}}$,\quad $\delta_{{\suup}{-}}$ \tabularnewline
  \hline
$\widetilde {\mathcal Q}_{{\suup}\dot{+}}$ &$0$&$\;\;\;\half$&  $\;\;\;\half$&$-\half$& $\tilde \delta_{{\suup}\dot{+}}=  E+2j_2-2R+r$
 & $\tilde \delta_{{\sudown}\dot{-}}$,\quad $\delta_{{\suup}{+}}$,\quad $\delta_{{\suup}{-}}$  \tabularnewline
 %
\hline
$\widetilde {\mathcal Q}_{{\sudown}\dot{-}}$ &$0$&$-\half$&  $-\half$&$-\half$& $\tilde \delta_{{ \sudown}\dot{-}}=  E-2j_2+2R+r$ 
& $\tilde \delta_{{\suup}\dot{+}}$,\quad $\delta_{{\sudown}{+}}$,\quad $\delta_{{\sudown}{-}}$  \tabularnewline
    \hline
$\widetilde {\mathcal Q}_{{ \sudown}\dot{+}}$ &$0$&$\;\;\;\half$&  $-\half$&$-\half$& $\tilde \delta_{{  \sudown}\dot{+}}=  E+2j_2+2R+r$
 & $\tilde \delta_{{\suup}\dot{-}}$,\quad $\delta_{{\sudown}{+}}$,\quad $\delta_{{\sudown}{-}}$  \tabularnewline
  \hline
  \end{tabular}
  \par  \end{centering}
  \caption{ \label{charges} For each supercharge ${\cal Q}$, we list its quantum numbers, the associated
  $\delta \equiv 2\left\{{\mathcal Q},{\mathcal Q}^\dagger\right\}$, and the other $\delta$s commuting with it.
  Here $I = \suup,\sudown$ are $SU(2)_R$ indices and
$\alpha = \pm$, $\dot \alpha = \pm$ Lorentz indices.  
 $E$  is the conformal dimension,  $(j_1, j_2)$ the Cartan generators of the $SU(2)_1 \otimes SU(2)_2$ isometry group, and $(R \, ,r)$, the Cartan generators 
  of  the  $SU(2)_R \otimes U(1)_r$ R-symmetry group.
  }\end{table}

 For   four-dimensional ${\cal N}=2$ SCFTs, which are non-chiral, different choices of ${\cal Q}$ lead to physically equivalent indices. 
The subalgebra of $SU(2,2|2)$ commuting with a single supercharge is $SU(1,1|2)$, which has rank three, so the ${\cal N}=2$  index depends on three superconformal fugacities.
In addition, there will  be fugacities associated with the flavor symmetries.
For definiteness
we choose ${\cal Q} = \widetilde{\cal Q}_{{\suup}\dot{-}}$. See table \ref{charges} for a summary of our notations. 
There are three supercharges commuting with $\widetilde {\mathcal Q}_{{\suup}\dot{-}}$
and $(\widetilde {\mathcal Q}_{{\suup}\dot{-}})^\dagger$:
\be
{\mathcal Q}_{{\suup}{-}}\, , \qquad
{\mathcal Q}_{{\suup}{+}}\, ,\qquad
\widetilde {\mathcal Q}_{{ \sudown}\dot{+}}
\, .
\ee 
A useful choice is to take as a basis for the Cartan generators of the commutant subalgebra  $SU(1,1|2)$
 the three $\delta$s of these supercharges. For each ${\cal Q}$ the associated $\delta$ is defined as
\be 
\delta \equiv 2\left\{{\mathcal Q},{\mathcal Q}^\dagger\right\} \,,
\ee 
and it has a non-negative real spectrum. We then write the index as 
\be\label{indA}
{\mathcal I}(\q, \p, \t)=\Tr(-1)^F\,
\q^{\half\delta_{{\suup}{-}}}\,
\p^{\half\delta_{{\suup}{+}}}\,
\t^{\half\tilde \delta_{{\sudown}\dot{+}}}\,
e^{-\beta\,\tilde \delta_{{\suup}\dot{-}}}\, .
\ee
 In table \ref{charges} we give the expressions of the $\delta$ charges in terms of  the more familiar
 Cartan generators $(E, j_1, j_2, R, r)$ of $SU(2,2 |2)$. This parametrization of the fugacities
  makes it easy to consider special limits with enhanced supersymmetry, which is our goal in this paper.\footnote{
Although at first glance the trace formula (\ref{indA}) may seem to depend symmetrically on four equivalent $\delta$s, this is not the case.
 The charge $\tilde \delta_{{\suup}\dot{-}}$
is special: the associated supercharge $\widetilde {\mathcal Q}_{{\suup}\dot{-}}$
 commutes with all the four $\delta$s,  but the supercharges associated to the other three $\delta$s do not.  This is then
 the index ``computed with respect to $\widetilde{\cal Q} _{{\suup}\dot{-}}$'', and it is independent of $\beta$, which we will usually omit.
  }
 Another very useful parametrization is in terms of fugacities $(p, q, t)$, related to $(\p, \q, \t)$ as
 \be
 p = \t \p\, , \quad  q = \t \q\, , \quad  t = \t^2 \,.
 \ee
This is the choice that corresponds to  the $(p,q)$ labels of the elliptic Gamma function~\cite{Spiridonov4}, and also, as we shall see, to the
 $(t,q)$ labels of Macdonald polynomials\footnote{Note that while the fugacities $(q, p)$ have exactly the same meaning in our previous papers~\cite{Gadde:2009kb,Gadde:2010en,Gadde:2010te}, 
 the  fugacity $t$  is different from the one introduced in \cite{Kinney:2005ej} and used in ~\cite{Gadde:2009kb,Gadde:2010en,Gadde:2010te}. We made this change of notations to make contact with the Macdonald literature, where $t$ has a canonical definition that one wishes to respect.}. In terms of these fugacities, the definition of the index reads
 \be\label{indqpt}
{\mathcal I}(p, q, t)& =& \Tr(-1)^F\,p^{\half\delta_{{\suup}{+}}} \,
q^{\half\delta_{{\suup}{-}}}\,
t^{R+r}\,
e^{-\beta'\,\tilde \delta_{{\suup}\dot{-}}}\,  
\ee
In appendix \ref{short} we review the shortening conditions of the ${\cal N}=2$ superconformal algebra and give the expression
of the index for the various short multiplets. Given the index of a SCFT, the formulae of appendix \ref{short} allow to determine its spectrum of short multiplets,
up to the usual recombination ambiguities (spelled out in section 5.2 of \cite{Gadde:2009dj}).

For a theory with a weakly-coupled description the index can be explicitly computed as a matrix integral,
\be
\label{index}
{\cal I}(V,\q,\p,\t) =\int\left[dU\right]\,
 \exp\left(\sum_{n=1}^\infty\frac{1}{n}\;\sum_{j} f^{\mathcal R_j}(\q^n,\p^n,\t^n)  \cdot \chi_{{\mathcal R_j}}(U^n,\,V^n)\right) \, .
\ee Here $U$ denotes an element of the gauge group,  with $\left[dU\right]$  the invariant Haar measure,
and $V$ an element   of the flavor group. 
 The  sum is over the different ${\cal N}=2$ supermultiplets appearing in the Lagrangian,
with  ${\mathcal R_j}$ the representation of the $j$-th multiplet under the flavor and  gauge groups  and $\chi_{\mathcal R_j}$
 the corresponding character. The Haar measure has the following property
\be
\int\left[dU\right]\,\prod_{j=1}^n \chi_{{\mathcal R_j}}(U)=\# {\text{of singlets in}\; } {\mathcal R_1}\otimes\dots\otimes {\mathcal R_n}\,.
\ee
The functions $f^{(j)}$ are the ``single-letter'' partition functions, $f^{(j)}= f^{{V}}$ or $f^{(j)} = f^{\half {H}}$ according to whether the $j$-th multiplet
is an ${\cal N}=2$ vector or ${\cal N}=2$ $\frac{1}{2}$-hypermultiplet.
\begin{table}
\begin{centering}
\begin{tabular}{|c|r|r|r|r|r|c|c|}
\hline
Letters & $  E$ & $j_1$ & $  j_2$ & $R$ & $r$ & $\mathcal{I}(\p, \q, \t)$   & $\mathcal{I}(p, q, t)$ \tabularnewline
  \hline
   \hline
$  \phi$ & $1$ & $0$ & $0$ & $0$ & $-1$ & $\p \q  $  & $pq/t$   \tabularnewline
  \hline
$  \lambda_{\suup\pm}$ & $  \frac{3}{2}$ & $  \pm  \frac{1}{2}$ & $0$ & $  \frac{1}{2}$ & $-  \frac{1}{2}$ & $-\p\t,\;-\q\t$  &  $-p$, $-q$ \tabularnewline
  \hline
$  \bar{\lambda}_{\suup\dot{+}}$  & $  \frac{3}{2}$ & $0$ & $  \frac{1}{2}$ & $  \frac{1}{2}$ & $  \frac{1}{2}$ & $-\t^2$ &  $-t$ \tabularnewline
  \hline
$  \bar{F}_{\dot{+}\dot{+}}$ & $2$ & $0$ & $1$ & $0$ & $0$ & $\p\q\t^2$  &  $pq$ \tabularnewline
  \hline
  $  \partial_{-\dot{+}}  \lambda_{\suup+}+  \partial_{+\dot{+}}  \lambda_{\suup-}=0$ & $  \frac{5}{2}$ & $0$ & $  \frac{1}{2}$ & $  \frac{1}{2}$ &
 $  -\frac{1}{2}$ & $\p\q\t^2$  & $pq$  \tabularnewline
  \hline
\hline
$q$ & $1$ & $0$ & $0$ & $  \frac{1}{2}$ & $0$ & $\t$  &  $\sqrt{t}$ \tabularnewline
  \hline
$  \bar{\psi}_{\dot{+}}$ & $  \frac{3}{2}$ & $0$ & $  \frac{1}{2}$ & $0$ & $-  \frac{1}{2}$ & $-\p\q\t$  & $-pq/\sqrt{t}$ \tabularnewline
  \hline
    \hline
$  \partial_{\pm\dot{+}}$ & $1$ & $  \pm  \frac{1}{2}$ & $  \frac{1}{2}$ & $0$ & $0$ & $\p\t,\;\q\t$   & $p$, $q$ \tabularnewline
\hline
\end{tabular}
\par  \end{centering}
  \caption{Contributions to the index from  ``single letters''.
  We denote by $(\phi, \bar \phi,  \lambda_{I,\alpha}, \bar\lambda_{I\,\dot \alpha},  F_{\alpha \beta}, \bar F_{\dot \alpha \dot \beta})$
 the components of the adjoint ${\cal N} = 2$ vector multiplet,  by $(q, \bar q, \psi_\alpha, \bar \psi_{\dot \alpha})$ the
 components  of the  ${\cal N} = 1$
chiral multiplet,  and by $\partial_{\alpha \dot \alpha}$ the spacetime derivatives.
}
\label{letters}
\end{table}
The ``single letters'' of an ${\mathcal N}=2$ gauge theory contributing to the index  obey $
 \tilde \delta_{{\suup}\dot{-}}=E - 2 j_2 - 2 R +r = 0$  
 and are enumerated in table \ref{letters}.
The first block of table~\ref{letters} shows the contributing letters from the ${\cal N} = 2$ vector multiplet,
including the equations of motion constraint. The second block shows the contributions from the half-hypermultiplet (or ${\cal N}= 1$ chiral multiplet). The
   last line shows the spacetime derivatives
contributing to the index. Since each field can be hit by an arbitrary number of derivatives, the derivatives give
a multiplicative contribution to the single-letter partition functions of the form
\be
\sum_{m=0}^\infty \sum_{n=0}^\infty (\q\t)^m\,(\p\t)^n  = \frac{1}{(1-\q\t)(1-\p\t)}\,.
\ee
The single-letter partition functions of the ${\cal N}=2$ vector and ${\cal N}=1$ chiral multiplets are thus given by
\be\label{letterpart}
f^{V}
&=&-\frac{\p\t}{1-\p\t}-\frac{\q\t}{1-\q\t}+\frac{\p\q-\t^2}{(1-\q\t)(1-\p\t)}\\
&=&-\frac{p}{1-p}-\frac{q}{1-q}+\frac{pq/t-t}{(1-q)(1-p)} \, ,\nonumber \\
f^{\half H}
&=&\frac{\t}{(1-\q\t)(1-\p\t)}(1-\q\p)
=\frac{\sqrt{t}- pq/\sqrt{t}}{(1-q)(1-p)}\,.
\ee
For general values of the three fugacities 
 the explicit expression for the index of a Lagrangian
theory is most elegantly
expressed~\cite{Dolan:2008qi} in terms of the elliptic Gamma functions (see~\cite{Spiridonov4}
for a nice review of these special functions). In this
paper however we consider reduced forms of the index and do not utilize the power of these special functions.
We  comment on the relation to  elliptic functions in the concluding section~\ref{discsec}.

\

\

\section{TQFT structure of the index }\label{TQFTsec}

Four-dimensional superconformal field theories  of ${\cal S}$ \cite{Gaiotto:2009we, Gaiotto:2009hg} arise
from partially-twisted compactification of the six-dimensional $(2,0)$ theory on a punctured Riemann surface ${\cal C}$.
The  complex-structure moduli of ${\cal C}$ are  identified with the exactly marginal 
couplings of the $4d$ SCFT, while the punctures are associated to flavor symmetries.  

Any punctured surface can be obtained, usually in more than one way, by gluing three-punctured spheres (pairs of pants) with cylinders.
The three-punctured spheres are then the elementary building blocks. They correspond to 
isolated $4d$ SCFTs with flavor symmetry $G_1 \otimes G_2 \otimes G_3$, where each
 factor $G_I$ is associated to one of the three punctures.\footnote{In this paper we  focus on class ${\cal S}$ theories that descend from the $(2,0)$ theory of type $A_{k-1}$.
Then the punctures are classified by the possible embeddings of $SU(2)$ into $SU(k)$ and $G_I \subset SU(k)$ is the commutant of the chosen embedding.}
The cylinders correspond to ${\cal N}=2$  vector multiplets, and the gluing operation 
 amounts to gauging a common $SU(k)$ symmetry of two punctures. The gluing parameter 
is interpreted as the complexified gauge coupling, with  zero coupling corresponding to an infinitely long cylinder --
a degeneration limit of the surface.
Different pairs-of-pants  decompositions of  the same  surface ${\cal C}$   correspond to different 
descriptions of the same SCFT, related by  generalized S-dualities. 

Since the index is independent of  the moduli, and is invariant under S-dualities, 
it is naturally viewed as a correlator in a $2d$ topological QFT living on ${\cal C}$ \cite{Gadde:2009kb}.
Let us review  how this works. 
We parametrize the index of a  three-punctured sphere
as $\mathcal{I}(\mathbf{a_1},\mathbf{a_2},\mathbf{a_3})$, where $\mathbf{a}_I$ are fugacities dual to the Cartan subgroup
of $G_I$: except in special cases these are {\it a priori} unknown functions.
On the other hand we can easily write down the 
``propagator'' associated to a cylinder,
\be
\eta(\mathbf{a}, \mathbf{b}) = \Delta({\mathbf a}) {\cal I}^V(\mathbf{a})\,  \delta(\mathbf{a}, \mathbf{b}^{-1})\,,
\ee
where  $\Delta({\mathbf a})$ is the Haar measure and ${\cal I}^V(\mathbf{a})$  the index of a vector multiplet, which is known explicitly.
The index of a generic theory of class ${\cal S}$ can be written in terms of the index of these elementary constituents.
As the simplest example, gluing two three-punctured spheres with one cylinder one obtains the
 index of a four-punctured sphere,
\be \label{following}
{\cal I} ( {\mathbf a_1},{\mathbf a_2},{\mathbf a}_3, {\mathbf{a}_4 })
&=& 
\oint [d \mathbf{a}] \oint [d \mathbf{b}] \;
 \mathcal{I}({\mathbf a_1},{\mathbf a_2},{\mathbf a})\, \eta(\mathbf{a}, \mathbf{b})   \,
\mathcal{I}({\mathbf b},{\mathbf a_3},{\mathbf a_4})\, \\
&=& \oint [d \mathbf{a}]\;\Delta({\mathbf a})\,\mathcal{I}({\mathbf a_1},{\mathbf a_2},{\mathbf a})\,{\mathcal I}^V({\mathbf a})\,
\mathcal{I}({\mathbf a}^{-1},{\mathbf a_3},{\mathbf a_4})\, , \nonumber
\ee  
where we have introduced the notation
\be
\oint [d \mathbf{a}] \equiv  \oint\prod_{i=1}^{k-1}\frac{da_i}{2\pi i a_i}   \,.
\ee
If we expand the index in a convenient basis of  functions 
$\{ f^\a({\mathbf a}) \}$,
labeled by $SU(k)$ representations $\{ \a \}$,\footnote{
For theories of type $A$,   $\{ f^\a({\mathbf a})\}$ are symmetric functions of their arguments, which are fugacities dual to the Cartan generators of $SU(k)$. More generally, for theories of type $D$ and $E$,
$\{ f^\a({\mathbf a})\}$ are invariant under the appropriate Weyl group.
}
we can associate to each  three-punctured sphere ``structure constants'' $C_{\a  \beta \gamma}$
and to each propagator a metric $\eta^{\a \beta}$,
\be
{\mathcal I}({\mathbf a},{\mathbf b},{\mathbf c}) & = &
\sum_{\alpha,\beta,\gamma} C_{\alpha\beta\gamma}
\,f^\alpha({\mathbf a})\,f^\beta({\mathbf b})\,f^\gamma({\mathbf c})\, \\
\eta^{\alpha \beta}& = &   
\oint [d \mathbf{a}] \oint [d \mathbf{b}] \;
 \eta(\mathbf{a}, \mathbf{b}) \,f^\alpha({\mathbf a})\,f^\beta({\mathbf b})\,.
\ee
 Invariance of the index
under the different ways to decompose the surface is tantamount of saying
that $C_{\a  \beta \gamma}$ and  $\eta^{\a \beta}$ define a two-dimensional topological QFT.\footnote{We are using this term
somewhat loosely. As axiomatized by Atiyah,  a TQFT is understood to have a finite-dimensional state-space,
while in our case the state-space will be infinite-dimensional. The best-understood example of a $2d$ topological theory with an infinite-dimensional
state-space is the zero-area limit of $2d$ Yang-Mills theory~\cite{Witten:1991we, Witten:1992xu} (see {\it e.g.}~\cite{Cordes:1994fc} for a comprehensive review). Happily, the $2d$ topological
theory associated to the index turns out to be closely related to  
$2d$ Yang-Mills.}
The crucial property is associativity, 
\be\label{asso}
C_{\a\beta\gamma}{C^{\gamma}}_{\delta\e}=C_{\a\delta\gamma}{C^{\gamma}}_{\beta\e}\, ,
\ee where indices are raised  with the metric $\eta^{\a\beta}$ and lowered with the inverse metric $\eta_{\a \beta}$.

It is very natural to choose the complete set of functions $\{ f^\a({\mathbf a}) \}$ to be orthonormal under the measure that appears in the propagator,
\be \label{orthrelnatural}
\oint [d \mathbf{a}] 
\; \Delta({\mathbf a})\, {\cal I}^V(\mathbf{a})\,  f^\a({\mathbf a})   f^\beta({\mathbf a}^{-1})   = \delta^{\alpha \beta}\,.
\ee
Then  
the metric $\eta^{\alpha \beta}$ is trivial, 
\be
\eta^{\alpha \beta}= \delta^{\alpha \beta} \, .
\ee
Condition (\ref{orthrelnatural}) still leaves
considerable freedom, as it is obeyed by infinitely many bases of functions related by orthogonal transformations. 
The real simplification arises if we can find
an {\it explicit} basis $\{ f^\a({\mathbf a})\}$, such that the structure constants
are {\it diagonal},  
\be\label{Ndiag}
C_{\a\beta\gamma}
 \neq 0\quad\to\quad \a=\beta=\gamma\,.
\ee 
Associativity  (\ref{asso}) is then automatic. 
For structure constants satisfying    (\ref{asso}) 
one can always find a basis in which they are diagonal: we  give a  detailed example of such a diagonalization procedure in appendix 
\ref{derHL} for the simplest limit of the index.  The challenge is to describe the basis in concrete form.

 In general the measure appearing in the propagator is complicated and no explicit set of orthonormal functions is available.
 We find it very useful to consider an ansatz 
 \be \label{clever}
 f^\a({\mathbf a})= {\cal K}({\mathbf a}) P^\a({\mathbf a})\, ,
 \ee
for some function ${\cal K}({\mathbf a})$. Clearly, from (\ref{orthrelnatural}), the functions $\{ P^\a({\mathbf a})\}$ are orthornormal under the new measure $ \hat \Delta({\mathbf a})$,
\be \label{orthrel}
\oint [d \mathbf{a}] \;
\hat \Delta({\mathbf a})\,  P^\a({\mathbf a})   P^\beta({\mathbf a^{-1}})   = \delta^{\alpha \beta}\,,\qquad \hat \Delta({\mathbf a}) \equiv {\mathcal I}^V({\mathbf a})\, {\mathcal K}({\mathbf a})^2\,\Delta({\mathbf a})\,.
\ee
(Recall that $\Delta({\mathbf a})$ always denotes the Haar measure).
The name of the game is to find a clever choice of ${\cal K}(\mathbf{a})$, for which   $\hat \Delta({\mathbf a})$ is a simple known measure
and the orthonormal basis $\{ P^\a({\mathbf a})\}$ an explicit set of functions such that (\ref{Ndiag}) holds.

Once the diagonal basis $\{ f^\alpha (\mathbf{a}) \}$ and the structure constant $C_{\alpha \alpha \alpha}$ are known, one can easily
calculate the index of the SCFT associated to the genus ${\frak g}$ surface with $s$ punctures. Such a surface can be built
by gluing $2 {\frak g} - 2 + s$ three-punctured spheres, so we have\footnote{Here for simplicity we are considering the case
where all external punctures  are ``maximal'', {\it i.e.} they have flavor symmetry $SU(k)$. The prescription for punctures with reduced symmetry is discussed in detail in sections  5, 6 and 7.} 
\be \label{Igs}
{\cal I}_{{\frak g}, s} (\mathbf{a}_1, \mathbf{a}_2, \dots ,\mathbf{a}_s ) = \sum_\alpha (C_{\alpha \alpha \alpha})^{2 {\frak g} - 2 + s} \,\prod_{I=1}^s f^\alpha (\mathbf{a}_I)\,.
\ee

In the rest of the paper we implement the following strategy. We start by considering
the generalized $SU(2)$ quivers. Since they have a Lagrangian description,   closed form expressions for the index (as matrix integrals) are readily available.
 We  then look for a basis of functions $\{ f^\alpha (\mathbf{a}) \}$
  that diagonalizes the structure constants. 
Fortunately, for each special limit
  of the index that we consider,  the diagonal basis is of  the form  (\ref{clever}), with
 $\{  P^\alpha (\mathbf{a}) \}$  well-known symmetric {\it polynomials}: Hall-Littlewood, Schur or Macdonald polynomials. (The first two are in fact special cases of Macdonald polynomials).
Since   these polynomials are defined for arbitrary rank, we can extrapolate from the $SU(2)$ case
and formulate compelling conjectures for the index of
 \textit{all} generalized quivers of type $A$. (This approach readily generalizes
 to all $ADE$ theories, but in this paper we focus on the $A$ series).
Finally we check our conjectures  against expected symmetry enhancements and
S-dualities.

\

\

\section{Limits of the index with additional supersymmetry}\label{limitssec}

We now consider several limits of the superconformal index, such that
the states contributing to it are
 annihilated by more than one supercharge. 
 Recall  that before taking any limit
 the index receives contributions only from states with 
 \be \label{basicdelta}
 \tilde \delta_{1 \dot -}= E-2 j_2 -2 R + r =0\,,
 \ee which are
 annihilated by $\widetilde {\cal Q}_{1 \dot -}$.
 We  tend to refer to the different limits of the index by the type of
symmetric  polynomials relevant
for their evaluation. In appendix~\ref{short} we discuss which short multiplets of the superconformal algebra
are counted by the index in each of these limits.

\

\subsubsection*{Macdonald index}

We first consider the  limit\footnote{An equivalent limit can be obtained by sending $\q$ to zero.} 
\be 
\p\to0\, , \qquad \q \, ,\t \; {\rm fixed}\,,
\ee
(which is the same as $p\to 0$ with $q$ and $t$ fixed). 
The limit is  well-defined since the power of $\p$ in the trace formula~\eqref{indA} is given by
$ \half\delta_{{\suup}{+}} \geq 0$.
The index is given by
\be
{\mathcal I}_M & = & \Tr_{M} (-1)^F\,           \q^{\half(E-2j_1-2R-r)}\, \t^{\half(E+2R+2j_2+r)}\, \\
 & = & \Tr_{M} (-1)^F\,           q^{\half(E-2j_1-2R-r)}\, t^{R+r}\, ,\nonumber
\ee 
where $\Tr_{M}$ denotes the trace restricted to states with $\delta_{{\suup}{+}}=E+2j_1-2R-r=0$. 
Such states are annihilated by ${\mathcal Q}_{{\suup}{+}}$. 
All in all ${\mathcal I}_M$ is a $\frac14$-BPS object receiving contributions only
from states annihilated by two supercharges, one chiral $({\mathcal Q}_{{\suup}{+}})$
and one anti-chiral $(\widetilde {\cal Q}_{1 \dot -})$.
The single letter partition functions of the half-hypermultiplet and the vector simplify to
\be 
f^{\half H}=\frac{\t}{1-\q\t}= \frac{\sqrt{t}}{1-q}\, ,\quad f^V=  \frac{-\t^2- \q\t}{1-\q\t}\,=\frac{-t-q}{1-q}\,.
\ee 

\

\subsubsection*{Hall-Littlewood index}

We further specialize the index by sending $\q \to 0$, 
so we are taking the limit 
\be
\p\, \to0\, ,\qquad \q \to 0\,, \qquad  \t \; {\rm fixed}\,,
\ee
(equivalently, $q\, ,p \to 0$ with $t$ fixed), which is   well-defined thanks to $\delta_{{\suup}{\pm}}\geq 0$.
The index is given by
\be
{\mathcal I}_{HL}=\Tr_{HL} (-1)^F\, \t^{\half(E+2R+2j_2+r)}=\Tr_{HL} (-1)^F\, \t^{2(E-R)} \,,
\ee
where $\Tr_{HL}$ denotes the trace restricted to states with  $\delta_{{\suup}{\pm}}=E\pm2j_1-2R-r=0$.
All in all, taking (\ref{basicdelta}) into account,
the states contributing to the index obey
\be
j_1 = 0 \, ,\qquad j_2 =r \,, \qquad E = 2R +r \,,
\ee
and are annihilated by three supercharges:  ${\mathcal Q}_{{\suup}{+}}$, ${\mathcal Q}_{{\suup}{-}}$ and  $\widetilde {\mathcal Q}_{{\suup}\dot{-}}$.

Let us consider the Hall-Littlewood  (HL) index for a theory with a Lagrangian description.
From table 2, we see that it gets contributions only from the scalar $q$ of the hypermultiplet and from
the fermion $\bar \lambda_{1\dot{+}}$ of the vector multiplet.
The single letter partition function of the half-hypermultiplet and the vector multiplet is then
\be 
f^{\half H}=\t,\quad f^V=-\t^2\,.
\ee
Remarkably, for generalized quivers with a sphere topology the computation of the HL index
 is  equivalent to the computation of the partition 
function over the Higgs branch discussed in~\cite{Benvenuti:2010pq,Hanany:2010qu} (the 
Hilbert series of the Higgs branch).\footnote{
A relation of a similar limit of the ${\mathcal N}=1$ index with the
counting problems discussed in~\cite{Gray:2008yu,Hanany:2008kn} was mentioned in~\cite{Spiridonov:2009za}.
We thank V. Spiridonov for bringing this reference to our attention.
}
This can be shown as follows. To compute the partition function of~\cite{Benvenuti:2010pq,Hanany:2010qu} 
for the Higgs branch of an
${\mathcal N}=2$  gauge theory one counts all the possible gauge invariant operators 
built from the scalar components of the hypermultiplets taking into account the 
F-term superpotential constraints. In an ${\mathcal N}=2$ gauge theory with $M$ $SU(2)$ gauge factors
the superpotential takes the form
\be
W=\sum_{i=1}^M \sum_{\alpha\in \{i\}} Q^{(\alpha)}_{a_i a_k a_l}\,{\Phi^{a_i}}_{b_i}\, Q^{(\alpha)\,b_i a_k a_l}\,, 
\ee where the summation over $i$ is over the gauged groups. The set $\{i\}$ is the set of (at most two) trifundamental 
hypermultiplets
transforming non-trivially under gauge group $i$. 
 The F-term constraints then read
\be \label{F}
Q^{(\alpha_1)}_{a_i a_k a_l}\,Q^{(\alpha_1)b_i a_k a_l}+Q^{(\alpha_2)}_{a_i a_m a_n}\,Q^{(\alpha_2)b_i a_m a_n}=0\,.
\ee If the  quiver diagram does not have loops, {\it i.e.} the corresponding
Riemann surface has a topology of a sphere, this is a set of $M$ independent constraints.
It then follows that the computation of this partition function is the same as the computation of the index.
Indeed, one associates a fugacity $\t$ for each scalar component of $Q$. The constraint (\ref{F}) is quadratic
in $Q$ and is in the adjoint representation of the gauge group. It is implemented 
by multiplying the unconstrained partition function with the following factor~\cite{Benvenuti:2010pq,Hanany:2010qu},
\be
\exp\left[-\sum_{n=1}^\infty\frac{1}{n}\t^{2n}\left(a^{2n}_i+a^{-2n}_i+1\right)\right]=
(1-\t^2)(1-\t^2\, a^2_i)(1-\t^2\, a^{-2}_i)\,.
\ee This factor is the same as the index of the letter $\bar \lambda_{1\dot{+}}$. Thus, one can think 
of the letter $\bar \lambda_{1\dot{+}}$ in the calculation of the index as playing the same role as the superpotential constraint in the calculation of the Higgs partition function!
This logic can be extended to higher-rank theories, where not all the building blocks have Lagrangian description, but
the Higgs branch can still be described in terms of operators obeying certain constraints. This concludes
the argument that the HL index is the same as the Higgs partition function for theories with sphere topology.
Our derivation also makes it clear that this correspondence fails for higher-genus theories. 

In~\cite{Benvenuti:2010pq}  non-trivial very explicit expressions for the Higgs branch partition 
function of the SCFTs with exceptional  flavor symmetry groups~\cite{Minahan:1996fg,Minahan:1996cj} were conjectured.
We will see that they are exactly reproduced by the HL index. 

\

\subsubsection*{Schur index}

The Schur index is defined by specializing the fugacities to
 $\q=\t$  with $\p$ arbitrary (equivalently $q = t$ with $p$ arbitrary).
  It reads
\be \label{schurTr}
{\mathcal I}_S=\Tr (-1)^F\, \p^{\half(E+2j_1-2R-r)}\,\q^{E-j_1+j_2}\,e^{-\beta(E-2j_2-2R+r)}  \,.
\ee 
By construction,  all 
charges  in the trace formula commute with the supercharge $\widetilde {\mathcal Q}_{{\suup}\dot{-}}$ ``with respect to which'' the index is evaluated.
From table 1, we observe that the charges in (\ref{schurTr})  {\it also}
 commute with  ${\mathcal Q}_{{\suup}{+}}$.
Thus the index receives contributions from states with $\delta_{{\suup}{+}}=
 \tilde \delta_{{\suup}\dot{-}}=0$ (the intersection of the cohomologies of  ${\mathcal Q}_{{\suup}{+}}$
 and of  $\widetilde {\mathcal Q}_{{\suup}\dot{-}}$) 
and it is independent of {\textit{both}} $\p$ and $\beta$. 
We can then write
\be
{\mathcal I}_S=\Tr (-1)^F\, \q^{2(E-R)} =  \Tr (-1)^F\, q^{E-R} \,.
\ee 
The Schur index can also be obtained as a special case of the Macdonald index  by setting $\q=\t$ (equivalently $q=t$);  we have just seen that
for $\q=\t$ the index becomes independent of $\p$ so the limit $\p\to0$ that we take to obtain the Macdonald index is immaterial. 

The single letter partition functions of the half-hypermultiplets and the vector multiplet are given by
\be 
f^{\half H}=\frac{\q}{1-\q^2}=\frac{\sqrt{q}}{1-q}\,,\quad f^V=\frac{-2\q^2}{1-\q^2}= \frac{-2q}{1-q}\,.
\ee 

The Schur index  is the same as the  index studied in \cite{Gadde:2011ik}, where we referred to it as the {\it reduced index}.

\

\subsubsection*{Coulomb-branch index}

Finally we consider the limit 
\be
\t\to0\,,  \qquad \q \, ,\p \; {\rm fixed}\, ,
\ee 
which is well-defined thanks to $\tilde \delta_{2 \dot +}  \geq 0$.
The trace formula becomes
\be
{\mathcal I}_C=
\Tr_{C}(-1)^F\,\p^{\half(E+2j_1-2R-r)}\,\q^{\half(E-2j_1-2R-r)}\,e^{-\beta(E-2j_2-2R+r)}\,,
\ee
where $ \Tr_{C}$ denotes the trace over the  states with 
$\tilde \delta_{2 \dot +} = E+2j_2+2R+r=0$, which are annihilated by $\widetilde {\mathcal Q}_{{\sudown}\dot{+}}$.
All in all, the index  gets contributions from states annihilated by two antichiral supercharges, $\widetilde {\mathcal Q}_{{\suup}\dot{-}}$ and
$\widetilde {\mathcal Q}_{{\sudown}\dot{+}}$. 

In this limit the single-letter partition function of the half-hypermultiplet and the vector multiplet are
\be \label{coulsinglet}
f^{\half H}=0,\qquad f^V=\p\q \equiv T\,.
\ee 
From the viewpoint of the the single-letter partition functions one can take an interesting less restrictive limit,
\be\label{CMlimit}
\t\,,\p\to0 \,, \qquad \q\to\infty\, \qquad {\rm with}\; \;\;Q\equiv \t  \q \;\;{\rm and} \;\; T\equiv \p\q\, \;\;{\rm fixed}.
\ee
In this limit we have
\be 
f^{\half H}=0,\qquad f^V=\frac{T-Q}{1-Q}\,.
\ee
We recover (\ref{coulsinglet}) for $Q \to 0$.
 In terms of the new fugacities $Q$ and $T$ the index reads 
\be 
{\mathcal I}_{CM}=
\Tr_{CM}(-1)^F\,
T^{\half(E+2j_1-2R-r)}\,Q^{\half(E+2j_2+2R+r)}\,,
\ee where $\Tr_{CM}$ denotes the trace restricted to states satisfying $E+2j_1+r=0$.
This index is well-defined 
for Lagrangian theories and for theories related to them by dualities.

\

We now describe the explicit  evaluation of  these special limits of the
 index for the SCFTs of class ${\cal S}$.

\

\

\section{Hall-Littlewood index}\label{betasec}

We begin with the Hall-Littlewood index,
\be
{\mathcal I}_{HL}(\t)=\Tr_{HL}(-1)^F\,\t^{2E-2R}\, ,
\ee
 where $\Tr_{HL}$ denotes the trace restricted to states with  $j_1=0$ and $E-2R-r=0$.
This is the limit that leads to the greatest simplifications.

\subsection{$SU(2)$  quivers}

Let us start from the  $SU(2)$ generalized quivers, for which the basic building blocks
are known explicitly. There is only one type of non-trivial puncture, the maximal puncture with  $SU(2)$ flavor symmetry.
The SCFT corresponding the three-punctured sphere, denoted by $T_2$ in \cite{Gaiotto:2009we},
is the  theory of free hypermultiplets
in the trifundamental representation
of $SU(2)$.  Its index is immediately evaluated,
\be
{\mathcal I} (a, b, c)=PE\left[\t \chi_1(a)\chi_1(b)\chi_1(c)\right]_{a,b,c,\t}
= \frac{1}{\prod_{s_a,s_b,s_c=\pm1}(1-\t\,a^{s_a}\,b^{s_b}\,c^{s_c})}\,,
\ee  where the fugacities $a$, $b$, and $c$ 
label the Cartans of the three $SU(2)$ flavor groups. The plethystic exponent $PE$ is defined as
\be
PE\left[f(x_i)\right]_{x_i}\equiv \exp\left(\sum_{n=1}^\infty \frac{1}{n}f(x_i^n)\right)\,.
\ee We will often omit the subscript $x_i$ in the expressions for $PE[\dots]$.  $\chi_1(a)$ is the character of fundamental representation of $SU(2)$. More generally the $SU(2)$ Schur polynomials
$\chi_\lambda$ are given by
\be
\chi_\lambda(a)=\frac{a^{-1-\lambda}-a^{1+\lambda}}{a^{-1}-a}\,.
\ee 
The propagator $\eta(a, b)$  is also easily evaluated:
\be
\eta(a, b) = \Delta(a) {\cal I}^V(a) \delta(a, b^{-1}) \, ,
\ee
where ${\cal I}^V(a)$ is the index of the vector multiplet,
\be
{\cal I}^V (a)= PE[-\t^2 \chi_2(a)]_{a, \t}=(1-\t^2)\,(1-\t^2\, a^2)\,(1-\t^2\, a^{-2})   \, ,
\ee
and $\Delta(a)$ the $SU(2)$ Haar measure,
\be 
\Delta(a) = \half(1-a^2)(1-\frac1{a^2})\,.
\ee
Following the strategy outlined in section 3, we look for a complete set of functions $\{ f^\lambda(a) \}$
orthonormal under the propagator measure  such that the structure constants are diagonal,
\be
{\mathcal I} (a, b, c) = \sum_{\lambda =0}^\infty C_{\lambda \lambda \lambda}  \, f^\lambda (a) f^\lambda (b) f^\lambda (c)\,. 
\ee
We describe this calculation in appendix A. 
We find the remarkable result
\be
 f^\lambda(a)&  =  &  {\cal K}(a) \;P_{HL}^{\lambda}(a,a^{-1}|\t) ,\\
C_{\lambda \lambda \lambda} &  = &\frac{  \sqrt{1-\t^2}\,(1 + \t^2) }{P_{HL}^{\lambda}(\t,\t^{-1}|\t)}\,.
\ee
Here $P_{HL}^\lambda$ are the $SU(2)$ Hall-Littlewood polynomials, 
\be \label{SU2HL}
P_{HL}^{\lambda}(a,a^{-1}|\t) & = & \chi_{\lambda}(a)-\t^2\chi_{\lambda-2}(a) \quad{\rm for}\; \lambda \geq 1 \,, \quad P^{\lambda=0}_{HL}(a,a^{-1}|\t) = \sqrt{1 + t^2} \,, 
\ee
which are orthonormal under the measure 
\be
\hat \Delta(a) = \Delta_{HL} (a) =   \frac{1}{2} \frac{(1-a^2)(1-a^{-2})}{(1-\t^2 a^2)(1-\t^2 a^{-2})}\,.
\ee
The requirement that $\{ f^\lambda(a) \}$ be orthonormal under the propagator measure $\Delta(a) {\cal I}^V(a)$  fixes the prefactor ${\cal K}(a)$,
\be
{\cal K}(a) =\left( \frac{\Delta_{HL}(a)}{\Delta(a) {\cal I}^V(a)} \right)^{\frac{1}{2}} = \frac{1}{\sqrt{1-\t^2}} \frac{1}{(1-\t^2 a^2)(1-\t^2 a^{-2})}\,.
\ee

We can now immediately write
down an explicit formula for the index of any  generalized $SU(2)$ quiver associated to a genus ${\frak g}$ Riemann surface with $s$ punctures.
From (\ref{Igs}), 
\be\label{compsu2}
{\mathcal I}_{{\frak g},s}(a_1, a_2, \dots, a_s)&=&\left(1-\t^2\right)^{{\frak g}-1}\left(1+\t^2\right)^{2{\frak g}-2+s}\cdot\\
&&
\sum_{\lambda=0}^\infty  \frac{1}{\left[P_{HL}^\lambda(\t,\t^{-1}|\;\t)\right]^{2{\frak g}-2+s}}
\prod_{I=1}^s \frac{P_{HL}^\lambda(a_I,a_I^{-1}|\;\t)}{(1-\t^2 a_I^2)(1-\t^2 a_I^{-2})}
\,.\nonumber
\ee 
In particular for genus ${\frak g}$ with no punctures the sum over the $SU(2)$ irreducible representations in~\eqref{compsu2}
can be explicitly performed and 
one gets
\be 
{\mathcal I}_{\frak g}^{(2)}=\frac{\left(1-\t ^2\right)^{\frak g -1}
 \left(\t ^{2\frak g-2 } +\left(1+\t ^2\right)^{\frak g-1 }
 \left(1-\t ^{2 \frak g-2 }\right)\right)}{ 1-\t ^{2 \frak g-2 } }\,.
\ee 

We observe that setting a flavor fugacity $a=\t$  we ``close'' the corresponding puncture. For example we can go from the three-punctured sphere
to the two-punctured sphere (=cylinder),
\be
{\mathcal I}(a_1,a_2,\t)\sim \sum_\lambda P^\lambda_{HL}(a_1,a_1^{-1}|\t)P^\lambda_{HL}(a_2,a_2^{-1}|\t) = \eta(a_1, a_2)\,.
\ee 
(There is an overall divergent proportionality  factor).
This procedure of (partially) closing punctures by trading (some of) the flavor fugacities
with $\t$ plays an important role, as it will allow us to construct the 
index for theories with arbitrary types of punctures. For $SU(k)$ theories the punctures are classified by 
the different embeddings of $SU(2)$ inside $SU(k)$~\cite{Gaiotto:2009we,Gaiotto:2009hg},
which are conveniently labelled by auxiliary
Young diagrams with $k$ boxes. For $SU(2)$ we get only two possibilities: (i) a row with two boxes 
corresponding to the ``maximal'' puncture with $SU(2)$ flavor symmetry, (ii) a column with two boxes 
corresponding to the absence of a puncture. 
For higher-rank theories the space of possibilities
will be more interesting.

\

\

\subsection{Higher rank: preliminaries}

For higher-rank quivers the situation is more complicated since the basic building blocks are given
by strongly-interacting SCFTs for which direct computations are not possible. 
However, the expressions that we obtained for the index of the $SU(2)$ quivers
can be naturally extrapolated to higher rank. The basic conjecture is that the 
set of functions $\{ f^\alpha(\mathbf{a}) \}$ that diagonalize the structure 
constants are related to  Hall-Littlewood polynomials for higher-rank as well.

The Hall-Littlewood (HL) polynomials associated to $U(k)$ are a set of orthogonal polynomials labeled
by Young diagrams with at most $k$ rows, $\lambda=(\lambda_1,\dots,\lambda_k)$, $\lambda_j \geq \lambda_{j+1}$. 
They are given by~\cite{Mac}
\be\label{HLdef}
P_{HL}^{\lambda}(x_1,\dots,x_k|\;\t)=
{\mathcal N}_\lambda(\t)\;\sum_{\sigma \in S_k}
x_{\sigma(1)}^{\lambda_1} \dots x_{\sigma(k)}^{\lambda_k}
\prod_{i<j}   \frac{  x_{\sigma(i)}-\t^2 x_{\sigma(j)} } {x_{\sigma(i)}-x_{\sigma(j)}}\,,
\ee 
and they are orthonormal
under the  measure
\be
\Delta_{HL}=\frac{1}{k!}\,\prod_{i\neq j}\frac{1-x_i/x_j}{1-\t^2x_i/x_j}\, .
\ee
The normalization ${\mathcal N}_\lambda(t)$ 
is given by
\be\label{normHL1}
{\mathcal N}^{-2}_{\lambda_1,...\lambda_k}(\t)=\prod_{i=0}^\infty \prod_{j=1}^{m(i)}\,
 \left(\frac{1-\t^{2j}}{1-\t^2}\right)\, ,
\ee where $m(i)$ is the number of rows in the Young diagram $\lambda=(\lambda_1,\dots,\lambda_k)$ of length $i$.
For $SU(k)$ groups we take Young diagrams with $\lambda_k=0$ and the product of $x_k$ in~\eqref{HLdef} is constrained as $\prod_{i=1}^kx_k=1$.

Let us also quote from the outset the  expression for the $SU(k)$ propagator,
 \be
 \eta (\mathbf{a}, \mathbf{b}^{-1}) = \Delta (\mathbf{a})  {\mathcal I}^{V}(\mathbf{a}) \delta(\mathbf{a}, \mathbf{b}^{-1}) \, ,
 \ee
 where $\Delta(\mathbf{a})$ is the $SU(k)$ Haar measure,
 \be  \label{Haarexpression}
 \Delta(\mathbf{a}) = \frac{1}{k!} \prod_{i \neq j}\left(1-\frac{a_i}{a_j} \right)\, , \qquad  \prod_{i}^k a_i = 1 \, ,
 \ee
and   ${\mathcal I}^{V}(\mathbf{a})$ the vector multiplet index,
\be
{\mathcal I}^{V}=\frac{1}{1-\t^2}\prod_{j, i=1}^{k}(1-\t^2{a}_j/{a}_i)\,.
\ee

\subsection{$SU(3)$ quivers --  the $E_6$ SCFT}

We now focus on the $SU(3)$  theories. There are two kinds of non-trivial punctures: the maximal puncture, associated to the Young diagram $(3,0,0)$,
which carries the full $SU(3)$ flavor symmetry;   the puncture associated with the Young diagram $(2,1,0)$, which carries $U(1)$ flavor symmetry. The elementary building blocks are the $333$ vertex
and the $331$ vertex, where $3$ and $1$ are shorthands for the $SU(3)$ and $U(1)$ punctures, respectively.

The $333$ vertex  corresponds to  the $E_6$ SCFT of~\cite{Minahan:1996fg},   denoted by
$T_3$ in \cite{Gaiotto:2009we}.
 A maximal subgroup of the $E_6$ 
flavor symmetry is given by $SU(3)^3$ and we  parametrize the Cartans of the three $SU(3)$s by ${\mathbf a}_I$.
Guided by the expression of the $T_2$ index obtained in the previous subsection, we conjecture 
that the index of $T_3$ is given by
\be\label{E6ind}
{\mathcal I}(\mathbf{a_1,a_2,a_3})&=&
\sum_{\lambda_1,\lambda_2}
\frac{{\cal A}(\t)}{P_{HL}^{\lambda_1,\lambda_2}(\t^2,\t^{-2},1|\;\t)}
\prod_{I=1}^3 {\mathcal K}({\mathbf a_I})\, P_{HL}^{\lambda_1,\lambda_2}(\mathbf{a}_I|\;\t)\,\\
{\mathcal K}({\mathbf a})&=& \frac{1}{1-\t^2}\prod_{i,j=1,\;i\neq j}^3\frac{1}{\left(1-\t^2a_i/a_j\right)}\,,\qquad \prod_{i=1}^3 a_i=1\,\\
{\cal A}(\t)&=&(1-\t^4)(1+\t^2+\t^4)\,.
\ee The function ${\cal K}(\mathbf{a})$ is fixed as always by (\ref{orthrel}), with $\hat \Delta = \Delta_{HL}$,
while the overall fugacity-independent normalization factor ${\cal A}(\t)$ was fixed by comparing with the known result for this index~\cite{Gadde:2010te}.
We expanded the above expression in power series in $\t$ and found a perfect match with~\cite{Gadde:2010te}.\footnote{All the expressions for the HL index 
we obtain here are geometric progressions which in principle can be explicitly summed. However,
for the purposes of this paper we often found it computationally more feasible and insightful
to perform perturbative checks to high order in expansion in $\tau$.}
In~\cite{Benvenuti:2010pq} an explicit expression was conjectured for the partition function over the Higgs branch
of the $E_6$ SCFT, which we argued in section~\ref{limitssec} to be equivalent to the Hall-Littlewood index. This
expression has a very simple form \cite{Benvenuti:2010pq},
\be
{\mathcal I}(\mathbf{z}_{E_6})=
\sum_{k=0}^\infty [0,k,0,0,0,0]_{\mathbf z}\,\t^{2k}\,,
\ee where $\mathbf z$ is an $E_6$ fugacity and $[0,k,0,0,0,0]_{\mathbf z}$ are the characters of the
irreducible representation of $E_6$  with Dynkin labels $[0,k,0,0,0,0]$. This expression is 
manifestly $E_6$ covariant while~\eqref{E6ind} is not: however, order by order in the $\t$-expansion
we find that the fugacities of $SU(3)^3$ combine to label representations of $E_6$ and we obtain perfect agreement.
We emphasize that for this to happen the overall factors 
${\mathcal K}({\mathbf a}_i)$ are absolutely crucial -- without taking them 
into account the flavor-symmetry enhancement  to $E_6$ does not occur.

We can define an {\textit{unrefined}} index by setting all the flavor fugacities to one. In this case the series
can be easily summed up in closed form and we obtain that the unrefined index is given by
\be
&&{\mathcal I}=\\
&&\frac{ 1+\t^{20}+ 55(\t^2+\t^{18})+890(\t^4+\t^{16}) +5886 (\t^6+\t^{14})
+17929 (\t^8+\t^{12})+26060 \t  ^{10}}
{\left(1+\t ^2\right)^{-1}\left(1-\t ^2\right)^{22}}\,,\nonumber	
\ee in complete agreement with~\cite{Benvenuti:2010pq}. 

The $331$ vertex corresponds to the SCFT of a free hypermultiplet
in the bifundamental of $SU(3)$ and charged under  $U(1)$. Its index is given by
\be\label{331}
{\mathcal I}(\mathbf{a_1,a_2},c)&=&PE\left[\t \chi_1({\mathbf a})\chi_1({\mathbf b})c\right]_{{\mathbf a},{\mathbf b},c}
PE\left[\t \chi_1({\mathbf a}^{-1})\chi_1({\mathbf b}^{-1})c^{-1}\right]_{{\mathbf a},{\mathbf b},c}\\
&&\qquad=\prod_{i,j=1}^3\frac{1}{1-\t a_i b_j c}\,\frac{1}{1-\t \frac{1}{a_i b_j c}},\qquad
\prod_{i=1}^3 a_i=\prod_{i=1}^3 b_i=1\nonumber\,.
\ee It can be rewritten by partially closing a puncture of the $E_6$ vertex (\ref{E6ind}), as
\be \label{331alt}
{\mathcal I}(\mathbf{a_1,a_2},c)=\frac{1-\t^6}{1-\t^2}\frac{{\mathcal K}({\mathbf a_1}){\mathcal K}({\mathbf a_2})}{(1-\t^3 c^3)(1-\t^3 c^{-3})}\,
\sum_{\lambda_1,\lambda_2}
\frac{P_{HL}^{\lambda_1,\lambda_2}(\t c,\t^{-1}c,c^{-2}|\;\t)}{P_{HL}^{\lambda_1,\lambda_2}(\t^2,\t^{-2},1|\;\t)}
\prod_{I=1}^2P_{HL}^{\lambda_1,\lambda_2}(\mathbf{a}_I|\;\t)\,.\nonumber\\
\ee The sum over representations here is a geometric progression
and can be easily performed establishing the equivalence of (\ref{331}) and  (\ref{331alt}) 
(in the process we have fixed the overall
 $\t$-dependent factor). 
\begin{figure}
\begin{center}
\includegraphics[scale=0.35]{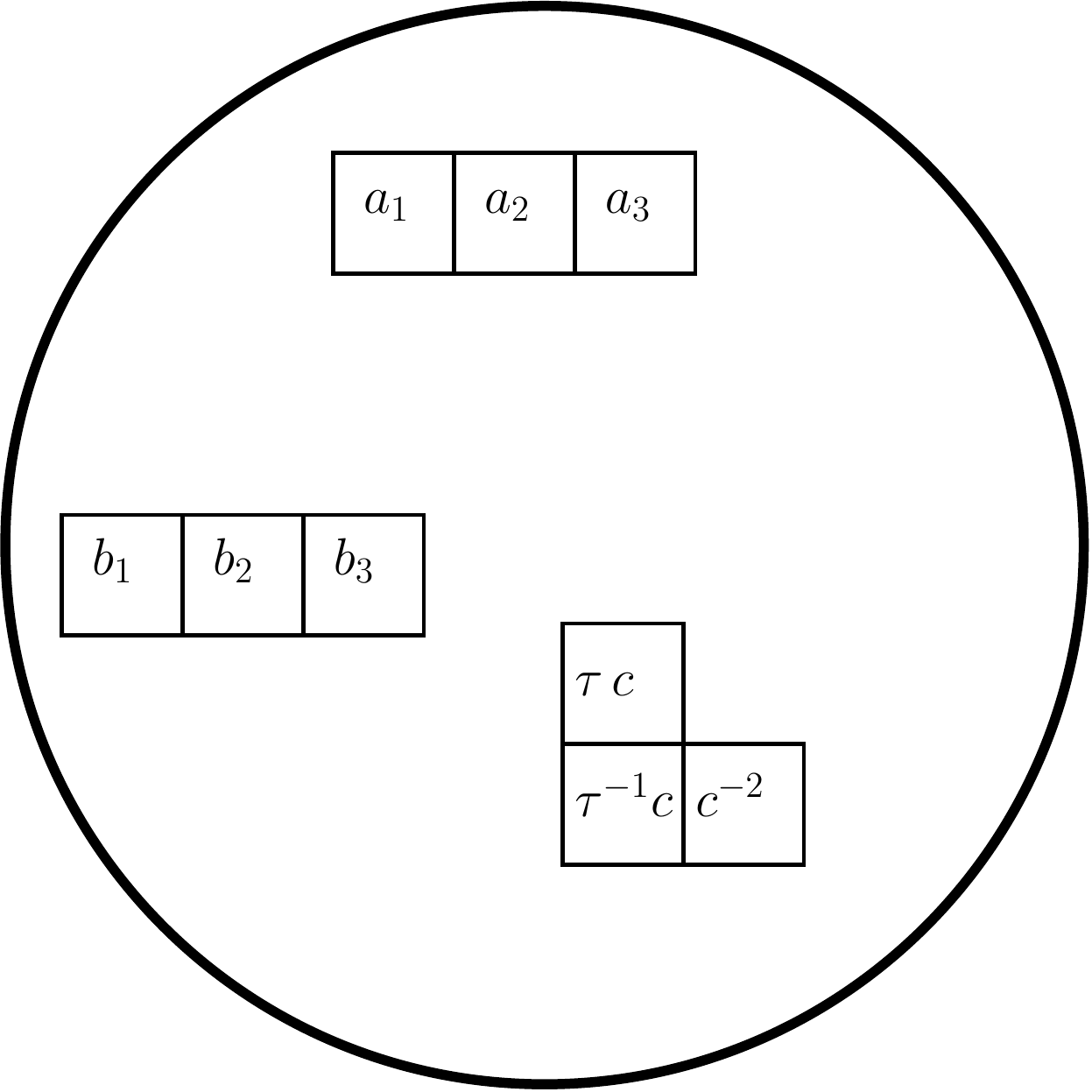}
\end{center}
\caption{Association of  flavor fugacities for the vertex corresponding to the $331$ of the
 $SU(3)$ quivers. Here $a_1a_2a_3=1$ and $b_1b_2b_3=1$.}
\end{figure}
\label{331fig}

We can use the above expressions to write the index of any $SU(3)$
quiver. Let us give again the example of the genus $\frak g$ theory,
\be
&&{\mathcal I}_{{\frak g}}^{(3)}=\\
&&\left(1-\t ^4\right)^{{\frak g} -1}\left(1-\t ^6\right)^{{\frak g} -1}+
\frac{\left(1 +2 \left(1+\t ^{-2}\right)^{{\frak g} -1} \left(\t ^{2-2{\frak g} }
-\t ^{2 {\frak g} -2}\right)\right)\t ^{4 ({\frak g} -1)}\left(1-\t ^2\right)^{2{\frak g} -2} }{ \left(\t ^{2-2{\frak g} }
-\t ^{2 {\frak g} -2}\right)^2}\,.\nonumber
\ee

We can subject (\ref{E6ind}) and (\ref{331alt}) to a further non-trivial check.
The channel-crossing duality of the four-punctured sphere with two $SU(3)$ and two $U(1)$
punctures corresponds to  Argyres-Seiberg duality~\cite{argyres-2007-0712}.
 In one channel we glue together two $331$ 
vertices along two 3 punctures, while in the other channel
the $333$ vertex (index of $T_3$) is (formally) glued to a $311$ vertex. Requiring equality of the two channels we find the index of the $311$ vertex,
\be
{\mathcal I}_{311}(\mathbf a, c,d)=
\frac{1-\t^6}{(1-\t^2)(1-\t^4)}\,
\frac{{\mathcal K}({\mathbf a})}
{(1-\t^3 c^3)(1-\t^3 c^{-3})(1-\t^3 d^3)(1-\t^3 d^{-3})}\,\\
\sum_{\lambda_1,\lambda_2}
\frac{P_{HL}^{\lambda_1,\lambda_2}(\t c,\t^{-1}c,c^{-2}|\;\t)P_{HL}^{\lambda_1,\lambda_2}(\t d,\t^{-1}d,d^{-2}|\;\t)
P_{HL}^{\lambda_1,\lambda_2}(\mathbf{a}|\;\t)}
{P_{HL}^{\lambda_1,\lambda_2}(\t^2,\t^{-2},1|\;\t)}
\,.\nonumber
\ee In the expression above the sum over representations
 diverges. The 311
should be regarded as a formal construct that
only makes sense as a part of the larger theory. It can be interpreted as
 implementing a $\delta$-function constraint 
on the flavor indices. The non-singular way to view
the gluing of $333$ vertex with $311$ vertex is as gauging an $SU(2)$ 
subgroup of $E_6$, as opposed to an $SU(3)$ subgroup~\cite{argyres-2007-0712}.
With this interpretation of the $311$ vertex,
equality of the two channels amounts to 
\be
&&(1-\t^2)\oint\frac{da}{4\pi i a} P_{HL}^{\lambda_1,\lambda_2}(a r, a^{-1}r,r^{-2}|\t)
\prod_{\s_1,\s_2,\s_3,\s_4,\s_5 = \pm1}\frac{1}{1-\t s^{\s_1} a^{\s_2} }
\frac{1}{1-\t^2 r^{3\s_3 } a^{\s_4} }(1-a^{2\s_5 })\nonumber\\
&&\qquad\qquad=
\frac{1-\t^6}{1-\t^4}
 \frac{\prod_{\s=\pm1}P_{HL}^{\lambda_1,\lambda_2}(\t \frac{s^{\s/3}}{r}, \t^{-1} \frac{s^{\s/3}}{r},\frac{s^{-2\s/3}}{r^{-2}}|\t)}
{P_{HL}^{\lambda_1,\lambda_2}(\t^2,\t^{-2},1|\t)\prod_{\s_1,\s_2=\pm1}(1-\t^3 s^{\s_1}/r^3)(1-\t^3 s^{\s_2}r^3)}\,.
\ee In the first line we gauge an $SU(2)$ subgroup of $E_6$ and couple it to a single hypermultiplet, and
in the second line a $311$ vertex is glued to $333$ vertex by gauging an $SU(3)$ flavor group. 
This is a non-trivial identity involving  HL polynomials which we have checked to very high order in a perturbative expansion in $\t$.

\subsection{A conjecture for the structure constants with generic punctures}\label{conjsec}

Extrapolating from the $SU(2)$ and $SU(3)$ cases, we are now
 formulate a complete conjecture for the index of all building blocks of $SU(k)$ quivers.
The building blocks are classified by a triple of Young  diagrams $(\Lambda_1,\Lambda_2,\Lambda_3)$.
We conjecture
\be\label{genconjHL}
{\mathcal I}_{\Lambda_1,\Lambda_2,\Lambda_3}=
\frac{\prod_{j=2}^k (1-\t^{2j})}{(1-\t^2)^{-k-2}}\prod_{I=1}^3\hat {\mathcal K}_{\Lambda_I}(\mathbf{ a}_I)
\sum_{{\lambda}}\frac{\prod_{I=1}^3P_{HL}^{{\lambda}}(\mathbf{a_I}(\Lambda_I)|\t)}
{P_{HL}^{{\lambda}}(\t^{k-1},\t^{k-3},\dots,\t^{1-k}|\t)}\,.
\ee Here the assignment of fugacities according to the Young diagram labelling the type of the puncture,
 $\mathbf{a}(\Lambda)$, is as illustrated in figure~2. The summation over $\lambda$ is over the Young diagrams with $k-1$
rows, $\lambda=(\lambda_1,\dots,\lambda_{k-1})$, $\lambda_j \geq \lambda_{j+1}$.
\begin{figure}[htbp]
\begin{center}
\includegraphics[scale=0.6]{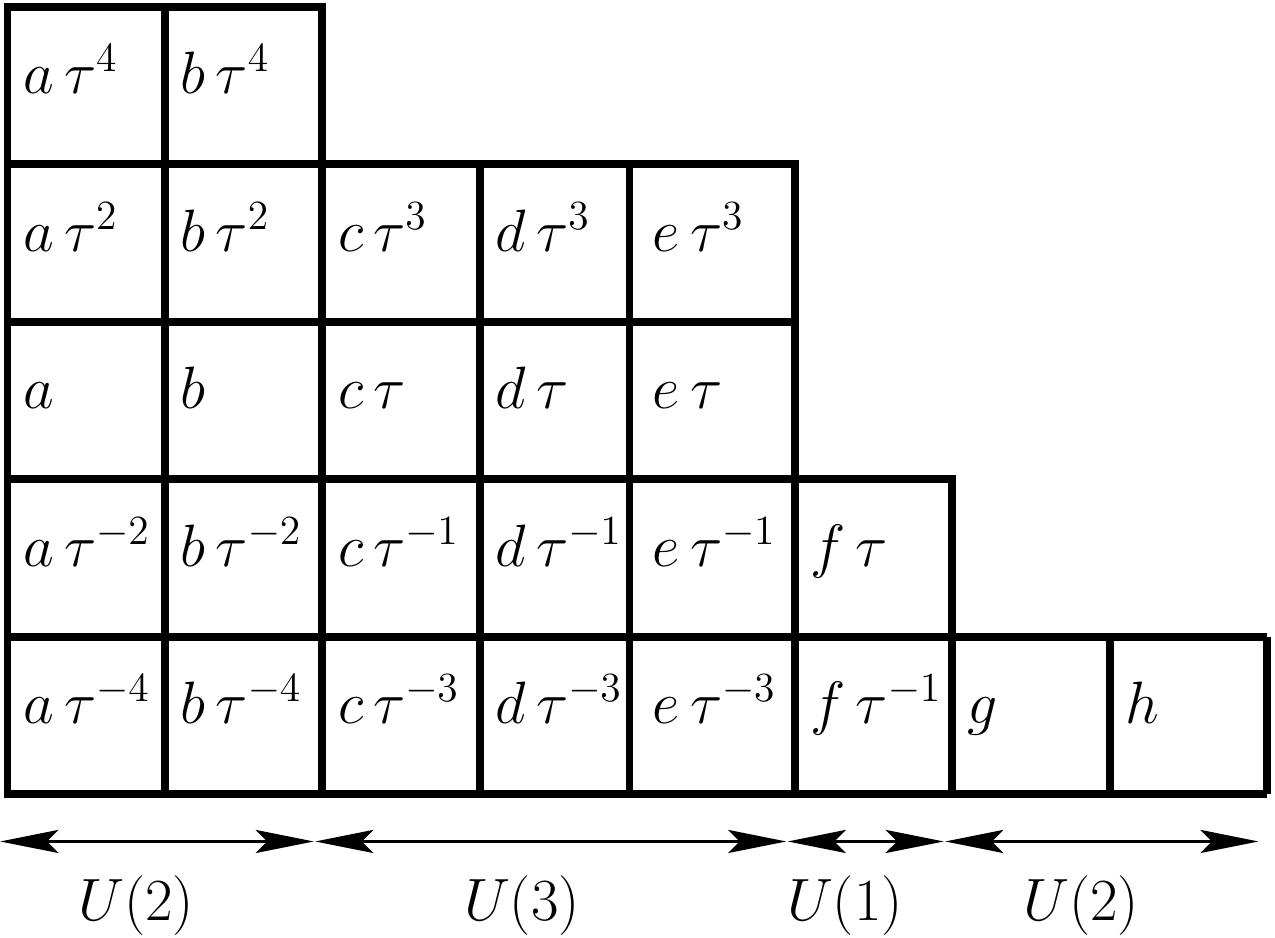}
\end{center}
\caption{Association of the flavor fugacities for a generic puncture. Punctures are classified
 by embeddings of  $SU(2)$ in $SU(k)$, so they are specified by the decomposition
 of the fundamental representation of $SU(k)$ into irreps of $SU(2)$, that is, by a partition of $k$.
 Graphically we represent the partition by an auxiliary Young diagram $\Lambda$ with $k$ boxes,
 read from left to right. In the figure we have the fundamental of $SU(26)$ decomposed as $\mathbf{5} + \mathbf{5} + \mathbf{4} + \mathbf{4} + \mathbf{4}  + \mathbf{2} + \mathbf{1} + \mathbf{1}$. 
 The commutant of the
embedding gives the residual flavor symmetry, in this case $S(U(3)\times U(2)\times U(2)\times U(1))$,
where the $S(\dots)$ constraint amounts to removing the overall $U(1)$.
 The $\t$ variable is viewed here as an $SU(2)$ fugacity, while the Latin variables are fugacities of the residual flavor symmetry.
The $S(\dots)$ constraint implies that the flavor fugacities satisfy
 $(ab)^{5}(cde)^4f^2gh=1$.}
\end{figure}\label{flavfig}
The factors $\hat {\mathcal K}_{\Lambda}(\mathbf{ a})$ are defined as
\be 
\hat {\mathcal K}_{\Lambda}(\mathbf{ a})=
\prod_{i=1}^{row(\Lambda)}\prod_{j, k=1}^{l_i}\frac{1}{1-{\frak a}^i_j\bar {\frak a}^i_k}\,.
\ee Here $row(\Lambda)$ is the number of rows in $\Lambda$ and $l_i$ is the length of $i$th row. The coefficients
${\frak a}^i_k$ are associated to the Young diagram as illustrated in figure~3. 
Our conjecture is consistent with the $SU(2)$ and $SU(3)$ cases 
 seen previously as well as with all other examples discussed below.

For three maximal punctures (the $T_k$ theory), \eqref{genconjHL} becomes
\be\label{Tkind}
{\mathcal I}_{T_k}(\mathbf{a_1,a_2,a_3})&=&
\sum_{\lambda_1\geq\lambda_2\geq...\geq\lambda_{k-1}}
\frac{{\cal A}(\t)}{P_{HL}^{\lambda_1,..,\lambda_{k-1}}(\t^{k-1},..,\t^{1-k}|\;\t)}
\prod_{I=1}^3 {\mathcal K}({\mathbf a_I})\, P_{HL}^{\lambda_1,..\lambda_{k-1}}(\mathbf{a}_I|\;\t)\,,\nonumber\\
{\mathcal K}({\mathbf a})&=& \frac{1}{(1-\t^2)^{\frac{k-1}2}}\prod_{i,j=1,\;i\neq j}^k\frac{1}{\left(1-\t^2a_i/a_j\right)}\,,\qquad \prod_{i=1}^k a_i=1\,,\\
{\cal A}(\t)&=&\frac{\prod_{j=2}^k (1-\t^{2j})}{(1-\t^2)^{\frac{k-1}{2}}}
\,.\nonumber
\ee
Let us illustrate the power of these TQFT expressions  by computing
the index of the genus $\frak{g}$ $SU(k)$ theory. It is given by
\be\label{genusHL}
{\mathcal I}^{(k)}_{\frak g}=\frac{\left(\prod_{j=2}^k (1-\t^{2j})\right)^{2\frak g-2}}{(1-\t^2)^{(k-1)(\frak g-1)}}
\sum_\lambda \frac{1}{P_{HL}^\lambda(\t^{k-1},\t^{k-3},\dots,\t^{1-k}|\t)^{2\frak g-2}}\,,
\ee where the summation is over all Young diagrams with $k-1$ rows, {\it i.e.} over the finite 
irreducible representations of $SU(k)$.

The sum over representations in ~\eqref{genconjHL} does not converge for arbitrary choices of the three Young diagrams $\Lambda_I$.
We have already encountered an example  in the last subsection: the 311 vertex of $SU(3)$ theories has a divergent expression. 
There is no actual SCFT  corresponding to the 311 vertex, but 
one can glue this vertex to a larger quiver and 
obtain meaningful results. There are cases however where the divergent vertex cannot appear as a piece of a larger 
quiver and thus the expression~\eqref{genconjHL} for its index does not have a clear physical interpretation. An example of such a vertex is the 
index of an $SU(6)$ theory with three $SU(3)$ punctures. We have checked in several cases  that a divergence in (\ref{genconjHL})
correlates with the fact that the
graded rank of the Coulomb branch  (as defined in~\cite{Chacaltana:2010ks}) of
the putative SCFT 
has negative components. This is an indication that associating field theories to such punctured 
surfaces may be delicate. Punctured surfaces 
of this type were recently considered in~\cite{Moore:2011ee} and subtleties 
associated with them addressed in~\cite{GMT}.
\begin{figure}[htbp]
\begin{center}
\includegraphics[scale=0.6]{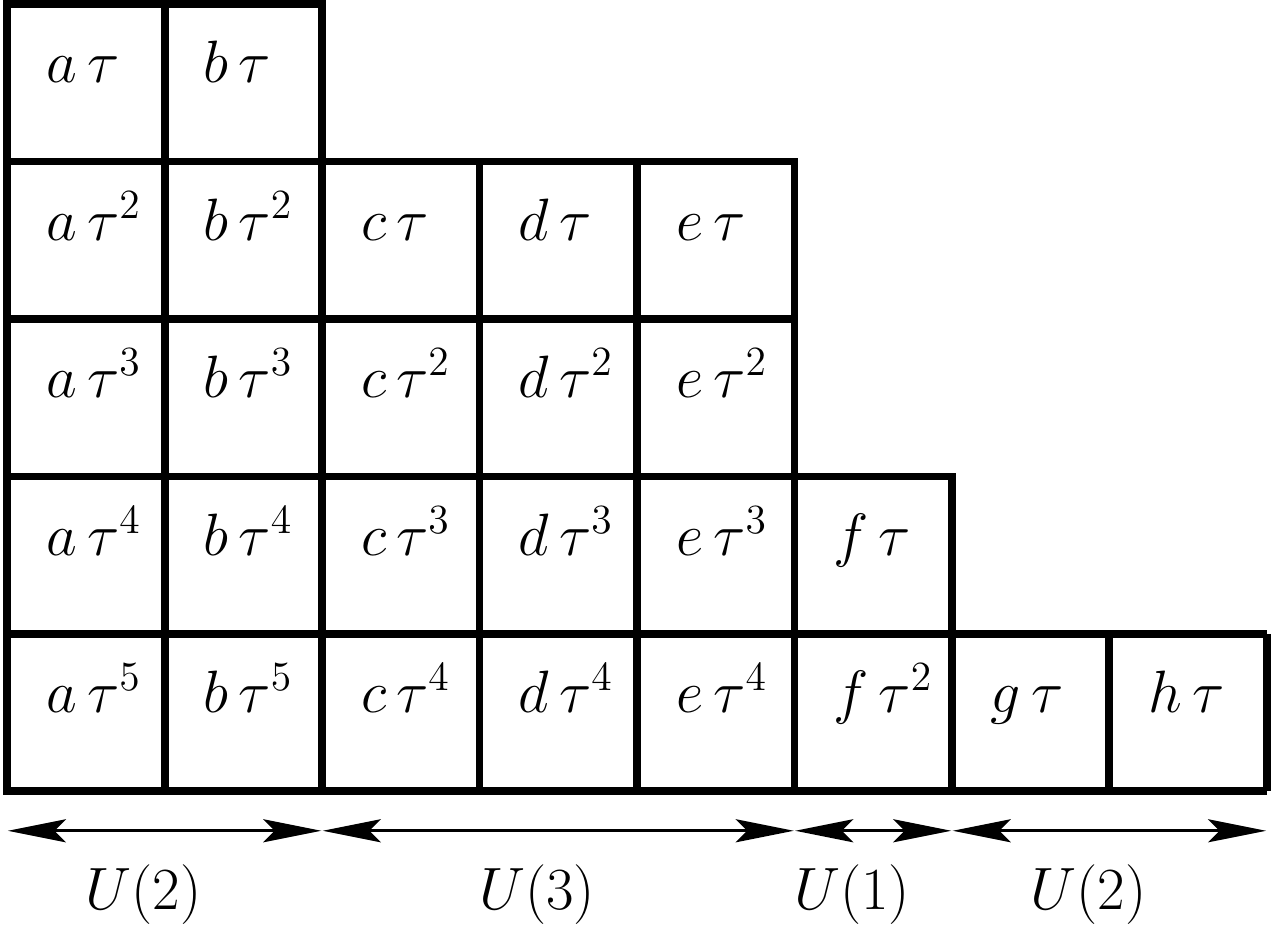}
\end{center}
\caption{The factors ${\frak a}^i_k$ associated to a generic Young diagram. The upper index is 
the row index and the lower is the column index. In $\bar{\frak a}^i_k$ one takes the inverse of flavor fugacities
while $\t$ is treated as real number. As before, the flavor fugacities in this example 
satisfy  $(ab)^{5}(cde)^4f^2gh=1$.}
\end{figure}\label{vectfig}

\

\subsection{$SU(4)$ quivers -- $T_4$ and  the $E_7$ SCFT}

Let us use the general expressions of the previous section to discuss some of the features of $SU(4)$ quivers.
First, from~\eqref{Tkind} we can compute the unrefined index of $T_4$,
\be 
{\mathcal I}_{T_4}=1+45 \t  ^2+128 \t  ^3+1249 \t  ^4+5504 \t  ^5+30786 \t  ^6+136832 \t  ^7+623991 \t  ^8+\dots\,.
\ee We present a closed form expression for it in appendix~\ref{T4app}.
Refining with the flavor fugacities one gets
\be 
{\mathcal I}_{T_4}=&&1
+\left[(\mathbf{ 15},1,1)+(1,\mathbf{ 15},1)+(1,1,\mathbf{ 15})\right]\t^2+
\left[(\mathbf{ 4},\mathbf{ 4},\mathbf{ 4})+(\mathbf{ \bar 4},\mathbf{ \bar 4},\mathbf{ \bar 4})\right]\t^3+\\
&&\left[1+(\mathbf{ 15},1,1)+(1,\mathbf{ 15},1)+(1,1,\mathbf{ 15})+(\mathbf{ 20},1,1)+(1,\mathbf{ 20},1)+(1,1,\mathbf{ 20})+\right.\nonumber\\
&&\left.+(\mathbf{ 15},\mathbf{ 15},1)+(1,\mathbf{ 15},\mathbf{ 15})+(\mathbf{ 15},1,\mathbf{ 15})+
(\mathbf{ 84},1,1)+(1,\mathbf{ 84},1)+(1,1,\mathbf{ 84})+\right.\nonumber\\
&&\left.+(\mathbf{ 6},\mathbf{6},\mathbf{ 6})\right]\t^4+\dots\,.\nonumber
\ee In terms of Young diagrams $\mathbf{ 84}=(4,2,2),\;\mathbf{ 6}=(1,1,0),\;\mathbf{ 20}=(2,2,0)$. The symmetric product of
the $\t^2$ term reproduces all the terms at the $\t^4$ order except for the $(\mathbf{ 6},\mathbf{6},\mathbf{ 6})$ term, and for the 
fact that two singlets are missing (the symmetric product contains three singlets while only one is present at order $\t^4$). 
We deduce that the $(\mathbf{ 6},\mathbf{6},\mathbf{ 6})$ state is 
an additional generator of the Higgs branch, and that there is a constraint allowing only
for one singlet in the symmetric product of the $\tau^2$ states to appear at $\tau^4$ order.
 Unlike the situation for the $E_6$
SCFT where the Higgs branch is generated by a single scalar transforming as  ${\mathbf {78}}$ of $E_6$~\cite{Gaiotto:2008nz}
 here one has new generators
appearing at higher orders in the $\t$ expansion and thus having different $E-R$ quantum numbers.

Next, we can partially close a puncture to obtain the index of 
the $441$ vertex. On one hand, the 441 vertex correspond to the free hypermultiplet SCFT in the bifundamental
of two $SU(4)$s and charged under the $U(1)$, so its index can be evaluated by direct counting,
\be \label{441}
{\mathcal I}(\mathbf{a_1,a_2},c)=\prod_{i,j}^4\frac{1}{1-\t a_i b_j c}\,\frac{1}{1-\t \frac{1}{a_i b_j c}},\qquad
\prod_{i=1}^4 a_i=\prod_{i=1}^4 b_i=1\,.
\ee
On the other hand, from (\ref{genconjHL}),
\be \label{441alt}
&&{\mathcal I}(\mathbf{a_1,a_2},c)=
\frac{1-\t^8}{(1-\t^2)^{}}\frac{{\mathcal K}({\mathbf a_1}){\mathcal K}({\mathbf a_2})}{(1-\t^4 c^4)(1-\t^4 c^{-4})}\,\\
&&\qquad\sum_{\lambda_1,\lambda_2,\lambda_3}
\frac{P_{HL}^{\lambda_1,\lambda_2,\lambda_3}(\t^2 c,c,\t^{-2}c,c^{-3}|\;\t)}
{P_{HL}^{\lambda_1,\lambda_2,\lambda_3}(\t^3,\t,\t^{-1},\t^{-3}|\;\t)}
\prod_{i=1}^2P_{HL}^{\lambda_1,\lambda_2,\lambda_3}(\mathbf{a_i}|\;\t)\,.\nonumber
\ee We have checked the equivalence of these two expressions perturbatively to very high order in $\t$.
\begin{center}
\begin{figure}
\begin{center}
\includegraphics[scale=0.35]{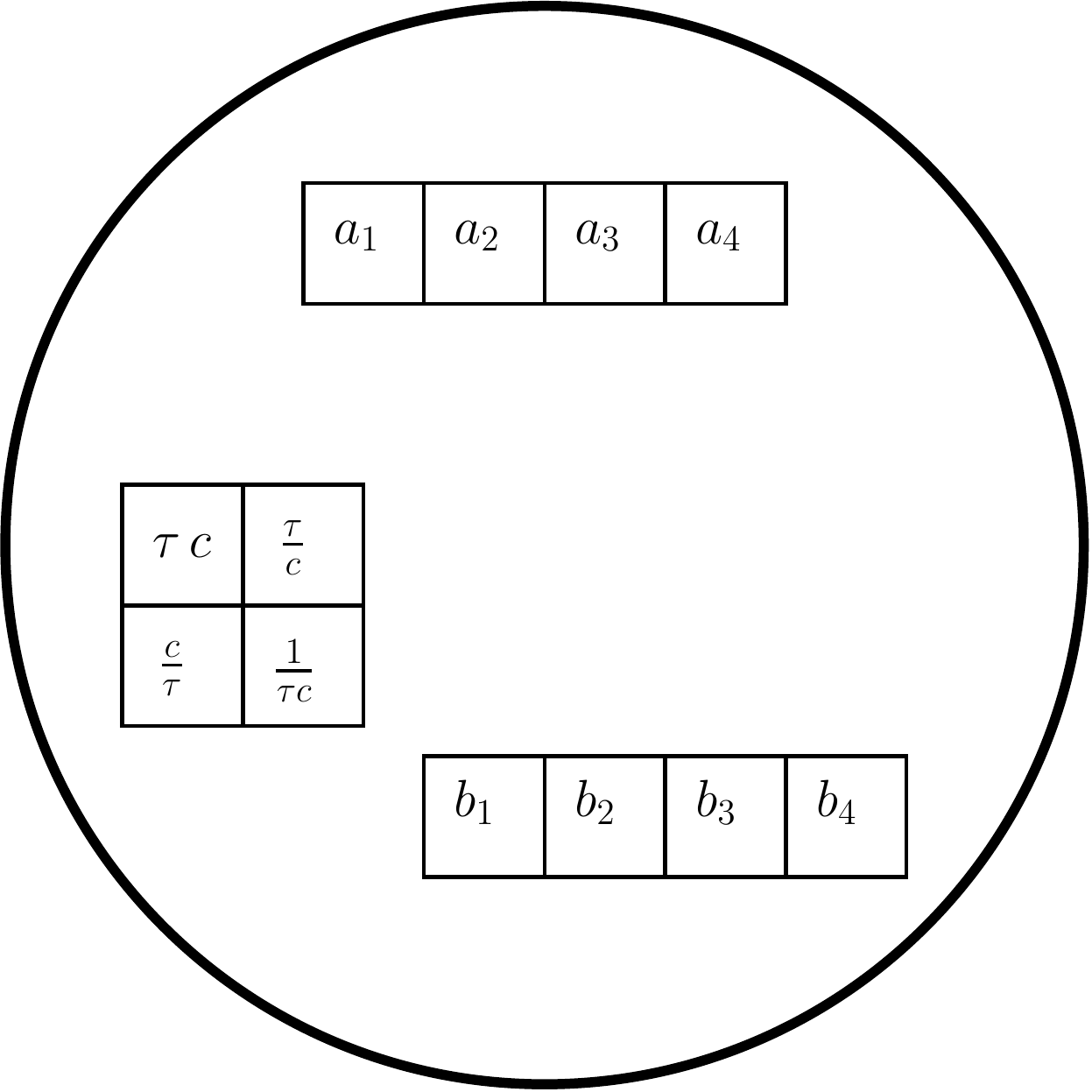}\\
\end{center}
\caption{Association of the flavor fugacities for the $E_7$ vertex.
Here 
$\prod_{i=1}^4b_i=\prod_{i=1}^4a_i=1$.}
\end{figure}
\label{E7fig}
\end{center}
Finally, let us look at the vertex with two maximal punctures and one puncture corresponding
to a square Young diagram, which carries an $SU(2)$ flavor symmetry, see figure~4. The flavor symmetry of this theory is known to 
enhance to $E_7$~\cite{argyres-2007-0712}.
From (\ref{genconjHL}), the Hall-Littlewood index of this SCFT is given by
\be\label{E7ind}
{\mathcal I}_{E_7}(\mathbf{a_1,a_2},c)&=&(1+\t^2+\t^4)(1+\t^4)
\frac{{\mathcal K}({\mathbf a_1}){\mathcal K}({\mathbf a_2})}{(1-\t^2 c^{\pm2})(1-\t^4 c^{\pm2})}\,\times\nonumber\\
&&\qquad \sum_{\lambda_1,\lambda_2,\lambda_3}
\frac{P_{HL}^{\lambda_1,\lambda_2,\lambda_3}(\t c,\frac{c}{\t},\frac{\t}c,\frac{1}{\t c}|\;\t)}
{P_{HL}^{\lambda_1,\lambda_2,\lambda_3}(\t^3,\t,\t^{-1},\t^{-3}|\;\t)}
\prod_{i=1}^2P_{HL}^{\lambda_1,\lambda_2,\lambda_3}(\mathbf{a_i}|\;\t)\,.
\ee In~\cite{Benvenuti:2010pq} an explicit expression for the Higgs partition function was conjectured,
\be
{\mathcal I}(\mathbf{z}_{E_7})=
\sum_{k=0}^\infty [k,0,0,0,0,0,0]_{\mathbf z}\,\t^{2k}\,,
\ee where $\mathbf z$ is an $E_7$ fugacity and $[k,0,0,0,0,0,0]_{\mathbf z}$ are the characters of the
irreducible representation of $E_7$  with Dynkin labels $[k,0,0,0,0,0,0]$.
We have checked also here ~\eqref{E7ind} is 
in complete agreement with~\cite{Benvenuti:2010pq}, and thus in particular is secretly $E_7$ covariant:
the check can be done analytically for the unrefined index and perturbatively in $\t$ to high order 
for the refined one.

The expression (\ref{E7ind}) can be also checked by the Argyres-Seiberg duality between $USp(4)$ theory
 coupled to six fundamental hypermultiplets and $E_7$ theory with an $SU(2)$ subgroup gauged~\cite{argyres-2007-0712}.
 The former has a weakly-coupled description and its index can be computed directly,
\begin{equation} \label{Iusp}
  \mathcal{I}_{USp(4)}=1+\chi^\rep{66}_{SO(12)}(u,v,w,x,y,z)\t^2+\cdots.
\end{equation}
Since there are six fundamental hypermultiplets the flavor group is $SO(12)$.
On the other hand, gauging an $SU(2)$ inside one $SU(4)$ subgroup of the $E_7$ index (\ref{E7ind}) gives
\begin{equation} \label{Ip}
\mathcal{I}=\oint\frac{de}{4\pi i e}\,(1-e^2)\,
(1-e^{-2})\,PE\left[-\t^2\chi_{2}(e)\right]_{t,e}\mathcal{I}_{E_7}(\mathbf{a},\{es,s/e,b/s,1/bs\},c)\,.
\end{equation}
We have checked perturbatively in $\t$ that (\ref{Iusp}) and (\ref{Ip}) coincide under the following identification of the fugacities:
\begin{equation}
  u\rightarrow \frac{a_1}{s},\quad
 v\rightarrow \frac{a_2}{s},\quad
 w\rightarrow \frac{a_3}{s},\quad
 x\rightarrow \frac{1}{a_1a_2a_3s},
\quad y\rightarrow bc,
\quad
 z\rightarrow \frac{b}{c}\,.
\end{equation}

\

\subsection{$SU(6)$ quivers --  the $E_8$ SCFT}

As our last example, we consider
the index of the $E_8$ SCFT~\cite{Minahan:1996cj}.
This  theory corresponds to a sphere with a maximal $SU(6)$ puncture and two non-maximal
punctures with $SU(3)$ and $SU(2)$ flavor symmetries, see figure~5. The group $SU(6)\times SU(3)\times SU(2)$
is a maximal subgroup of $E_8$.
Following the general prescription~\eqref{genconjHL} the index of $E_8$ SCFT is given by
\be\label{E8ind}
&&{\mathcal I}_{E_8}(\mathbf{a},b_1,b_2,c)=
\frac{(1-\t^8)(1-\t^{10})(1-\t^{12})}{(1-\t^2)^{1/2}(1-\t^4)^{4}(1-\t^6)^{}}\times\\
&&\frac{{\mathcal K}({\mathbf a})}{(1-\t^2 c^{\pm2})(1-\t^4 c^{\pm2})(1-\t^6 c^{\pm2})\prod_{i\neq j}
(1-\t^2 b_i/b_j)(1-\t^4 b_i/b_j)}\,\times\nonumber\\
&&\;\;\; \sum_{\lambda_1,\dots,\lambda_5\equiv{\mathbf{\lambda}}}
\frac{P_{HL}^{\mathbf{\lambda}}(\t b_1,\t b_2,\t b_3,\frac{b_1}{t},\frac{b_2}{\t},\frac{b_3}{\t}|\;\t)
P_{HL}^{\mathbf{\lambda}}(\t^2 c,c,\frac{c}{\t^{2}},\frac{\t^{2}}{c},\frac1c,\frac1{\t^{2}c},|\;\t)
P_{HL}^{\mathbf{\lambda}}(\mathbf{a_i}|\;\t)}
{P_{HL}^{\mathbf{\lambda}}(\t^5,\t^3,\t,\t^{-1},\t^{-3},\t^{-5}|\;\t)}
\,.\nonumber
\ee  In~\cite{Benvenuti:2010pq} it was conjectured that the Higgs partition function has the following
$E_8$ covariant expansion,
\be \label{E8covariant}
{\mathcal I}(\mathbf{z}_{E_8})=
\sum_{k=0}^\infty [k,0,0,0,0,0,0,0]_{\mathbf z}\,\t^{2k}\,,
\ee where $\mathbf z$ is an $E_8$ fugacity and $[k,0,0,0,0,0,0,0]_{\mathbf z}$ are the characters of the
irreducible representation of $E_8$  with Dynkin labels $[k,0,0,0,0,0,0,0]$.
We have again checked equivalence of~\eqref{E8ind} and (\ref{E8covariant}) in  the $\t$-expansion,
though in this case due to computational complexity we could perform the expansion only up to order $\t^8$.
The size of representations of $E_8$ contributing to the index grows very fast with  the order of $\t$, {\it {\it e.g.}} 
the unrefined index is given by
\be
{\mathcal I}_{E_8}=
1+248\, \t  ^2+27000\, \t  ^4+1763125\, \t  ^6+79143000\, \t  ^8+\dots\,.
\ee
\begin{center}
\begin{figure}
\begin{center}
\includegraphics[scale=0.4]{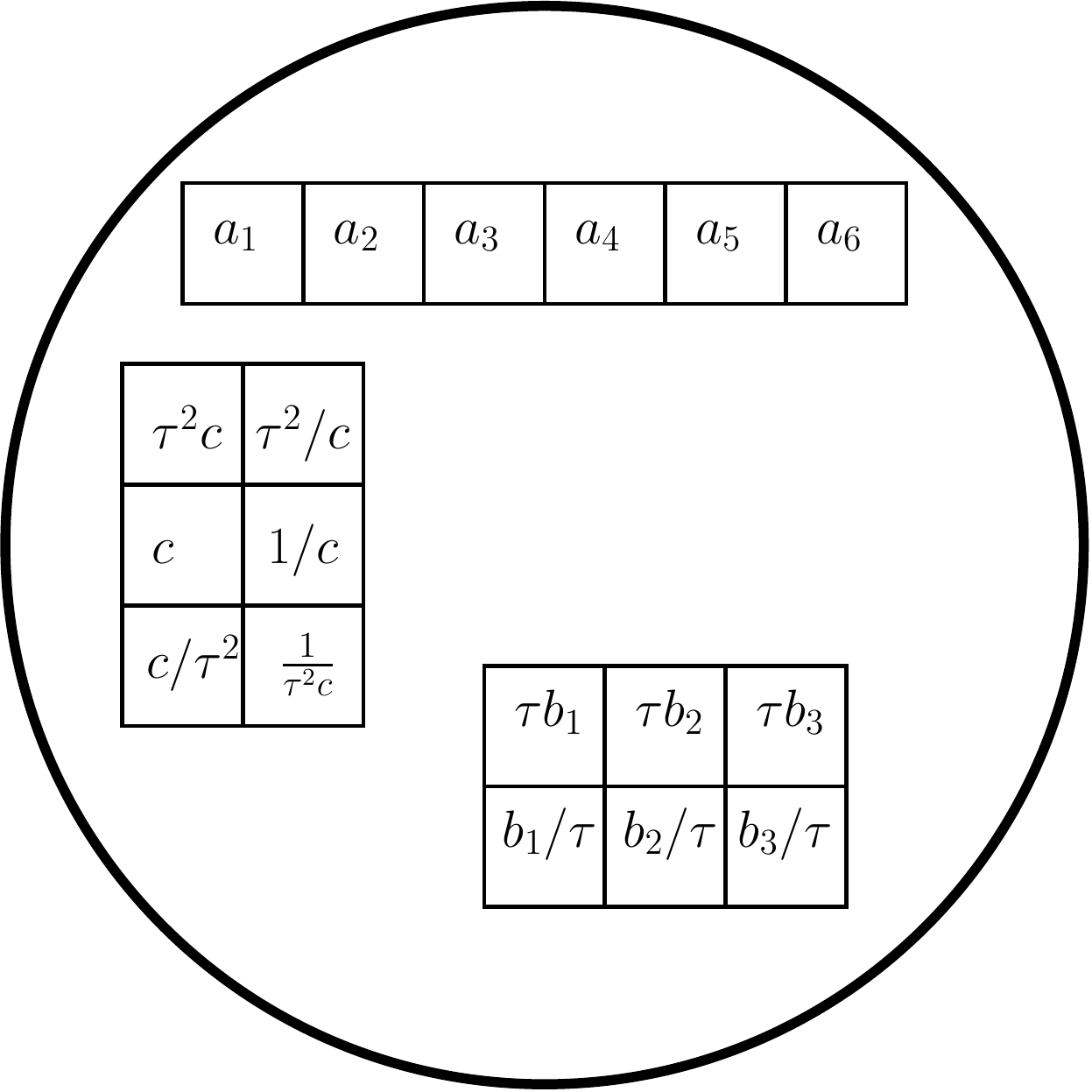}\\
\end{center}
\caption{Association of the flavor fugacities for the $E_8$ vertex. Here 
$\prod_{i=1}^3b_i=\prod_{i=1}^6a_i=1$.}
\end{figure}
\label{E8fig}
\end{center}

\

\subsection{Large $k$ limit }

It is not difficult to evaluate the large $k$ limit of the HL index of $SU(k)$ generalized quivers.\footnote{We thank
 Davide Gaiotto and Juan Maldacena for discussions on  issues related to this section.}
For instance, for the index of the theory corresponding to a genus ${\frak g}$ 
surface without punctures~\eqref{genusHL},
\be\label{largeKg}
{\mathcal I}^{(k\to\infty)}_{\frak g}=\prod_{j=2}^\infty(1-\t^{2j})^{{\frak g}-1}=
PE\left[-({\frak g}-1)\frac{\t^4}{1-\t^2}\right]\,.
\ee In appendix~\ref{largeKsec}
we give a short derivation of this expression. In the large $k$ limit only the singlet in the sum over the representations
of~\eqref{genusHL} contributes. 
Since (\ref{largeKg}) is of order one for large $k$ it is expected to be matched
by counting the appropriate supergravity modes in the dual
AdS background~\cite{Gaiotto:2009gz}. We can also compute the 
index of the $T_k$ theories  in the large $k$ limit,
\be\label{TkLargeK}
{\mathcal I}_{T_{k\to\infty}}(\mathbf{a_1,a_2,a_3})&=&
\prod_{j=2}^\infty\frac{1}{1-\t^{2j}}\,
\prod_{I=1}^3\prod_{j\neq i}^\infty\frac{1}{1-\t^2a^I_i/a^I_j}\,\\
&=&PE\left[\frac{\t^4}{1-\t^2}\right]\,
\prod_{I=1}^3 PE\left[\t^2\sum_{i\neq j}a^I_i/a^I_j\right]\,.\nonumber
\ee From here the large $k$ index of any generalized quiver  is trivial to compute;
in particular~\eqref{largeKg} can be obtained by gluing together the index of~\eqref{TkLargeK}.
Unrefining the index of $T_k$ by setting all the flavor fugacities $a^I_j=1$ we see that it has 
a non-trivial  $k$ dependence for large $k$ limit. Taking the plethystic log of~\eqref{TkLargeK} 
(that is,  considering the index of single-particle states) we find
\be
{\mathcal I}^{s.p.}_{T_{k\to\infty}}= 3\,\t^2\,(k^2-k)+\frac{\t^4}{1-\t^2}+O\left(\frac{1}{k}\right)\,.
\ee The term of order $k^2$ 
on the right-hand-side comes from states in the adjoint representation of the flavor group,
while the 
 term of order $k$ comes from neutral states. 
At least some of $O(k^2)$  states  in the adjoint representation must correspond to modes of the AdS gauge fields  that couple to
 the flavor currents of the boundary theory.
It would be interesting to check whether all the $O(k^2)$ and $O(k)$ states can
be accounted for by supergravity states.  
 If not,  the extra states could arise
as non-perturbative states in the bulk geometry ({\it {\it e.g.}} wrapped branes or black holes).
In all cases studied so far the index  is of order one in the large $k$ limit and thus cannot capture the non-perturbative states of the bulk theory
~\cite{Kinney:2005ej,Nakayama:2005mf,Nakayama:2006ur,Gadde:2010en}.
This is not a contradiction, since the index only counts protected states with signs. The index
 vanishes on combinations of short multiplets that can in principle recombine into long ones,
 even when such kinematically-allowed recombination do not actually happen~\cite{Kinney:2005ej}.
 However, for linear quivers (in particular for the $T_k$ theories) the HL index has the meaning of a Hilbert series over 
the Higgs branch, so it is expected to  capture {\textit{all}} the relevant $\frac{3}{8}$-BPS 
states of the dual theory.
We leave the very interesting comparison with the bulk theory for future research.

\

\section{Schur index}\label{qsec}

We   turn  to the Schur index,
\be
{\mathcal I}_S=\Tr(-1)^F\,q^{E-R}\, ,
\ee
which is the same as  the {\it reduced index} considered in \cite{Gadde:2011ik}.
Let us first recall the expression for the $SU(k)$ propagator. It is of the usual form 
\be
 \eta (\mathbf{a}, \mathbf{b}^{-1}) = \Delta (\mathbf{a})  {\mathcal I}^{V}(\mathbf{a}) \delta(\mathbf{a}, \mathbf{b}^{-1}) \, ,
\ee
where $\Delta(\mathbf{a})$ is the Haar measure (\ref{Haarexpression}), and ${\cal  I}^V(\mathbf{a})$ the index
of the vector multiplet, given by
\be
{\mathcal I}_q^{V}(\mathbf{a})=PE\left[\frac{-2q}{1-q}\chi_{adj}(\mathbf{a})\right]_{q,\mathbf{a}}
\,.
\ee
The set of functions $\{ f_q^\lambda (\mathbf{a})  \}$ that diagonalize the structure constants 
are proportional to the Schur polynomials~\cite{Gadde:2011ik},
\be
 f_q^\lambda (\mathbf{a}) = {\cal K}_q(\mathbf{a}) \;\chi^\lambda (\mathbf{a}) \,.
\ee
The Schur polynomials are orthonormal under the Haar measure, so in this case $\hat \Delta (\mathbf{a}) = \Delta(\mathbf{a})$
(recall (\ref{clever})) and the factor ${\cal K}_q(\mathbf{a})$ is given by
\be
{\cal K}_q(\mathbf{a}) = \frac{1}{[{\mathcal I}_q^{V}(\mathbf{a})]^{\frac{1}{2}}} \,.
\ee
Generalizing our results in~\cite{Gadde:2011ik}, we
 conjecture the following expression for the Schur index of a three-punctured sphere with generic punctures,
\be\label{qgen}
{\mathcal I}_{\Lambda_1,\Lambda_2,\Lambda_3}=
\frac{(q;q)^{k+2}}{\prod_{j=1}^{k-1} (1-q^{j})^{k-j}}\prod_{I=1}^3\hat {\mathcal K}_{\Lambda_I}(\mathbf{ a}_I)
\sum_{{\lambda}}\frac{\prod_{I=1}^3\chi^{{\lambda}}(\mathbf{a_I}(\Lambda_I))}
{\chi^{{\lambda}}(q^{\frac{k-1}2},q^{\frac{k-3}2},\dots,q^{\frac{1-k}2})}\,.
\ee 
Here the sum is over the finite-dimensional irreducible representations of $SU(k)$.
The assignment of fugacities according to the Young diagram, $\mathbf{a}(\Lambda)$,  is again as in figure~2, with $\t \to q^{1/2}$. 
The Pochhammer symbol $(a;\;b)$ is defined by
\be
(a;\;b)=\prod_{i=0}^\infty(1-a\,b^i)\,.
\ee
The character of the representation corresponding to Young diagram $\lambda=(\lambda_1,\dots,\lambda_{k-1},0)$ is
given by a Schur polynomial,
\be
\chi^{{\lambda}}(\mathbf a)=\frac{\det(a_i^{\lambda_j+k-j})}{\det (a_i^{k-j})}\, .
\ee
 The $\hat {\mathcal K}_\Lambda$ prefactors are given by
\be 
\hat {\mathcal K}_{\Lambda}(\mathbf{ a})=
\prod_{i=1}^{row(\Lambda)}\prod_{j, k=1}^{l_i}PE\left[\frac{{\frak a}^i_j\bar {\frak a}^i_k}{1-q}\right]_{{\frak a}_i,q}\, ,
\ee where $row(\Lambda)$ is the number of rows in $\Lambda$ and $l_i$ is the length of $i$th row. The coefficients
${\frak a}^i_k$ are associated to the Young diagram again as in figure~3, with $\t \to q^{1/2}$.
Note that the quantity appearing in the denominator of~\eqref{qgen} is the quantum dimension of the representation ${\lambda}$
of $SU(k)$,
\be
\text{dim}_{q}{\lambda}=\chi^{{\lambda}}(q^{\frac{k-1}2},q^{\frac{k-3}2},\dots,q^{\frac{1-k}2})\,.
\ee 
For $SU(2)$ the quantum dimension is also known as the $q$-number $[\lambda]_q$. 

We have subjected (\ref{qgen}) to similar checks as the one described for the Hall-Littlewood index, finding complete agreement with expectations;
a few such checks were reported in~\cite{Gadde:2011ik}.
Let us only mention here the basic identity following from compatibility of (\ref{qgen})
with the  index of the $SU(2)$ trifundamental hypermultiplet, 
\be\label{qIndSU2}
&&PE\left[\frac{q^{1/2}}{1-q}\left(a_1+\frac1{a_1}\right)\left(a_2+\frac1{a_2}\right)\left(a_3+\frac1{a_3}\right)\right]_{a_i,\t}=\\
&&\qquad (q;q)^{3}(q^2;q)\prod_{i=1}^3 PE\left[\frac{q}{1-q}\left(a_i^2+a_i^{-2}+2\right)\right]_{a_i,\t}\,
\sum_{\lambda=0}^\infty\frac{\prod_{i=1}^3\chi^{\lambda}(a_i,a_i^{-1})}
{\chi^{\lambda}(q^\half,q^{-\half})}\,.\nonumber
\ee A proof of this identity is outlined in appendix~\ref{su2proofsec}.

\newpage

\section{Macdonald index}\label{higgssec}

We are now ready to combine and generalize the results of the two previous sections.
The Hall-Littlewood and Schur polynomials are special cases of a two-parameter
family of polynomials discovered by Macdonald~\cite{Mac}. One naturally expects  Macdonald
polynomials to be relevant for the calculation of the index in a two-dimensional slice of the full three-dimensional
fugacity space. The precise confirmation of this idea is our main result. Identifying the correct
slice is by no means obvious, but at this point
it will come as no great surprise that it is given by the limit that we have called the Macdonald index
in section 4,
\be \label{mac}
{\mathcal I}_M
 =  \Tr_{M} (-1)^F\,           q^{E-2R-r}\, t^{R+r} =  \Tr_{M} (-1)^F\,           q^{-2j_1}\, t^{R+r}\, ,
 \ee
where $\Tr_{M}$ denotes the trace restricted to states with $\delta_{{\suup}{+}}=E+2j_1-2R-r=0$.
For $q=t$
Macdonald polynomials reduce to Schur polynomials, while for $q=0$ they reduce to Hall-Littlewood polynomials.
By design, the Macdonald trace formula  (\ref{mac}) reproduces respectively the Schur and Hall-Littlewood trace formulae  in the same limits.

Our basic ansatz is that the complete set of functions $\{ f_{q,t}^\lambda (\mathbf{a})\}$
that diagonalize the structure constants are proportional to
 Macdonald polynomials with parameters $q$ and $t$,
 \be
 f_{q,t}^\lambda (\mathbf{a}) = {\cal K}_{q,t}(\mathbf{a})\, P^\lambda(\mathbf{a}|q,t)\, .
\ee
The Macdonald polynomials~\cite{Mac}\footnote{
Macdonald polynomials appear in physics in
many different contexts. Some recent papers on subjects related 
to ${\mathcal N}=2$ gauge theories that discuss Macdonald polynomials 
are~\cite{Awata:2009ur,Schiappa:2009cc,Mironov:2011dk}.
}
$\{ P^\lambda (\mathbf{a}) \}$ are defined as the set of polynomials labeled by Young diagrams $\lambda$, orthonormal
under the measure
\be\label{macme}
\Delta_{q,t}({\mathbf a})=
\frac{1}{k!}\,PE\left[-\frac{1-t}{1-q}\,\left(\chi_{adj}({\mathbf a})-k+1\right)\right]_{q,t,{\mathbf a}}=\frac{1}{k!}\prod_{n=0}^\infty\prod_{i\neq j}\frac{1-q^{n}a_i/a_j}{1-t\,q^{n}a_i/a_j}\,\,,
\ee and having the  expansion
\be
P^\lambda={\mathcal N}_\lambda(q,t)\left\{m_\lambda+\sum_{\mu<\lambda}h_{\lambda\mu}(q,t)\,m_\mu\right\}\,.
\ee Here we define
\be
m_{\lambda=(\lambda_1,..,\lambda_k)}(\mathbf a)=\sum_{\sigma\in S'_k} \prod_{i=1}^k a_i^{\sigma(\lambda_i)}\,,
\ee where $S'_k$ denotes the set of distinct permutations of $(\lambda_1,...,\lambda_k)$.

The factor  ${\cal K}_{q,t}(\mathbf{a})$ is again fixed by requiring orthonormality
of $\{ f_{q,t}^\lambda (\mathbf{a})\}$ under the propagator measure. The propagator takes the standard form
\be
 \eta (\mathbf{a}, \mathbf{b}^{-1}) = \Delta (\mathbf{a})  {\mathcal I}^V(\mathbf{a}) \delta(\mathbf{a}, \mathbf{b}^{-1}) \, ,
\ee
where as always $\Delta(\mathbf{a})$ is the Haar measure (\ref{Haarexpression}), while
  the index of the vector multiplet is in this case 
 given by
\be
{\mathcal I}^V_{q,t}(\mathbf{a})=PE\left[\frac{-q-t}{1-q}\chi_{adj}(\mathbf{a})\right]_{q,\mathbf{a}}
\,.
\ee
We then have
\be
{\cal K}_{q,t}(\mathbf{a}) &=&\left( \frac{\Delta_{q,t}({\mathbf a})}{\Delta(\mathbf{a})\, {\cal I}^V_{q,t}(\mathbf{a})} \right)^{\frac{1}{2}} \,.
\ee

We can finally  state our main conjecture.
The Macdonald index of the $SU(k)$ quiver theory associated to a sphere with three punctures
of generic type is

\be\label{macgen}
\boxed{{\mathcal I}_{\Lambda_1,\Lambda_2,\Lambda_3}=
(t;q)^{k+2}\prod_{j=2}^{k}\frac{(t^j;q)}{ (q;q)}\prod_{I=1}^3\hat {\mathcal K}_{\Lambda_I}(\mathbf{ a}_I)
\sum_{{\lambda}}\frac{\prod_{I=1}^3P^{{\lambda}}(\mathbf{a}_I(\Lambda_I)|q,t)}
{P^{{\lambda}}(t^{\frac{k-1}{2}},t^{\frac{k-3}{2}},\dots,t^{\frac{1-k}{2}}|q,t)}\,}\, .
\ee

 The assignment of fugacities according to the Young diagram $\mathbf{a_i}(\Lambda_i)$ is  again as in figure~2,
with $\t\to t^{1/2}$.
 The $\hat {\mathcal K}$ prefactors are
\be
\hat {\mathcal K}_{\Lambda}(\mathbf{ a})=
\prod_{i=1}^{row(\Lambda)}\prod_{j, k=1}^{l_i}PE\left[\frac{{\frak a}^i_j\bar {\frak a}^i_k}{1-q}\right]_{{\frak a}_i,q}\, ,
\ee
with the coefficients ${\frak a}^i_k$ associated to the Young diagram again as in figure~3, with $\t\to t^{1/2}$.
It is immediate to check that (\ref{macgen}) reduces to the HL and Schur expressions in the respective limits.
For three maximal punctures~\eqref{macgen} becomes,
\be  \label{Tkmac}
{\mathcal I}_{T_k}(\mathbf{a_1,a_2,a_3})&=&
\sum_{\lambda_1\geq\lambda_2\geq...\geq\lambda_{k-1}}
\frac{{\cal A}(q,t)}{P^{\lambda_1,..,\lambda_{k-1}}(t^{\frac{k-1}2},..,t^{\frac{1-k}2}|\;q,\;t)}
\prod_{I=1}^3 {\mathcal K}_{q,t}({\mathbf a_I})\, P^{\lambda_1,..\lambda_{k-1}}(\mathbf{a}_I|\;q,\;t)\,,\nonumber\\
{\cal A}(q,t)&=&PE\left[\half(k-1)\frac{t-q}{1-q}\right]\,
\prod_{j=2}^{k}(t^j;q)\,.
\ee
For $k=2$,  this expression must agree with  the index of the hypermultiplet in the trifundamental 
representation of $SU(2)$, 
\be
&&PE\left[\frac{t^{1/2}}{1-q}\left(a_1+\frac1{a_1}\right)\left(a_2+\frac1{a_2}\right)\left(a_3+\frac1{a_3}\right)\right]_{a_i,q,t}=\\
&&\qquad \frac{(t;q)^{4}(t^2;q)}{(q;q)}\prod_{i=1}^3 PE\left[\frac{t}{1-q}\left(a_i^2+a_i^{-2}+2\right)\right]_{a_i,q,t}\,
\sum_{\lambda=0}^\infty\frac{\prod_{i=1}^3P^{\lambda}(a_i,a_i^{-1}|q,t)}
{P^{\lambda}(t^{\frac{1}{2}},t^{-\frac{1}{2}}|q,t)}\,.\nonumber
\ee We have verified this identity in the $t$ and $q$ expansions. It helps that for $SU(2)$  one can write
an explicit form for the Macdonald polynomials,
\be
P^{\lambda}(a,a^{-1}|q,t)={\mathcal N}_\lambda(q,t)\,\sum_{i=0}^\lambda \prod_{j=0}^{i-1}\frac{1-t\,q^j}{1-q^{j+1}}
 \prod_{j=0}^{\lambda-i-1}\frac{1-t\,q^j}{1-q^{j+1}}\,a^{2i-\lambda}\,,
\ee where ${\mathcal N}_\lambda(q,t)$ is a normalization constant
rendering the Macdonald polynomials orthonormal under the measure~\eqref{macme}.
More generally, equating the index for the $(nn1)$ vertex from (\ref{macgen}) with the index of a hypermultiplet in the bifundamental representation of  $SU(k)$
and charged under  $U(1)$, we obtain the identity
\be
&&PE\left[\frac{t^{1/2}}{1-q}\left(c\sum_{i,j=1}^ka_ib_j+\frac1c\sum_{i,j=1}^ka^{-1}_ib^{-1}_j\right)\right]_{a,b,c,q,t}=
\frac{(t;q)^{k}}{(q;q)^{k-1}}(t^k;q)\,\times\\
&&\quad PE\left[\frac{t}{1-q}\sum_{i,j=1}^k a_i\,a_j^{-1}\right]_{a,q,t}\,
PE\left[\frac{t}{1-q}\sum_{i,j=1}^k b_i\,b_j^{-1}\right]_{b,q,t}\,
PE\left[\frac{t^{\frac{k}{2}}}{1-q}(c^{k}+c^{-k})\right]_{c,q,t}\,
\times\nonumber\\
&&\quad \sum_{{\lambda}}\frac{P^{{\lambda}}(c\, t^{\frac{k-2}{2}},c\, t^{\frac{k-4}{2}},\dots,c\, t^{\frac{2-k}{2}},c^{1-k}|q,t)
P^{{\lambda}}(a_i|q,t)P^{{\lambda}}(b_i|q,t)}
{P^{{\lambda}}(t^{\frac{k-1}{2}},t^{\frac{k-3}{2}},\dots,t^{\frac{1-k}{2}}|q,t)}\,. \nonumber
\ee
It would be interesting to have an analytic proof of these identities.
\begin{table}
\begin{center}
\begin{tabular}{|m{0.4in}|m{1.in}|c|}
\hline
\footnotesize{Symbol} & \footnotesize{Surface} & \footnotesize{Value} \\
\hline
\hline
&&\\
$C_{\a\b\g}$ & \epsfig{file=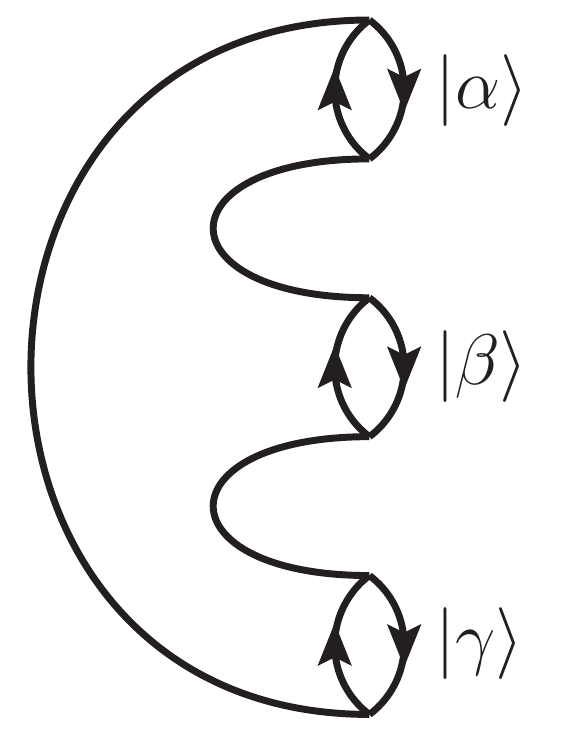,scale=0.4}  &
$\frac{{\mathcal A}(q,t)}{dim_{q,t}(\a)}
\;\delta_{\a\beta}\;\delta_{\a\gamma}$\\
\hline
&&\\
 $V^{\a}$ & \epsfig{file=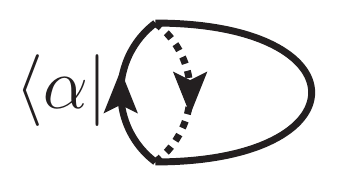,scale=0.4}  &
$\frac{dim_{q,t}(\a)}{{\cal A}(q,t)}$\\
\hline
&&\\
$\eta^{\a\b}$ & \epsfig{file=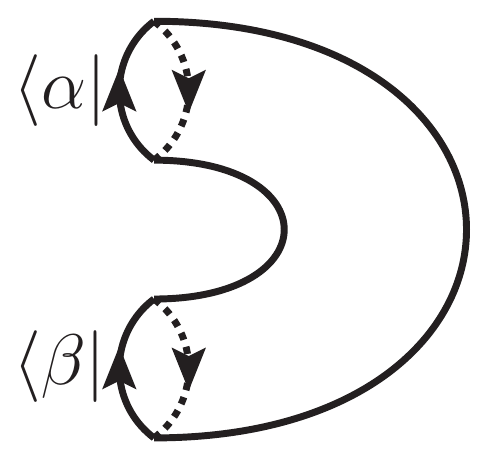,scale=0.4} &
$\delta^{\a\beta}$\\
\hline
\end{tabular}
\end{center}
\caption{\label{strucs1}The structure constants, the cap, and the metric
for the TQFT of the Macdonald index.%
 }
\end{table}

From (\ref{Igs}) we can readily  calculate the index of the genus ${\frak g}$ theory with $s$ punctures,
\be
{\mathcal I}_{{\frak g},s} (\mathbf{a}_I; q,t)=\prod_{j=2}^{k}(t^j;q)^{2{\frak g}-2+s}\frac{(t;q)^{(k-1)(1-{\frak g})+s}}
{(q;q)^{(k-1)(1-{\frak g})}}\;
\sum_{\lambda}
\frac{\prod_{i=1}^s \hat {\mathcal K}_{\Lambda_i}({\mathbf a_i})\;P^\lambda({\mathbf a_i}(\Lambda_i)|q,t)}
{\left[P^\lambda(t^{\frac{k-1}{2}},t^{\frac{k-3}{2}},\dots,t^{\frac{1-k}{2}}|q,t)\right]^{2{\frak g}-2+s}}\,.\nonumber\\
\ee
Let us dwell upon this result. Let us first consider the genus ${\frak g}$ partition function
(no punctures) in the Schur limit, $q=t$. We can write it as
\be\label{genGYM}
{\mathcal I}_{{\frak g}}(q)=\left[(q;q)^{2{\frak g}-2}\right]^{k-1}\;S_{00}(q)^{2-2{\frak g}}\;
\sum_{\lambda}
\frac{1}
{\left[dim_q(\lambda)\right]^{2{\frak g}-2}}\, . 
\ee Here $S_{00}$ is the partition function of  $SU(k)$  level ${\ell}$ Chern-Simons theory on $S^3$  if
we formally  identify $q=e^{\frac{2\pi i}{{\ell}+k}}$,
\be
S_{00}(q)=\prod_{j=2}^{k}\frac{(q;q)}{(q^j;q)}\,.
\ee 
The expression~\eqref{genGYM}, up to the simple  factor $\left[(q;q)^{2{\frak g}-2}\right]^{k-1}$,
 is the genus ${\frak g}$
partition function of  $q$-deformed $2d$ Yang-Mills  theory in the zero area limit~\cite{Aganagic:2004js},
which is in fact the same as  the partition function of  $SU(k)$ level ${\ell}$ Chern-Simons theory
on ${\Surface}_{\frak g}\times S^1$ with $q=e^{\frac{2\pi i}{{\ell}+k}}$~\cite{Aganagic:2004js}.\footnote{More precisely, $q$-deformed Yang-Mills theory in the zero area limit can be viewed as an analytical continuation
of Chern-Simons theory, or equivalently of the G/G WZW model (see~\cite{Blau:1993hj} for a review of the latter),
to non-integer rank $\ell$. }
If we  reintroduce  punctures,  the index is related to a correlator
 the $q$-deformed $2d$ Yang-Mills theory; the relation involving both a
 flavor independent factor and   flavor-dependent factors $\hat {\mathcal K}_\Lambda$ associated to the punctures.
We have recovered in more generality the relation found in~\cite{Gadde:2011ik} between the  Schur index and  $2d$ $q$-deformed Yang-Mills theory.\footnote{Ordinary $2d$ Yang-Mills theory~\cite{Witten:1991we, Witten:1992xu} is obtained by sending $q\to1$. From the index perspective,
because of the additional overall factors, this is a singular limit. However, with  proper regularization
this limit can be understood as  reducing the $4d$ index to a $3d$ partition function~\cite{Dolan:2011rp,Gadde:2011ia,Imamura:2011uw}. 
See also~\cite{Benini:2011nc} for yet another 3d/4d connection.}

In the more general case of $q\neq t$ the genus ${\frak g}$ partition function 
  can be written as
\be\label{QTYM}
{\mathcal I}_{{\frak g}}(q,t)=\left[(t;q)^{{\frak g}-1}\;(q;q)^{{\frak g}-1}\right]^{k-1}\;\hat S_{00}(q,t)^{2-2{\frak g}}\;
\sum_{\lambda}
\frac{1}
{\left[dim_{q,t}(\lambda)\right]^{2{\frak g}-2}}\,,
\ee where the generalized quantum dimension is given by
\be
dim_{q,t}(\lambda)=P^\lambda(t^{\frac{k-1}{2}},t^{\frac{k-3}{2}},\dots,t^{\frac{1-k}{2}}|q,t)\,\ee
and we have defined
\be
\hat S_{00}(q,t)=\prod_{j=2}^{k}\frac{(t;q)}{(t^j;q)}\,.
\ee
This result appears to be closely related to the
{\it refinement} of Chern-Simons theory  recently discussed by Aganagic and Shakirov~\cite{Aganagic:2011sg}.
 Up to overall factors,  ${\mathcal I}_{{\frak g}}(q,t)$ is equal to the partition function of refined Chern-Simons 
on ${\Surface}_{\frak g}\times S^1$.
In terms of the Chern-Simons matrix model the refinement of ~\cite{Aganagic:2011sg} amounts 
to changing the matrix integral measure from Haar to Macdonald. We can thus identify
the $2d$  theory whose correlators give the 
Macdonald index  as the theory  obtained from $q$-Yang-Mills theory by deforming in the same way the path integral measure.
It would be interesting to find a more conventional Lagrangian description of this $2d$ theory, for example the deformed measure  could arise by integrating out some matter fields.
It would also be desirable to have a better understanding of the  flavor-independent factors needed to relate  $2d$ Yang-Mills ($q$-deformed or $(q,t)$-deformed)
to the index.  They can be formally associated to a decoupled TQFT with a single operator (the identity). Perhaps this decoupled
TQFT plays a similar role as the decoupled $U(1)$ factor in the AGT correspondence \cite{Alday:2009aq}.

\section{Coulomb-branch index}\label{coulsec}

Finally we consider the index
\be\label{CoulIndAagain}
{\mathcal I}_{CM}(T, Q)=
\Tr_{CM}(-1)^F\,T^{\half(E+2j_1-2R-r)}\,Q^{\half(E+2j_2+2R+r)}\,,
\ee 
where $\Tr_{CM}$ stands for the trace over states with $E+2j_1+r=0$. 
This limit of the full index makes sense
for theories with a Lagrangian description, since the single-letter partition functions have well-defined expressions,
\be
f^{\half H}=0,\qquad f^V=\frac{T-Q}{1-Q}\,.
\ee 
Theories connected to Lagrangian theories by dualities also have a well-defined ${\cal I}_{CM}(T, Q)$.
As discussed in section 4,
the further limit $Q\to 0$ leads to the ${\cal I}_C(T)$ index, which is guaranteed to be well-defined for any ${\cal N}=2$ SCFT.

We refer to  (\ref{CoulIndAagain}) as the ``Coulomb-branch'' index, or Coulomb index for short, because in a Lagrangian theory it receives contributions
only from the $\bar {\cal E}$-type short-multiplets (see Appendix \ref{short}), whose bottom components are the gauge-invariant operators that parametrize
the Coulomb branch, for example 
\be \label{Trphik}
\mbox{Tr}\,\phi^{2},\mbox{ Tr}\,\phi^{3},\ldots,\mbox{ Tr}\,\phi^{k}\,\ee
for a  theory with $SU(k)$ gauge group. Since the hypermultiplets do not contribute,
the Coulomb index is independent of the flavor fugacities and the TQFT structure is very simple.
 The structure constants associated to a three-punctured sphere depend only on $T$ and $Q$,
  and so does the propagator, since the gauge-group matrix integral can be carried
  out independently of what the propagator connects to. The index of a quiver is then just
  a product over the indices of its constituents (propagators and vertices).

The index of 
a vector multiplet in the adjoint representation
of a  simple gauge group ${\mathcal G}$ is
\be\label{indCoul}
{\mathcal I}^V_{({\mathcal G})}(Q,T)=\oint_{{\mathbb T}^{r_{\mathcal G}}} \prod_{i=1}^{r_{\mathcal G}}\frac{da_i}{2\pi i}\Delta_{\mathcal G}({\mathbf a})\,
\exp\left[-\sum_{n=1}^\infty \frac{1}{n}\frac{Q^n-T^{n}}{1-Q^n}\chi_{adj}^{({\mathcal G})}({\mathbf a^n})\right]\, ,
\ee 
where $r_{\mathcal G}$ is the rank of ${\mathcal G}$ and 
 $\Delta_{\mathcal G}({\mathbf a})$ the Haar measure,
 \be
\Delta_{\mathcal G}({\mathbf a})=\frac{1}{|W_{\mathcal G}|}\exp\left[-\sum_{n=1}^\infty\frac{1}{n} (\chi_{adj}({\mathbf a}^n)-r_{\mathcal G})\right]\,,
\ee with $|W_{\mathcal G}|$  the order of the Weyl group.
We recognize the integrand in (\ref{indCoul}) as  the Macdonald measure~\eqref{macme} with parameters $Q$ and $T$.
The integral can be evaluated explicitly thanks to Macdonald's celebrated constant-term identities  (see {\it e.g}.~\cite{Mac,cherednik2005double} for  pedagogic expositions and~\cite{Kir} for a brief review),
\be \label{result}
{\cal I}_{{\cal G}}^{V}=PE\left[\,\mbox{rank}({\cal G})\tilde{{\cal I}}_{1}
+\sum_{\alpha\in R^{+}}\tilde{{\cal I}}_{1+{\Surface}_{\alpha}}-\tilde{{\cal I}}_{{\Surface}_{\alpha}}\right],
\qquad\qquad{\Surface}_{\alpha}\equiv\sum_{\beta\in R^{+}}\frac{(\alpha,\beta)}{(\beta,\beta)} \, ,
\ee
where $R^+$ is the collection of positive roots of ${\mathcal G}$ and
\be
\tilde {\cal I} _{\ell}=T^{\ell-1}\frac{T-Q}{1-Q} \,.
\ee
We recognize $\tilde{{\cal I}}_{\ell}$ as the index of the $\bar {\cal E}_{-\ell(0,0)}$ superconformal  multiplet,
which satisfies the shortening condition $E = \ell$ (see appendix~\ref{short}).
By a Lie-algebraic identity,  (\ref{result}) can be rewritten more succinctly as~\cite{Kir}
\be\label{evalMac}
{\cal I}_{{\cal G}}^{V}=PE\left[\sum_{j\in \mbox{exp}({\cal G})} \tilde{\cal I}_{j+1}\right]\,,
\ee
where exp$({\cal G})$ stands for the set of exponents of the Lie group ${\cal G}$. 
This result has an immediate physical interpretation.
The Coulomb index is saturated by the $\bar {\cal E}$-multiplets, whose bottom components are the  gauge-invariant
operators made of $\phi$s. The single-particle index (the argument of the plethystic exponential in (\ref{evalMac}))
then counts the independent gauge-invariant operators made of $\phi$s,  which are  in 1-1 correspondence with the Casimirs of the group,
that is with exp$({\cal G})$.
For example, for ${\cal G} = SU(k)$, $\mbox{exp}({\cal G})= \{ 1, 2,\dots, k-1 \}$, and we see that the Coulomb index counts the independent single-trace operators (\ref{Trphik})
that parametrize the Coulomb branch.
Turning the logic around, we can  view this as a ``physical'' (or perhaps, combinatorial)
 proof of Macdonald's constant term identities.
The integral over the Macdonald measure~\eqref{indCoul}  counts gauge-invariant words  built 
from certain letters of the vector multiplet; 
from superconformal 
representation theory we can identify which short multiplets are relevant for this counting problem, and
deduce \eqref{evalMac}.

\begin{figure}[htbp]
\begin{center}
\includegraphics[scale=0.6]{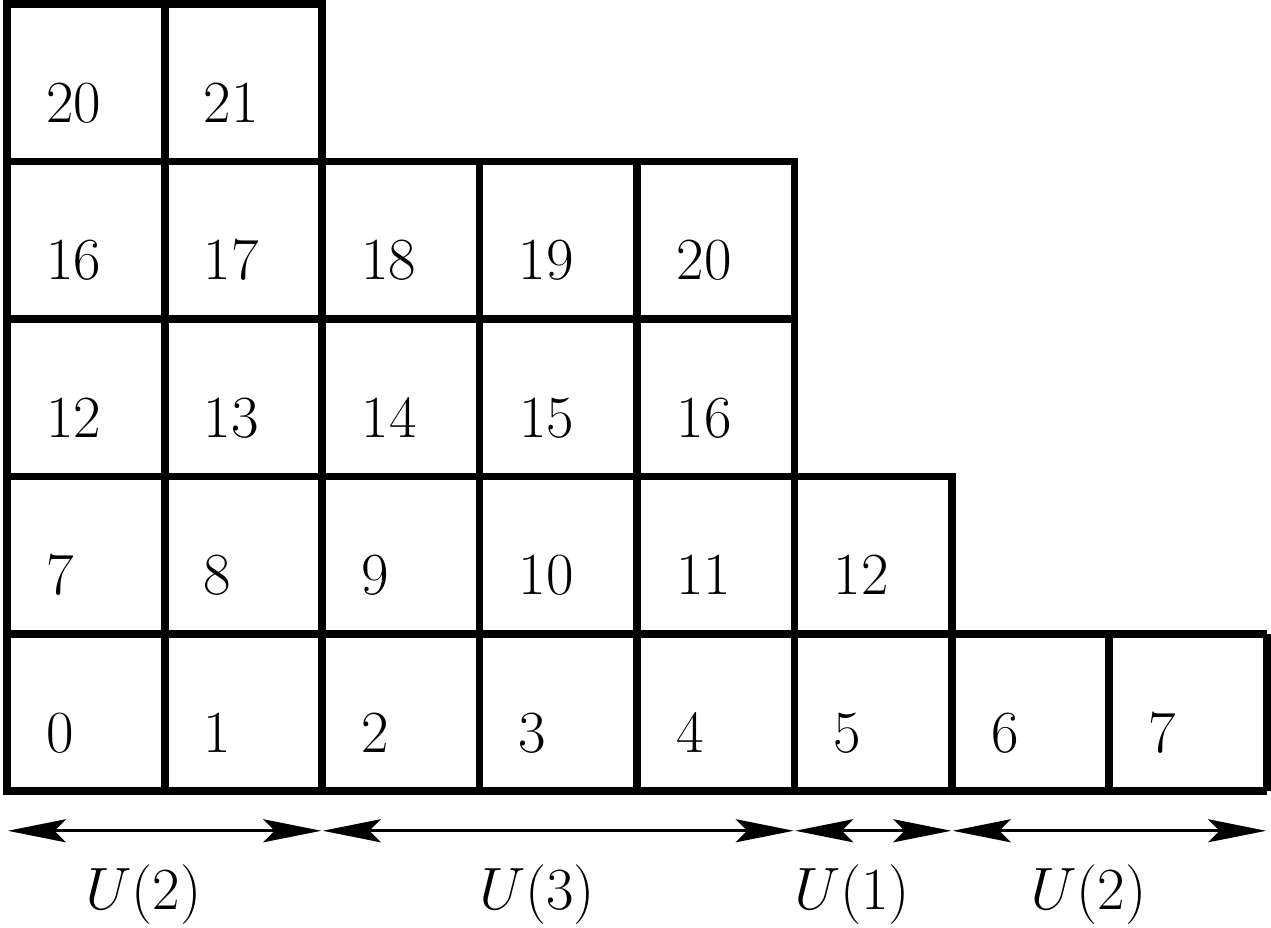}
\end{center}
\caption{The bottom left box is assigned $0$. The assigned integer increases from left to right. As we move up, the first
  box of  each row is assigned the same number as the last box in the  row below.}
\end{figure}\label{coulfig}

Though the TQFT structure for the Coulomb index is very simple, it is not entirely trivial.
We can deduce the Coulomb index of strongly-coupled theories by using dualities,
and check that different routes to obtain the index give the same result.
For example, using Argyres-Seiberg
duality~\cite{argyres-2007-0712}
\be
{\cal I}_{E_{6}}=\frac{{\cal I}_{SU(3)}^{V}}{{\cal I}_{SU(2)}^{V}}=PE[\tilde{{\cal I}}_{3}] \, ,
\ee
which is the expected result since the Coulomb branch of the $E_6$ 
SCFT is generated by an operator with $E=|r| = 3$.
Strongly coupled SCFTs are sometimes
obtained using S-dualities in more than one way~\cite{Argyres:2007tq}  but all the dualities
yield the same index,  for example
\begin{eqnarray}
{\cal I}_{E_{6}} & = & \frac{{\cal I}^V_{SU(3)}}{{\cal I}^V_{SU(2)}}=\frac{{\cal I}^V_{SU(4)}}{{\cal I}^V_{USp(4)}}=PE[\tilde{{\cal I}}_{3}]\,,\nonumber\\
{\cal I}_{E_{7}} & = & \frac{{\cal I}_{SU(4)}^{V}}{{\cal I}_{SU(3)}^{V}}=\frac{{\cal I}_{USp(4)}^{V}}{{\cal I}_{SU(2)}^{V}}
=\frac{{\cal I}_{SO(7)}^{V}}{{\cal I}_{G_{2}}^{V}}=\frac{{\cal I}_{SO(8)}^{V}}{{\cal I}_{SO(7)}^{V}}=PE[\tilde{{\cal I}}_{4}]\,,\\
{\cal I}_{E_{8}} & = & \frac{{\cal I}_{SU(6)}^{V}}{{\cal I}_{SU(5)}^{V}}=\frac{{\cal I}_{USp(6)}^{V}}{{\cal I}_{SO(5)}^{V}}=PE[\tilde{{\cal I}}_{6}]\,. \nonumber
\end{eqnarray}
The index of the $T_{k}$ theory is also obtained easily from the
generalized Argyres-Seiberg duality,
\be
{\cal I}_{T_{k}}=\frac{({\cal I}_{SU(k)}^{V})^{k-2}}{\prod_{j=2}^{k-1}{\cal I}_{SU(j)}^{V}}=
PE\left[\sum_{j=3}^{k}(j-2)\tilde{{\cal I}}_{j}\right]\,.
\ee
This is again as expected,  since the Coulomb branch of the $T_{k}$
theory is spanned by $(j-2)$ operators with $E = |r | = j$, for
 $j=3,\ldots,k$ (see {\it {\it e.g.}}~\cite{Benini:2009gi}).

Extrapolating from these examples let us conjecture  the Coulomb index of the theory corresponding
 to a sphere with three generic punctures. For a general puncture $I$ in the $A_{k-1}$ theory,
 we associate the set of $k$ numbers $\{p_{j}^{(I)}:j=1,\ldots k\}$ from the corresponding auxiliary Young diagram.
  The assignment is illustrated in figure~6.
The Coulomb branch index of the theory corresponding
to a sphere with three punctures $p^{(1)},p^{(2)},p^{(3)}$ is then
\be
{\cal I}_{p^{(1)},p^{(2)},p^{(3)}}=PE\left[d_j \;\tilde{{\cal I}}_{j}\right] \, ,\qquad d_j \equiv \sum_{j=2}^{k}(1-2j+p_{j}^{(1)}+p_{j}^{(2)}+p_{j}^{(3)})\,.
\ee
The dimension $d_j$ of the Coulomb branch spanned by operators with $E=|r| = j$
 agrees with the dimension of the space of meromorphic $j$-differentials on the Riemann 
surface having  poles of order at most $p^{(I)}_j$ at puncture $I$ \cite{Gaiotto:2009we, Chacaltana:2010ks}.

Let us finally observe that the Coulomb index~\eqref{indCoul} discussed in this section 
can also be interpreted as the index of ${\mathcal N}=4$ SYM in a certain limit of 
the ${\mathcal N}=4$ superconformal fugacities, such that the index of the ${\mathcal N}=4$ vector multiplet reduces  
 to the index of ${\mathcal N}=2$ vector multiplet. The authors of~\cite{Spiridonov:2010qv}
 noticed the appearance of the Macdonald
measure in this context.

\

\section{Discussion}\label{discsec}

Let us briefly summarize and discuss our results. 
We  have defined and studied several limits of the  ${\cal N}=2$ superconformal index.
They are characterized by enhanced supersymmetry and
 depend at most on two superconformal fugacities, out of the possible three. 
We have given a  prescription to calculate these limits for {all}
 $A$-series superconformal quivers of class ${\cal S}$, even when they lack a Lagrangian description.
Thanks
to the topological QFT structure of the index, it suffices to
find a formula 
for the elementary three-valent building blocks. For the $SU(2)$ quivers,
which do have a Lagrangian description, the building blocks
 can be written in terms of algebraic objects that admit
 a natural extrapolation to higher rank, leading to a compelling general conjecture that 
 passes many tests. These objects are the Macdonald polynomials, 
 tailor-made for our purposes as they depend on two fugacities, and
for which a beautiful general theory is already available.  We expect the generalization
of our results to the $D$-series quivers of  class ${\cal S}$ (and possibly to the $E$-series as well)
to be straightforward.

The  TQFT that calculates the index of the $A_{k-1}$ quivers
 is closely related to two-dimensional Yang-Mills theory with gauge group $SU(k)$.
An immediate qualitative hint, of course, is that the state-space of the index TQFT  is the space of irreducible $SU(k)$ representations.
As first discussed in~\cite{Gadde:2011ik}, and confirmed here in more generality,
there is in fact a precise quantitative correspondence between the limit of the index that we have dubbed the
``Schur index'', which depends on a single fugacity $q$,
  and 
 correlators of $q$-deformed $2d$ Yang-Mills theory~\cite{Aganagic:2004js} in the zero-area limit. In turn, the zero-area limit of $q$-deformed $2d$ Yang-Mills  on the Riemann surface
 ${\cal C}$ can be viewed as an analytic continuation of 
 Chern-Simons theory on ${\cal C} \times S^1$~\cite{Aganagic:2004js}.

 Recently, a ``refinement''  of Chern-Simons theory on three-manifolds admitting a  circle action 
 was defined in~\cite{Aganagic:2011sg},  via the relation
with topological string theory and its embedding into M-theory. 
Taking the three-manifold to be of the form ${\cal C}\times S^1$, and reducing on the $S^1$, one obtains an indirect definition
of  ``refined $q$-deformed Yang-Mills theory'' on ${\cal C}$, which depends on two parameters $q$ and $t$.
(The definition is indirect because unlike the purely $q$-deformed case no Lagrangian description is available for the refined theory.)
  The refinement essentially amounts 
to trading Schur polynomials with Macdonald polynomials, and  we have found a precise relation between
 our  $(q,t)$ ``Macdonald index'' and correlators of this $(q,t)$-Yang-Mills theory.
 It is natural to ask whether this is pointing to a direct connection between topological
string theory and the superconformal index. At first sight the geometries involved 
appear to be quite different, since  to obtain the superconformal index we must consider  the $(2,0)$ theory on $S^3 \times S^1 \times {\cal C}$,
with  appropriate twists induced by the fugacities, 
while in the setup of \cite{Cecotti:2010fi, Aganagic:2011sg} the relevant geometry is $(\mathbb{C}  \times S^1 \times M_3)_{q,t}$,
where one may take  $M_3= S^1 \times {\cal C}$ (we refer to the cited papers for a proper explanation). 
Moreover while the index admits a further refinement for a total of three fugacities, it seems difficult to introduce a third parameter
 in the framework of \cite{Cecotti:2010fi, Aganagic:2011sg} while preserving supersymmetry.
Nevertheless, at least for the special case of the Macdonald index, there should be a deeper way to understand
the striking similarity of the two results.

An obvious direction for future work is the generalization of our results to the full three-parameter index.
The  Haar measure together with the index of the ${\mathcal N}=2$ vector multiplet combine to \cite{Dolan:2008qi,Gadde:2010te}
\be
\frac{1}{k!}\,\prod_{i,j=1,i\neq j}^k\frac{1}{\Gamma\left(x_i/x_j;\;q,p\right)\Gamma\left(t\,x_i/x_j;\;q, p\right)}\,, 
\ee where  $\Gamma(z;p,q)$ is the elliptic Gamma function
\be
\Gamma(z;p,q)=\prod_{i,j=0}^\infty\frac{1-p^{i+1}q^{j+1}/z}{1-p^{i}q^{j}z}\,.
\ee 
A natural speculation is that 
the functions $f^\lambda_{p,q,t} (\mathbf{a})$ 
that diagonalize
the structure constants of the full index should be proportional to 
 elliptic extensions of   
the Macdonald polynomials,
to which they should reduce
in the limit $p\to 0$ (or $q\to 0$). Various proposals for elliptic Macdonald functions 
have appeared in the mathematical literature, see {\it e.g}.~\cite{Etingof:1994az,rains, rains2}.
We can in fact formulate a more precise conjecture, motivated
by the relation between two-dimensional gauge theories and integrable
quantum mechanical models of Calogero-Moser (CM) type, see {\it e.g}.~\cite{Gorsky:1993pe,Minahan:1993mv,Gorsky:1993dq,Gorsky:1994dj,gorsky,Gorsky:2000px}. 
 The reduction of  ordinary $2d$ Yang-Mills theory
to one dimension yields the rational (non-relativistic) CM model~\cite{Gorsky:1993pe}. One can consider
the trigonometric and elliptic generalizations of the non-relativistic model, 
as well as their relativistic cousins
(the relativistic versions are also known as  Ruijsenaars-Schneider (RS) models).
The relativistic trigonometric model (trigonometric RS)
depends on two parameters $(q,t)$,
 has Macdonald polynomials as its eigenfunctions, and
 is closely related to the two-dimensional G/G WZW model\footnote{
 See~\cite{gorsky} for a review and~\cite{Gerasimov:2006zt}
for  recent relevant work.} or equivalently to Chern-Simons theory on ${\cal C} \times S^1$.
At the summit of this hierarchy  is the elliptic relativistic model (elliptic RS), which depends on three parameters, analogous to $(p,q,t)$ of the full index.
Our  conjecture is then that the symmetric
functions relevant for the computation of the full index
are the eigenfunctions
of the elliptic RS model. Not too much is known about them, see~\cite{R} for a  review.\footnote{
Quantum mechanical integrable models have been recently related to the problem
of counting vacua of  ${\mathcal N}=2$ supersymmetric theories in
 the $\Omega$-background~\cite{Nekrasov:2009rc,Nekrasov:2010ka}.  See also~\cite{SpirCM, Spiridonov:2010em}
 for connections of elliptic Gamma functions to integrable systems.}

Perhaps the most interesting open problem is to give a ``microscopic'' derivation of the two-dimensional TQFT of the index 
from the six-dimensional $(2,0)$ theory.
A promising shortcut, which exploits the mentioned connection between $2d$ gauge theories and $1d$ Calogero-Moser models,
 is along the following lines.  Consider the $(2,0)$ theory on $S^3 \times S_{(1)}^1 \times {\cal C}_{{\frak g},s}$.
The Riemann surface ${\Surface}_{{\frak g},s}$ can be viewed as a circle, $S_{(2)}^1$, times a graph  $I_{{\frak g},s}$
By  first reducing the $(2,0)$ theory on $S_{(2)}^1$ (note that there is no twist  around this circle)
 one obtains $5d$ super Yang-Mills on $S^3 \times S^1_{(1)} \times I_{{\frak g},s}$. We propose that the further reduction of $5d$ SYM on  
$S_{(1)}^1\times S^3$ (with the  fugacity twists) yields the elliptic RS model on the graph  $I_{{\frak g},s}$,
with appropriate boundary conditions at the $s$ external punctures and at the internal junctures.
 In a suitable  limit, which corresponds to taking   $S_{(1)}^1$  to be small,
the $4d$ index becomes the $3d$ partition function~\cite{Dolan:2011rp,Gadde:2011ia,Imamura:2011uw},
and our proposal reduces to the one of~\cite{Nishioka:2011dq} (see also~\cite{Benvenuti:2011ga}). 
These authors show how to interpret such $3d$ partition functions
as  overlaps of  quantum mechanical wave functions. We are suggesting that a similar
idea may apply to the $4d$ index, and  that the relevant quantum mechanical
model is the elliptic RS model. Work is in progress along these lines.

\

\

\noindent{\bf Acknowledgments}:~
We would like to thank
 M.~Aganagic, C.~Beem, F.~van~de~Bult, T.~Dimofte, D.~Gaiotto, S.~Gukov,  D.~Jafferis, A.~Kirillov~jr, J.~Maldacena,
 Y.~Nakayama, N.~Nekrasov, A.~Okounkov, H.~Ooguri, E.~Rains, B.~van~Rees, Y.~Tachikawa, and E.~Witten for very useful discussions.
The research of SSR was supported in part by NSF grant PHY-0969448 and he would like to thank  the Aspen Center for Physics, where part of this work was conducted
with the  support of the National Science Foundation under Grant No. 1066293. LR thanks the Galileo Galilei Institute
for hospitality and the INFN for partial support during the completion of this work.
AG would like to thank Tata Institute  for Fundamental Research for hospitality
during the final stages of this project.
This work was supported in part by NSF grant PHY-0969739.  

\appendix

\

\

\section{Construction of the diagonal expression for the $SU(2)$ HL index}\label{derHL}

In this appendix we diagonalize the structure constants of the $SU(2)$ quivers
in the $\q \to 0$, $\p \to 0$ limit.
With hindsight, we have dubbed this limit the Hall-Littlewood (HL) index, since 
the diagonal  functions  turn out to be closely related to the Hall-Littlewood polynomials. This 
is precisely what we show in this appendix.

 For  $SU(2)$, the SCFT associated to  three-punctured sphere is the 
the free hypermultiplet in the trifundamental representation. In the limit of interest, its index reads
\be \label{indexapp}
{\mathcal I}(a, b, c)  =PE\left[\t \chi_1(a)\chi_1(b)\chi_1(c)\right]_{a,b,c,\t}
= \frac{1}{\prod_{s_a,s_b,s_c=\pm1}(1-\t\,a^{s_a}\,b^{s_b}\,c^{s_c})}\, ,
\ee  where the fugacities $a$, $b$, and $c$ 
label the Cartans of the three $SU(2)$ flavor groups. The index of the vector multiplet 
and the $SU(2)$ Haar measure combine to 
\be \label{natural}
\Delta(a) {\cal I}^V(a, \tau) = (1- \t^2) \Delta_{\t^2,\t^4}(a) \, ,
\ee
 where $\Delta_{\t^2,\t^4}(a)$  is the Macdonald measure~\eqref{macme} with $q=\t^2$
and $t=\t^4$,
\be\label{measure2}
\Delta_{\t^2,\t^4}(a)=\half(1-a^2)(1-\frac{1}{a^2})(1-\t^2 a^2)(1-\frac{\t^2}{a^2})\,.
\ee 
The corresponding Macdonald polynomials $P^\lambda(a,a^{-1};q,t)$, normalized to be orthonormal under (\ref{natural}),
are\footnote{This normalization is only used in this appendix. In the rest of the paper Macdonald
polynomials are taken to have unit norm with respect to the Macdonald measure.}
\be\label{macs2}
P^\lambda(a;\t^2,\t^4)&=&
\frac{\t}{\sqrt{1-\t^2}\left(1-\frac{1}{a^2}\t^2\right) \left(1- a^2\t ^2\right)}
\sqrt{\chi_\lambda(\t)\chi_{\lambda+2}(\t)}
\left\{
\frac{\chi_\lambda(a)}{\chi_\lambda(\t)}-\frac{\chi_{\lambda+2}(a)}{\chi_{\lambda+2}(\t)}
\right\}\,.\nonumber\\
\ee
By choosing $\{ P^\lambda(a,a^{-1};q,t) \}$ as a basis, the metric of the TQFT is then trivial, $\eta^{\lambda \mu}= \delta^{\lambda \mu}$. On the other hand,
the projection of ${\mathcal I}(a, b, c)$  into the basis functions gives the structure constants $C_{\mu\nu\lambda}$,
\be
{\mathcal I}(a, b, c)=\sum_{\mu,\nu,\lambda=0}^\infty C_{\mu\nu\lambda}P^\mu(a;\t^2,\t^4) P^\nu(b;\t^2,\t^4) P^\lambda(c;\t^2,\t^4)\,.
\ee 
We find that while the structure constants are not diagonal, they take  a relatively simple ``upper triangular'' form. 
The only non-vanishing
coefficients are
\be \label{nonvan}
C_{\lambda \lambda \lambda}  \equiv \Psi_\lambda \, , \qquad C_{ \lambda \mu \mu}  = C_{\mu \lambda \mu} = C_{\mu \mu \lambda}  \equiv \varOmega_\lambda \quad {\rm for} \; \mu < \lambda \, , \; (-1)^{\lambda + \mu} = 1\, ,
\ee
where
\be\label{psiomega}
&&\Psi_\lambda(\t)=
\frac{  \sqrt{1-\t^2}  }  { \sqrt{ \chi_{\lambda+2}(\t)  }   }  \left(
\frac{\t^{-1}+\t}{\sqrt{\chi_{\lambda}(\t)}}-\t^{\lambda+3}\sqrt{\chi_\lambda(\t)}
\right)\,,\\
&&\varOmega_\lambda(\t)=
\sqrt{1-\t^2}(\t^{-1}+\t) \frac{1}{\sqrt{\chi_\lambda(\t)\chi_{\lambda+2}(\t)}}\nonumber\,.
\ee
Associativity is easy to check.  It is trivial for most choices of external states, 
the one interesting case being 
 the four-point function $\mu\mu\nu\nu$ with $\mu<\nu$ and having the same parity (both even or both odd).
 Equality of the two channels reads
 \be
\sum_{\lambda\geq \nu, (-1)^{\lambda+\mu}=1}C_{\mu\mu\lambda} C_{\lambda \nu \nu}=\left[C_{\mu\mu\nu}\right]^2\,,
\ee 
which amounts to (no sum on $\nu$)
\be\label{asscons}
\sum_{\lambda>\nu,(-1)^{\lambda+\mu}=1}\varOmega_{\lambda}(\t)^2+\varOmega_{\nu}(\t)\Psi_{\nu}(\t)=\varOmega_{\nu}(\t)^2\,.
\ee One can verify that this property is satisfied for the particular values of the coefficients given in~\eqref{psiomega}.\footnote{
One needs the identity $\sum_{k=0}^\infty\frac{1}{\sinh\a(2k+3)\sinh\a(2k+1)}=\frac{e^{-\a}}{2\sinh^2\a\cosh\a}$
and induction on $\nu$.
}

Let us now perform an orthogonal transformation that diagonalizes the structure constants. 
From~\eqref{nonvan} we see even and odd Macdonald polynomials do not mix with each other and thus
can carry our the diagonalization separately for each parity; 
the discussion below is restricted to the even parity case for definiteness. The Latin letter indices below, $j,\dots$,
run over the integers and correspond to half the value of the Greek indices used above. 

We define real symmetric matrices $N_i$ as\footnote{Note that since the metric is trivial, $\eta^{ij} = \delta^{ij}$,
the upper or lower position of the indices is immaterial.}
\be
\left(N_i \right)_{jk} \equiv C_{ijk}\, .
\ee Associativity  implies that they commute, $[N_i, N_j]=0$,
so they can be simultaneously 
diagonalized. Recall that the structure of each matrix $N_j$ is 
\be
{(N_j)}_{ik}=\left\{
\begin{array}{lcr}
 i<j,\;i=k &\quad &\Omega_j\\
 i=k=j&\quad&\varPsi_j\\
 i>j,\;k=j&\quad&\quad\Omega_i\\
k>j,\;i=j&\quad&\quad\Omega_k\\
\text{other}&\quad &0
\end{array}
\right.
\ee
The non-zero eigenvalues of this matrix are $\Omega_j$ with multiplicity $j$, and $\varPsi_j-\Omega_j$ with multiplicity one.
The unique eigenvector with eigenvalue $\varPsi_j-\Omega_j$ is
\be
{\mathbf e}_{j+1}=(0,\,\dots,\,0,\,\varPsi_{j}-\Omega_{j},\,\Omega_{j+1},\,\Omega_{j+2},\,\dots\,)\,,
\ee where there are $j$ zeros in the beginning of the vector. Note that the ${\mathbf e}_j$s are orthogonal to each other,
\be
{\mathbf e}_{j+1}\,\cdot {\mathbf e}_{k+1}=(\varPsi_j-\Omega_j)\Omega_j+\sum_{i>j}\left[\Omega_i\right]^2=0\,,
\ee where we took $j>k$ without loss of generality and used the associativity constraint \eqref{asscons}.
Moreover, the vectors ${\mathbf e}_i$ turn out to be eigenvectors of \textit{all} the matrices $N_j$,
\be
&i< j&\qquad :\qquad N_j\,\cdot {\mathbf e}_{i+1}=\Omega_j\;{\mathbf e}_{i+1}\,,\\
&i= j&\qquad :\qquad N_j\,\cdot {\mathbf e}_{i+1}=(\varPsi_j-\Omega_j)\;{\mathbf e}_{i+1}\,,\nonumber\\
&i> j&\qquad :\qquad N_j\,\cdot {\mathbf e}_{i+1}=0\,.\nonumber
\ee This can be shown from the definitions with the help of the associativity constraint \eqref{asscons}.
To complete this set of vectors to a basis we have to add one more vector, orthogonal to all ${\mathbf e}_j$,
\be
{\mathbf e}_0 = (\Omega_1,\,\Omega_2,\,\dots\,)\,.
\ee  This is  an eigenvector of all the matrices $N_j$ with eigenvalue $\Omega_j$.
We have thus managed to diagonalize the matrices $N_i$. In the diagonal
basis $\{ {\mathbf e}_j \}$ the matrices are given by (we use hatted indices to represent  components in the new basis)
\be
(N_j)_{\hat i \hat k}=\left\{
\begin{array}{lcc}
 j> \hat i\, ,  &\quad &\Omega_j \, \delta_{\hat i \hat j}\\
 \hat i=j\, ,&\quad&(\varPsi_j-\Omega_j) \,\delta_{\hat i \hat j} \\
 j< \hat i,\; &\quad&0 
\end{array}
\right.
\ee Finally we perform the orthogonal  transformation to the new basis  also for the matrix label $j$ of $N_j$,
and find constants in the new basis read
\be
{ C}_{\hat j \hat i \hat k}=\frac{1}{n_{\hat j}}     \sum_l({\mathbf e}_{\hat j})_l\,\cdot{(N_l)}_{\hat i \hat k}\,,
\ee 
where $n_{\hat j}$ is the normalization of ${\mathbf e}_{\hat j}$,
\be
 n_{\hat j}   &= & \sqrt{{\mathbf e}_{\hat j}\cdot {\mathbf e}_{\hat j}} = \t^{2\hat j} \sqrt{1-\t^2}\,\;\quad \quad\quad \;{\rm for} \; \hat j > 0 \, ,\\
n_{\hat 0} & = & \sqrt{{\mathbf e}_{\hat 0}\cdot {\mathbf e}_{\hat 0}}= \sqrt{(1-\t^2)(1+\t^2)}\,. \nonumber
\ee 
A little calculation gives
\be
 C_{\hat i\hat  i \hat i } =n_{\hat i}  \, ,
\ee 
and zero for the other choices of the indices. So far we have restricted attention to even parity (in terms of the original Greek labels).
The case of odd parity works along completely parallel lines.

We can now explicitly compute the functions that diagonalize the structure constants,
 by contacting the normalized vectors ${\mathbf e}_{\mu}/n_{\mu}$ with the Macdonald polynomials~\eqref{macs2}.
A useful identity is $(\lambda >0)$ 
\be
\sum_{\mu=\lambda, \, (-1)^{\lambda+\mu}=1}^\infty P^{\mu}\,\Omega_{\mu}=\frac{1+ \t^2}{\left(1-\t^2a^2\right)\left(1-\t^2/a^2\right)}
\frac{\chi_{\lambda}(a) }{\chi_{\lambda}(\t)}\,.
\ee One finds that  the diagonal basis is  given by
\be
f^\lambda (a, \t) &=& 
\frac{1}{\sqrt{1-\t^2} } \frac{1}{\left(1-\t^2a^2\right)\left(1-\t^2/a^2\right)}
\left\{\chi_{\lambda}(a)- \t^{2}\chi_{\lambda-2}(a) \right\} \quad {\rm for} \;\; \lambda >0\, ,
\\
f^0 (a,\t) & = &\frac{1}{\sqrt{1-\t^2} } \frac{1}{\left(1-\t^2a^2\right)\left(1-\t^2/a^2\right)} \sqrt{1 + \t^2} \,. \nonumber
\ee 
 It is straightforward to verify that this basis is orthonormal under the measure~(\ref{natural}). Remarkably,
 the functions $f^\lambda(a, \t)$ are proportional to the $SU(2)$ Hall-Littlewood polynomials $P_{HL}^\lambda(a, a^{-1} | \t)$, see (\ref{SU2HL}),
 with a $\lambda$-independent proportionality factor ${\cal K}(a, \t)$.

Finally we can write the diagonalized form for the index,
\be
&&{\mathcal I}(a_1,a_2,a_3)=\\
&&\qquad\frac{1}{1-\t^2}
\prod_{i=1}^3\frac{1}{\left(1-\t^2a_i^2\right)\left(1-\t^2/a_i^2\right)}
\left\{(1+\t^2)^2+\sum_{\lambda=1}^\infty \t^{\lambda}\prod_{i=1}^3\left(\chi_{\lambda}(a_i)-\t^2\chi_{\lambda-2}(a_i)\right)\right\}\,.
\nonumber
\ee
The equality of this expression with (\ref{indexapp}) can be proven directly by elementary means since the sum above is a geometric sum.
By noting that
\be
\chi_{\lambda}(\t)-\t^2\chi_{\lambda-2}(\t)=\t^{-\lambda}(1+\t^2)\,,
\ee 
and recalling the definition (\ref{SU2HL}) of the HL polynomials  we can also write 
\be\label{ind222}
{\mathcal I}(a_1,a_2,a_3)=
\frac{1+\t^2}{1-\t^2}\prod_{i=1}^3\frac{1}{\left(1-\t^2a_i^2\right)\left(1-\t^2/a_i^2\right)}
\sum_{\lambda=0}^\infty \frac{1}{P^{HL}_\lambda(\t,\t^{-1}|\;\t)}\prod_{i=1}^3P^{HL}_\lambda(a_i,a_i^{-1}|\;\t)\,.\nonumber\\
\ee 

\

\section{Index of short multiplets of $\NN=2$ superconformal algebra}\label{short}

A generic long multiplet ${\cal A}_{R,r(j_1,j_2)}^{E}$ of the $\NN=2$
superconformal algebra is generated by the action of the eight Poincar\'e supercharges
$\QQ$ and $\tilde{\QQ}$ on a superconformal primary, which by definition is
 annihilated by all  conformal supercharges ${\cal S}$. If  some combination of
the  $\QQ$s  also annihilates the primary, the corresponding multiplet
is shorter and the conformal dimensions of all its members are protected against quantum corrections.
The shortening 
conditions for the $\NN=2$ superconformal algebra were studied in
 \cite{Dobrev:1985qv,Dobrev:1985qz,Dolan:2002zh}.
 We follow the nomenclature of \cite{Dolan:2002zh}, whose classification scheme is summarized in  table \ref{shortening}.
Let us take a moment to explain the notation.
The state $|R,r\rangle^{h.w.}_{(j_1,j_2)}$ is the highest weight state with
$SU(2)_{R}$ spin $R >0$, $U(1)_{r}$ charge $r$,
which can have either sign, and Lorentz quantum numbers  $(j_1,j_2)$.
The multiplet  built on this state is  denoted as $\mathcal{X}_{R,r(j_1,j_2)}$,
where the letter $\mathcal{X}$ characterizes the shortening condition.
The left column of table  \ref{shortening} labels
the condition. 
A superscript on the label  corresponds to the index $\II =1,2$ of the
supercharge that kills the primary:
for example ${\cal B}_{\suup}$ refers
to ${\cal Q}_{\suup\alpha}$. Similarly a ``bar'' on the label refers to the conjugate condition: for example
$\bar{\BB}_{\sudown}$ corresponds to $\tilde Q_{\sudown \, \dot \alpha}$ annihilating the state;
this would result in the short anti-chiral multiplet $\bar{\BB}_{R,r(j_1,0)}$, obeying $E = 2 R -r$.
Note that conjugation reverses the signs of $r$, $j_1$ and $j_2$ in the expression of the conformal dimension.

\begin{table}
\begin{centering}
\begin{tabular}{|c|l|l|l|l|}
\hline 
\multicolumn{4}{|c|}{Shortening Conditions} & Multiplet\tabularnewline
\hline
\hline 
$\BB_{\suup}$  & $\QQ_{\suup\alpha}|R,r\rangle^{h.w.}=0$  & $j_1=0$ & $E=2R+r$  & $\BB_{R,r(0,j_2)}$\tabularnewline
\hline 
$\bar{\BB}_{\sudown}$  & $\tilde{\QQ}_{\sudown \dot{\alpha}}|R,r\rangle^{h.w.}=0$  & $j_2=0$ & $E=2R-r$  & $\bar{\BB}_{R,r(j_1,0)}$\tabularnewline

\hline 
$\EE$  & $\BB_{\suup}\cap\BB_{\sudown}$  & $R=0$  & $E=r$  & $\EE_{r(0,j_2)}$\tabularnewline
\hline 
$\bar \EE$  & $\bar \BB_{\suup}\cap \bar \BB_{\sudown}$  & $R=0$  & $E=-r$  & $\bar \EE_{r(j_1,0)}$\tabularnewline
\hline 
$\hat{\BB}$  & $\BB_{\suup}\cap\bar{B}_{\sudown}$  & $r=0$, $j_1,j_2=0$  & $E=2R$  & $\hat{\BB}_{R}$\tabularnewline
\hline
\hline 
$\CC_{\suup}$  & $\e^{\alpha\beta}\QQ_{\suup\beta}|R,r\rangle_{\alpha}^{h.w.}=0$  &  & $E=2+2j_1+2R+r$  & $\CC_{R,r(j_1,j_2)}$\tabularnewline
 & $(\QQ_{\suup})^{2}|R,r\rangle^{h.w.}=0$ for $j_1=0$  &  & $E=2+2R+r$  & $\CC_{R,r(0,j_2)}$\tabularnewline
\hline 
$\bar \CC_{\sudown}$  & $\e^{\dot\alpha\dot\beta}\tilde\QQ_{\sudown\dot\beta}|R,r\rangle_{\dot\alpha}^{h.w.}=0$  &  & $E=2+2 j_2+2R-r$  & $\bar\CC_{R,r(j_1,j_2)}$\tabularnewline
 & $(\tilde\QQ_{\sudown})^{2}|R,r\rangle^{h.w.}=0$ for $j_2=0$  &  & $E=2+2R-r$  & $\bar\CC_{R,r(j_1,0)}$\tabularnewline
\hline 
  & $\CC_{\suup}\cap\CC_{\sudown}$  & $R=0$  & $E=2+2j_1+r$  & $\CC_{0,r(j_1,j_2)}$\tabularnewline
\hline 
  & $\bar\CC_{\suup}\cap\bar\CC_{\sudown}$  & $R=0$  & $E=2+2 j_2-r$  & $\bar\CC_{0,r(j_1,j_2)}$\tabularnewline
\hline 
$\hat{\CC}$  & $\CC_{\suup}\cap\bar{\CC}_{\sudown}$  & $r=j_2-j_1$  & $E=2+2R+j_1+j_2$  & $\hat{\CC}_{R(j_1,j_2)}$\tabularnewline
\hline   & $\CC_{\suup}\cap\CC_{\sudown}\cap\bar{\CC}_{\suup}\cap\bar{\CC}_{\sudown}$  & $R=0, r=j_2-j_1$ & $E=2+j_1+j_2$  & $\hat{\CC}_{0(j_1,j_2)}$\tabularnewline
\hline
\hline 
$\DD$  & $\BB_{\suup}\cap\bar{\CC_{\sudown}}$  & $r=j_2+1$  & $E=1+2R+j_2$  & $\DD_{R(0,j_2)}$\tabularnewline
\hline 
$\bar\DD$  & $\bar\BB_{\sudown}\cap{\CC_{\suup}}$  & $-r=j_1+1$  & $E=1+2R+j_1$  & $\bar\DD_{R(j_1,0)}$\tabularnewline
\hline 
 & $\EE\cap\bar{\CC_{\sudown}}$  & $r=j_2+1,R=0$  & $E=r=1+j_2$  & $\DD_{0(0,j_2)}$\tabularnewline
\hline
  & $\bar\EE\cap{\CC_{\suup}}$  & $-r=j_1+1,R=0$  & $E=-r=1+j_1$  & $\bar\DD_{0(j_1,0)}$\tabularnewline
\hline
\end{tabular}
\par\end{centering}
\caption{\label{shortening}Shortening conditions
and short multiplets for the  $\NN=2$ superconformal algebra.}
\end{table}

The superconformal index counts with signs the protected states of the theory, up to equivalence 
relations that set to zero all sequences of short multiplets that may in principle recombine 
into long multiplets. 
The recombination rules for ${\cal N}=2$ superconformal algebra are \cite{Dolan:2002zh}
\begin{eqnarray}
{\cal A}_{R,r(j_1,j_2)}^{2R+r+2j_1+2} & \simeq & \CC_{R,r(j_1,j_2)}\oplus\CC_{R+\frac{1}{2},r+\frac{1}{2}(j_1-\frac{1}{2},j_2)}
\label{recomb2}\,,\\
{\cal A}_{R,r(j_1,j_2)}^{2R-r+2 j_2+2} & \simeq & \bar\CC_{R,r(j_1,j_2)}\oplus\bar\CC_{R+\frac{1}{2},r-\frac{1}{2}(j_1,j_2-\frac{1}{2})}
\label{eq:3rd recomb}\,,\\
{\cal A}_{R,j_1-j_2(j_1,j_2)}^{2R+j_1+j_2+2} & \simeq & \hat{\CC}_{R(j_1,j_2)}\oplus\hat{\CC}_{R+\frac{1}{2}(j_1-\frac{1}{2},j_2)}
\oplus\hat{\CC}_{R+\frac{1}{2}(j_1,j_2-\frac{1}{2})}\oplus\hat{\CC}_{R+1(j_1-\frac{1}{2},j_2-\frac{1}{2})}\,.
\label{recomb1}
\end{eqnarray}
The ${\cal C}$, $\bar {\cal C}$ and $\hat {\cal C}$ multiplets
 obey certain ``semi-shortening'' conditions,  while ${\cal A}$ multiplets are generic long multiplets. 
 A long multiplet whose conformal dimension is exactly at the unitarity threshold can be decomposed  into shorter multiplets according to (\ref{recomb2},\ref{eq:3rd recomb},\ref{recomb1}).
  We can formally regard any  multiplet obeying some shortening condition (with the exception of the $\EE$ ($\bar \EE$) types, and $\bar{{\cal D}}_{0(j_{1},0)}$ (${\cal D}_{0(0,j_{2})}$) types)
 as a multiplet of  type ${\cal C}$, $\bar {\cal C}$ or $\hat \CC$  by allowing the spins $j_1$ and $j_2$, whose natural range is over the non-negative
 half-integers,  to  take the value $-1/2$ as well.   The translation is as follows:
\be
\CC_{R,r(-\frac{1}{2},j_2)}\simeq\BB_{R+\frac{1}{2},r+\frac{1}{2}(0,j_2)},\quad \bar\CC_{R,r(j_1,-\frac{1}{2})}\simeq\bar\BB_{R+\frac{1}{2},r-\frac{1}{2}(j_1,0)}\,,
\label{translation}
\ee
\be
\hat{\CC}_{R(-\frac{1}{2},j_2)}\simeq\DD_{R+\frac{1}{2}(0,j_2)},\qquad\quad\qquad\hat{\CC}_{R(j_1,-\frac{1}{2})}\simeq\bar{\DD}_{R+\frac{1}{2}(j_1,0)}\,, 
\ee
\be
\hat{\CC}_{R(-\frac{1}{2},-\frac{1}{2})}\simeq \DD_{R+\frac{1}{2}(0,-\frac{1}{2})} \simeq
\bar{\DD}_{R+\frac{1}{2}(-\frac{1}{2},0)} \simeq \hat{\BB}_{R+1}\,.
\ee
Note how these rules flip statistics: a multiplet with bosonic primary ($j_1+ j_2$ integer) is turned
into a multiplet with fermionic primary ($j_1 +  j_2$ half-odd), and vice versa.  
With these conventions, the rules (\ref{recomb2}, \ref{eq:3rd recomb}, \ref{recomb1}) are the most general recombination rules. 
The  $\EE$ and $\bar \EE$ multiplets never recombine.

The index of the $\CC$ and $\EE$ type multiplets vanishes identically (the choice of supercharge with respect
to which the index is computed, ${\cal Q} =\widetilde  {\cal Q}_{1 \dot -}$,
breaks the symmetry between $\CC$ ($\EE$) and $\bar \CC$ ($\bar \EE$) multiplets).
  The index of all remaining short multiplets can be specified by listing the index of $\bar\CC, \hat \CC$ , $\bar\EE$, ${\cal D}_{0(0,j_{2})}$, and $\bar {\cal D}_{0(j_{1},0)}$ multiplets,

\be
{\cal I}_{\bar\CC_{R,r(j_1, j_2)}}&=&
-(-1)^{2(j_1+ j_2)}\t^{2+2R+2 j_2}\p^{ j_2-r}\q^{ j_2-r}\frac{(1-\p\q)(\t-\p)(\t-\q)}{(1-\p\t)(1-\q\t)}\chi_{2j_1}\(\sqrt{\frac{\p}{\q}}\)\,,\nonumber\\
{\cal I}_{{\hat \CC}_{R(j_1, j_2)}}&=&
(-1)^{2(j_1+ j_2)}\frac{\t^{3+2R+2 j_2}\p^{j_1+\frac{1}{2}}\q^{j_1+\frac{1}{2}}(1-\p\q)}{(1-\p\t)(1-\q\t)}
\left(\chi_{2j_1+1}\left(\sqrt{\frac{\p}{\q}}\right)
-\frac{\sqrt{\p\q}}{\t}\chi_{2j_1}\left(\sqrt{\frac{\p}{\q}}\right)\right)\,,\nonumber\\
{\cal I}_{\bar\EE_{r(j_1,0)}}&=& (-1)^{2j_1} \p^{-r-1}\q^{-r-1} \frac{(\t-\p)(\t-\q)}{(1-\p\t)(1-\q\t)}\chi_{2j_1}\(\sqrt{\frac{\p}{\q}}\)\,,\label{indexE}\,\nonumber\\
{\cal I}_{\bar{{\cal D}}_{0(j_{1},0)}} & = & \frac{(-1)^{2j_{1}}(\sigma\rho)^{j_{1}+1}}{(1-\sigma\tau)(1-\rho\tau)}\times\nonumber\\
&&\qquad\qquad\left((1+\tau^{2})\chi_{2j_{1}}\left(\sqrt{\frac{\sigma}{\rho}}\right)-\frac{\tau}{\sqrt{\sigma\rho}}\chi_{2j_{1}+1}\left(\sqrt{\frac{\sigma}{\rho}}\right)-\tau\sqrt{\sigma\rho}\chi_{2j_{1}-1}\left(\sqrt{\frac{\sigma}{\rho}}\right)\right)\,,
\nonumber\\
{\cal I}_{{\cal D}_{0(0,j_{2})}} & = & \frac{(-1)^{2j_{2}+1}\tau^{2j_{2}+2}}{(1-\sigma\tau)(1-\rho\tau)}(1-\sigma\rho)\,.
\ee
where the Schur polynomial $\chi_{2j}\left(\sqrt{\frac{\p}{\q}}\right)$ gives the character of the spin $j$ representation of $SU(2)$.

Let us evaluate the interesting limits of the index studied in this paper on individual multiplets.

\subsubsection*{Macdonald index}
This index is obtained from the general index in the limit $\p\to0$. The 
index of the short multiplets in this limit is given by
\be
{\cal I}_{\bar\CC_{R,r(j_1, j_2)}}&=&0\,,\nonumber\\
{\cal I}_{{\hat \CC}_{R(j_1, j_2)}}&=&(-1)^{2(j_1+ j_2)}\frac{\t^{3+2R+2 j_2}\q^{2j_1+1}}{(1-\q\t)}\,,\\
{\cal I}_{\bar\EE_{r(j_1,0)}}&=& 0\,,\nonumber\\
{\cal I}_{\bar{{\cal D}}_{0(j_{1},0)}} & = & (-1)^{2j_{1}+1}\frac{\tau\rho^{2j_{1}+1}}{(1-\rho\tau)}\,,
\qquad
{\cal I}_{{\cal D}_{0(0,j_{2})}}  =  (-1)^{2j_{2}+1}\frac{\tau^{2j_{2}+2}}{(1-\rho\tau)}\,.\nonumber
\ee
While taking the limit of the $\bar \CC$ and $\bar \EE$ multiplet index we have used $j_2-j_1>r$ and $-r>j_1+1$ respectively.
 The first inequality follows from the  bound $\delta_{\suup - }\geq 0$ along with $\tilde \delta_{\suup \dot -}=0$ and the second
 one can be obtained by evaluating $\delta_{\suup - }\geq 0$ on the first descendant of the primary of the $\bar \EE$ multiplet.

\subsubsection*{Hall-Littlewood index}

This index is obtained from the Macdonald index by further taking the limit $\q\to0$.
The index of the short multiplets is
\be
{\cal I}_{\bar\CC_{R,r(j_1, j_2)}}&=&0\,,\nonumber\\
{\cal I}_{{\hat \CC}_{R(j_1, j_2)}}&=&-(-1)^{2 j_2}\t^{3+2R+2 j_2}\delta_{j_1,-\frac{1}{2}}\,,\\
{\cal I}_{\bar\EE_{r(j_1,0)}}&=& 0\,,\nonumber\\
{\cal I}_{\bar{{\cal D}}_{0(j_{1},0)}} & = & 0\,,\qquad
{\cal I}_{{\cal D}_{0(0,j_{2})}} =  (-1)^{2j_{2}+1}\tau^{2j_{2}+2}\,.\nonumber
\ee

\subsubsection*{Schur Index}
We take the limit $\t\to \q$. In this limit, the index becomes independent of $\p$ and the short multiplets give
\be
{\cal I}_{\bar\CC_{R,r(j_1, j_2)}}&=&0\,,\nonumber\\
{\cal I}_{{\hat \CC}_{R(j_1, j_2)}}&=&(-1)^{2(j_1+ j_2)}\frac{\t^{4+2(R+j_1+ j_2)}}{(1-\t^2)}\,,\\
{\cal I}_{\bar\EE_{r(j_1,0)}}&=&0\,,\nonumber\\
{\cal I}_{\bar{{\cal D}}_{0(j_{1},0)}} & = & (-1)^{2j_{1}+1}\frac{\tau^{2j_{1}+2}}{(1-\tau^{2})}\,,\qquad
{\cal I}_{{\cal D}_{0(0,j_{2})}}  =  (-1)^{2j_{2}+1}\frac{\tau^{2j_{2}+2}}{(1-\tau^{2})}\,.\nonumber
\ee

\subsubsection*{Coulomb Index}
Finally we take $\t \to 0$. In this limit only the $\bar\EE$ multiplet have a non-vanishing index  
\be
{\cal I}_{\bar\CC_{R,r(j_1, j_2)}}&=&0\,,\nonumber\\
{\cal I}_{{\hat \CC}_{R(j_1, j_2)}}&=&0\,,\\
{\cal I}_{\bar\EE_{r(j_1,0)}}&=& (-1)^{2j_1} (\p\,\q)^{-r} \chi_{2j_1}\(\sqrt{\frac{\p}{\q}}\)\,,\nonumber\\
{\cal I}_{\bar{{\cal D}}_{0(j_{1},0)}} & = & (-1)^{2j_{1}}(\sigma\rho)^{j_{1}+1}\chi_{2j_{1}}\left(\sqrt{\frac{\sigma}{\rho}}\right)\,,\qquad
{\cal I}_{{\cal D}_{0(0,j_{2})}}  =  0\,.\nonumber
\ee
The ${\mathcal N}=2$ vector multiplet is the direct sum of ${\cal D}_{0 (0,0)}$ and  ${\bar {\cal D}_{0(0,0)}}$, indeed
~\eqref{coulsinglet} is simply ${\bar {\cal D}_{0(0,0)}}$.\footnote{Note that in this limit 
${\cal I}_{\bar\EE_{-1(0,0)}}={\cal I}_{\bar{{\cal D}}_{0(0,0)}}$.
This is also true in the less restrictive Coulomb limit~\eqref{CMlimit}.}
In a Lagrangian theory, the only possible ${\bar{\cal D}}$
multiplets have $j_1 =0$, and are obtained from the  ${\bar{\cal D}_{0(0,0)}}$ half of the ${\cal N}=2$ vector multiplet.
In the less restrictive limit of $\p,\t\to0$ and $\q\to\infty$ the index of some of the short multiplets could 
potentially diverge. However, for Lagrangian theories the only contributing multiplets are
 ${\bar\EE_{r(0,0)}}$ multiplets  arising from tensor products of the ${\bar {\cal D}_{0(0,0)}}$
from the vector multiplet, whose index is finite.

\

\section{Large $k$ limit of the genus $\frak g$ HL index}\label{largeKsec}
In this appendix we give some details about the large $k$ limit of 
the HL index for  $SU(k)$ quivers corresponding to genus $\frak g$
surface with no punctures.  For finite $k$, the index is given by~\eqref{genusHL},
\be
{\mathcal I}^{(k)}_{\frak g}=\frac{\left(\prod_{j=2}^k (1-\t^{2j})\right)^{2\frak g-2}}{(1-\t^2)^{(k-1)(\frak g-1)}}
\sum_\lambda \frac{1}{P^{HL}_\lambda(\t^{k-1},\t^{k-3},\dots,\t^{1-k}|\t)^{2\frak g-2}}\,,
\ee   The denominator in the sum above is explicitly given by~\cite{Mac},
\be
P^{HL}_\lambda(\t^{k-1},\t^{k-3},\dots,\t^{1-k}|\t)=
{\mathcal N}_\lambda(\t)\,\t^{\sum_{i=1}^{k-1}\left(2i-k-1\right)\lambda_i}\,
\prod_{i=1}^{k}\frac{1-\t^{2i}}{1-\t^2}\,,
\ee where ${\mathcal N}_\lambda(\t)$ is given in~\eqref{normHL1},
\be
{\mathcal N}^{-2}_{\lambda_1,...\lambda_k}(\t)=\prod_{i=0}^\infty \prod_{j=1}^{m(i)}\,
 \left(\frac{1-\t^{2j}}{1-\t^2}\right)\, .
\ee Here $m(i)$ is the number of rows in the Young diagram $\lambda=(\lambda_1,\dots,\lambda_k)$ of length $i$. We need   to evaluate 
\be 
{\mathcal I}^{(k)}_{\frak g}&=&(1-\t^2)^{(k-1)(\frak g-1)}\,
\sum_{\lambda_1\geq\lambda_2\geq\dots\geq\lambda_{k-1}}
{\mathcal N}^{2-2{\frak g}}_{\lambda_1,\dots,\lambda_{k-1},0}\;
\tau^{-(2{\frak g}-2)\sum_{i=1}^{k-1}(2i-k-1)\lambda_i}=\\ 
&=&(1-\t^2)^{(k-1)(\frak g-1)}\,
\sum_{\eta_1,\eta_2,\dots,\eta_{k-1}=0}^\infty
{\mathcal N}^{2-2{\frak g}}_{\eta_1,...,\eta_{k-1}}\;
\tau^{(2{\frak g}-2)\sum_{i=1}^{k-1}(k-i)\,i\,\eta_i}\,,\nonumber
\ee where $\lambda_i=\sum_{j=1}^{k-i}\eta_{k-j}$.
In the large $k$ limit terms with non-zero $\eta_i$  vanish since we always assume $|\t|\ll1$.
Thus, the only contribution to the sum at leading order for   large $k$  is from the term with 
{\textit{all}} $\eta_i=0$,
\be 
{\mathcal I}^{(k\to\infty)}_{\frak g}&=&
\lim_{k\to\infty}\;(1-\t^2)^{(k-1)(\frak g-1)}{\mathcal N}^{2-2{\frak g}}_{\lambda_1=0,\dots,\lambda_{k-1}=0,0}=\\
&=&\prod_{j=2}^\infty(1-\t^{2j})^{{\frak g}-1}=
PE\left[-({\frak g}-1)\frac{\t^4}{1-\t^2}\right]\,.\nonumber
\ee
The same logic applies also to the large $k$ limit of the $T_k$ theories:
 the singlet is the only term contributing to the index at leading order.
 
\

\section{The unrefined HL index of $T_4$}
\label{T4app}
Using the conjecture of section~\ref{conjsec} we can write an explicit expression for the unrefined
index of the $T_4$ theory. We find
\be
{\mathcal I}_{T_4}=\frac{(1-\t^4)(1-\t^6)(1-\t^8)}{(1-\t^2)^{42}} 
\,
\sum_{\lambda_1\geq\lambda_2\geq\lambda_3\geq0}
\frac{\left(P^{HL}_{\lambda_1,\lambda_2,\lambda_3}(1,1,1,1|\beta)\right)^3}{P^{HL}_{\lambda_1,\lambda_2,\lambda_3}(\t^3,\t,\t^{-1},\t^{-3}|\beta)}\,.
\ee The sum over the representation can be explicitly evaluated to give 
\be
{\mathcal I}_{T_4}=\frac{1-\t}{(1-\t^2)^{13}(1-\t^3)^{17}(1-\t^4)^{13}} \,{\mathcal P}_{86}(\t)\, ,
\ee where ${\mathcal P}_{86}(\t)$ is a polyndromic polynomial of degree $86$ in $\t$ with  coefficients 
given in table~\ref{coefsT4}. The degree of the singularity when $\t\to1$ has a physical meaning:
since the Hall-Littlewood index computes the Hilbert series of the Higgs branch this is the 
complex dimension of the Higgs branch. For $T_4$ the HL index predicts the dimension to be $42$, in agreement with~\cite{Gaiotto:2009gz,Benini:2009gi}.
\begin{table}
\begin{centering}
\begin{tabular}{|l|l|l|l|}
\hline 
1&$\t$& 33$\t^{2}$&144$\t^3$\tabularnewline
873$\t^{4}$& 4169$\t^{5}$&19486$\t^{6}$&80693$\t^{7}$\tabularnewline
319237$\t^{8}$&1165632$\t^{9}$&4024927$\t^{10}$&13054735$\t^{11}$\tabularnewline
40137244$\t^{12}$&116876141$\t^{13}$&323853313$\t^{14}$&854555364$\t^{15}$\tabularnewline
2153519932$\t^{16}$&5188980328$\t^{17}$&11978372385$\t^{18}$&26521974729$\t^{19}$\tabularnewline
56409853881$\t^{20}$&115373040784$\t^{21}$&227178289971$\t^{22}$&431064583235$\t^{23}$\tabularnewline
788945072797$\t^{24}$&1393870863434$\t^{25}$&2379094134408$\t^{26}$&3925581861006$\t^{27}$\tabularnewline
6265884973841$\t^{28}$&9680331918067$\t^{29}$&14483072164070$\t^{30}$&20994033528147$\t^{31}$\tabularnewline
29497595795349$\t^{32}$&40188148151858$\t^{33}$&53110900086737$\t^{34}$&68104402838959$\t^{35}$\tabularnewline
84760383950971$\t^{36}$&102408879854636$\t^{37}$&120143187852325$\t^{38}$&136883008184825$\t^{39}$\tabularnewline
151478220483799$\t^{40}$&162834262989902$\t^{41}$&170047651342244$\t^{42}$&172521386089030$\t^{43}$\tabularnewline
\hline
\end{tabular}
\par\end{centering}
\caption{\label{coefsT4} The coefficients of ${\mathcal P}_{86}(\t)$. The coefficient of $\t^{86-k}$ is equal to the coefficient of
$\t^k$\,.}
\end{table}

\

\section{Proof of the $SU(2)$ Schur index identity}\label{su2proofsec}

In this appendix we prove the basic $SU(2)$ Schur index identity~\eqref{qIndSU2},
\be\label{qIndSU2b}
&&\frac{PE\left[\frac{q^{1/2}}{1-q}(a_1+\frac1{a_1})(a_2+\frac1{a_2})(a_3+\frac1{a_3})\right]_{a_i,q}}
{(q;q)^{3}(q^2;q)\prod_{i=1}^3 PE\left[\frac{q}{1-q}(a_i^2+a_i^{-2}+2)\right]_{a_i,q}}=
\sum_{\lambda=0}^\infty\frac{\prod_{i=1}^3\chi_{\lambda}(a_i,a_i^{-1})}
{\chi_{\lambda}(q^{\frac{1}{2}},q^{-\frac{1}{2}})}\,.
\ee 
The strategy is to study the analytic properties of this expression and show that the left- and right-handed sides  have the
same poles and residues. Let us first define
\be
x=\frac{a_1}{a_2a_3},\quad y=\frac{a_2}{a_1a_3},\quad z=\frac{a_3}{a_2a_1},\quad u=a_1a_2a_3,\quad x y z u=1\,,
\ee where $a_i$ are $SU(2)$ fugacities. We also define
\be
(a)\equiv(a;\;q)_\infty\equiv\prod_{i=0}^\infty(1-aq^i)\,.
\ee
We will use square brackets $[ \; ]$ to denote ordinary brackets (that delimit expressions).
Then, using
\be
\sum_{n=1}^\infty \frac{x^n}{n}=-\log(1-x)\,,
\ee
the LHS of~\eqref{qIndSU2b} is given by
\be
LHS=\frac{[1-q](q)^{2}(qxy)(qxz)(qxu)(qyz)(qyu)(qzu)}{(q^{1/2}x)(q^{1/2}/x)(q^{1/2}y)(q^{1/2}/y)(q^{1/2}z)(q^{1/2}/z)(q^{1/2}u)(q^{1/2}/u)}\,.
\ee Let us study the analytic properties of this expression as a function of $x$ (the expression is symmetric in $x,y,z,u$).
We have poles whenever $x=q^{1/2-l}$ with integer $l$ (positive, zero or negative). At $x\to 0,\,\infty$ we have accumulation of poles.
Let us for concreteness compute the residue with positive
$l$ 
\be
Res_{LHS}=\frac{[1-q](q)^{2}(q^{3/2-l}y)(q^{3/2-l}z)(q/(yz))(qyz)(q^{1/2+l}/z)(q^{1/2+l}/y)}
{(q^{1-l})'(q^{l})(q^{1/2}y)(q^{1/2}/y)(q^{1/2}z)(q^{1/2}/z)(q^{l}/(yz))(q^{1-l}yz)}\,.
\ee Here $(q^{1-l})'$ is $(q^{1/2}x)$ evaluated at ${x=q^{1/2-l}}$ with the  vanishing
factor removed. Now we have
\be
&&\frac{(q^{1/2-l+1}y)(q^{1/2+l}/y)}{(q^{1/2}y)(q^{1/2}/y)}=
\frac{[-y]^{l}}{q^{l^2/2}}\frac{1}{1-q^{1/2-l}y}\,.
\ee From here we get
\be
Res_{LHS}&=&\frac{[1-q](q)^{2} [yz]^l\prod_{i=0}^{l-2}(1-q^{1+i}/(yz))}
{(q^{1-l})'(q^{l})q^{l^2}[1-q^{1/2-l}y][1-q^{1/2-l}z]\prod_{i=0}^{l-1}(1-q^{-i}yz)} 
=
\frac{q^{-1/2}-q^{1/2}}
{A}\,,
\ee where
\be
A=x-\frac{1}{x}+y-\frac{1}{y}+z-\frac{1}{z}+u-\frac{1}{u}\,.
\ee
Let us now look on the RHS of~\eqref{qIndSU2b}, which can be written as
\be
RHS=\frac{q^{-1/2}-q^{1/2}}{A}\sum_{n=1}^\infty\frac{q^{n/2}}{1-q^n}(x^n-\frac{1}{x^n}+y^n-\frac{1}{y^n}+z^n-\frac{1}{z^n}+u^n-\frac{1}{u^n})\,.
\ee We again want to compute residues in $x$. To see the poles we write
\be\label{scidaux}
\sum_{i=1}^\infty\frac{q^{n/2}}{1-q^n}x^n =\sum_{i=0}^\infty\sum_{n=1}^\infty q^{n(1/2+i)}\,x^n=
\sum_{i=0}^\infty \frac{q^{1/2+i}\,x}{1-q^{1/2+i}\,x}\,.
\ee Thus again the poles are at $x=q^{1/2-l}$ for any integer $l$ (we have also  same expression as~\eqref{scidaux} with $x\to 1/x$).
The residue here is easily computed to give
\be
Res_{RHS}= \frac{q^{-1/2}-q^{1/2}}{A}\,.
\ee
All in all, the LHS and RHS have the same poles and residues.

\newpage

\bibliography{sdualityMAC}
\bibliographystyle{JHEP}

\end{document}